\documentclass{jfm}
\usepackage{amsfonts,amsmath,amssymb}
\usepackage{color}
\usepackage{multiJFMequations}
\usepackage{accents}

\usepackage[normalem]{ulem}

\usepackage[outerbars]{changebar}
\newcommand*{\red}[1]{\textcolor{black}{#1}}
\newcommand*{\blue}[1]{\textcolor{black}{#1}}

\newcommand*{\be}{\begin{equation}}
\newcommand*{\ee}{\end{equation}}
\newcommand*{\bse}{\begin{subequations}}
\newcommand*{\ese}{\end{subequations}}
\newcommand*{\bme}{\begin{multiequations}}
\newcommand*{\eme}{\end{multiequations}}
\newcommand*{\se}{\singleequation}

\newcommand*{\te}{\tripleequation}

\newcommand*{\Xint}[1]{\mathchoice
{\XXint\displaystyle\textstyle{#1}}%
{\XXint\textstyle\scriptstyle{#1}}%
{\XXint\scriptstyle\scriptscriptstyle{#1}}%
{\XXint\scriptscriptstyle\scriptscriptstyle{#1}}%
\!\int}
\newcommand*{\XXint}[3]{{\setbox0=\hbox{$#1{#2#3}{\int}$}
\vcenter{\hbox{$#2#3$}}\kern-.5\wd0}}

\newcommand*\dashint{\Xint-}

\newcommand*{\XXnotinfty}[3]{{\setbox0=\hbox{$#1{#2#3}{\to}$}
\vcenter{\hbox{$#2#3$}}\kern-.5\wd0}}

\renewcommand*\parallel{/\mbox{\hspace{-1 mm}}/}

\newcommand*{\ds}{\displaystyle}
\providecommand*{\dfrac}[2]{\ds\frac{#1}{#2}}

\renewcommand*{\Re}{\mbox{Re}}
\renewcommand*{\Im}{\mbox{Im}}
\renewcommand*{\tilde}{\widetilde}
\renewcommand*{\hat}{\widehat}
\renewcommand*{\bar}{\overline}

\newcommand*\bfcdot{\bm{\cdot}}

\renewcommand*{\sp}[2]{{#1}\bfcdot{#2}}
\newcommand*{\vp}[2]{{#1}\times{#2}}
\newcommand*{\od}[2]{\dfrac{{\mathrm d}{#1}}{{\mathrm d}{#2}}}
\newcommand*{\pd}[2]{\dfrac{\partial{#1}}{\partial{#2}}}

\renewcommand*{\epsilon}{\varepsilon}
\renewcommand*{\Lambda}{\varLambda}
\renewcommand*{\Upsilon}{\varUpsilon}
\renewcommand*{\Phi}{\varPhi}
\renewcommand*{\Psi}{\varPsi}
\renewcommand*{\Omega}{\varOmega}
\renewcommand*{\Theta}{\varTheta}

\newcommand*{\bm}{\boldsymbol}
\newcommand*{\Omegav}{\bm{\Omega}}

\newcommand*{\dR}{{\mathrm d}}\newcommand*{\eR}{{\mathrm e}}\newcommand*{\iR}{{\mathrm i}}\newcommand*{\nR}{{\mathrm n}}\newcommand*{\oR}{{\mathrm o}}\newcommand*{\tR}{{\mathrm t}}\newcommand*{\uR}{{\mathrm u}}
\newcommand*{\IR}{{\mathrm I}}\newcommand*{\KR}{{\mathrm K}}\newcommand*{\OR}{O}
\newcommand*{\PR}{{\mathrm P}}
\newcommand*{\nv}{{\bm{n}}}\newcommand*{\uv}{{\bm{u}}}\newcommand*{\vv}{{\bm{v}}}\newcommand*{\xv}{{\bm{x}}}

\newcommand*{\aS}{{\sf{a}}}\newcommand*{\wS}{{\sf{w}}}\newcommand*{\xS}{{\sf{x}}}
\newcommand*{\DS}{{\sf{D}}}\newcommand*{\JS}{{\sf{J}}}\newcommand*{\KS}{{\sf{K}}}\newcommand*{\LS}{{\sf{L}}}\newcommand*{\WS}{{\sf{W}}}\newcommand*{\ZS}{{\sf{Z}}}
\newcommand*{\CC}{{\mathcal C}}\newcommand*{\DC}{{\mathcal D}}\newcommand*{\FC}{{\mathcal F}}\newcommand*{\GC}{{\mathcal G}}\newcommand*{\HC}{{\mathcal H}}\newcommand*{\PC}{{\mathcal P}}\newcommand*{\SC}{{\mathcal S}}\newcommand*{\WC}{{\mathcal W}}

\newcommand*{\rB}{{\breve r}}\newcommand*{\sB}{{\breve s}}
\newcommand*{\stwoB}{{\sB^2}}\newcommand*{\sthreeB}{{\sB^3}}
\newcommand*{\sminusoneB}{{\sB^{\,-1}}}
\newcommand*{\uB}{{\breve u}}\newcommand*{\vB}{{\breve v}}\newcommand*{\wB}{{\breve w}}
\newcommand*{\xB}{{\breve x}}\newcommand*{\yB}{{\breve y}}\newcommand*{\zB}{{\breve z}}
\newcommand*{\xtwoB}{{\xB^2}}\newcommand*{\xfourB}{{\xB^4}}
\newcommand*{\ztwoB}{{\zB^2}}\newcommand*{\zthirdB}{{\zB^{1/3}}}

\newcommand*{\phiB}{{\breve \phi}}\newcommand*{\psiB}{{\breve \psi}}\newcommand*{\omegaB}{{\breve \omega}}
\newcommand*{\omegaBC}{{\omegaB_\CC}}\newcommand*{\omegaBCP}{{\omegaB_\CC^{\,\prime}}}\newcommand*{\omegaBCzero}{{\omegaB_{\CC 0}}}
\newcommand*{\xvB}{{\breve \xv}}\newcommand*{\uvB}{{\breve \uv}}\newcommand*{\vvB}{{\breve \vv}} 

\newcommand*{\kT}{{\tilde k}}\newcommand*{\xT}{{\tilde x}}\newcommand*{\zT}{{\tilde z}}
\newcommand*{\deltap}{{\delta_+}}\newcommand*{\deltam}{{\delta_-}}\newcommand*{\deltapm}{{\delta_\pm}}\newcommand*{\deltas}{{\delta_s}}\newcommand*{\deltaA}{{\delta_\Phi}}

\newcommand*{\xs}{{\xS}}

\newcommand*{\inC}{{\iR\nR}}
\newcommand*{\outC}{{\oR\uR\tR}}
  \newcommand*{\stary}{\setbox0=\hbox{$\star$}%
  \raise.3em\box0}

\begin{document}

\newtheorem{lemma}{Lemma}
\newtheorem{corollary}{Corollary}

\shorttitle{On the Equatorial Ekman Layer} 
\shortauthor{F. Marcotte et al} 

\title{\red{On} the Equatorial Ekman Layer}
\author
 {
 Florence Marcotte\aff{1}
  Emmanuel Dormy\aff{1}{\footnote{Current address: \red{D\'epartement de Math\'ematiques et Applications,
CNRS UMR-8553, \'Ecole Normale Sup\'erieure, 45 rue d'Ulm, 75005, Paris, France.}}}
  \and 
  Andrew Soward\aff{2}  
  \corresp{\email{andrew.soward@ncl.ac.uk}}
  }
\affiliation
{
\aff{1}
MAG (ENS/IPGP), LRA, D\'epartement de Physique, Ecole Normale Sup\'erieure, 24, rue Lhomond, F-75231 Paris Cedex 05, France
\aff{2}
School of Mathematics and Statistics, Newcastle University, Newcastle upon Tyne, NE1 7RU, UK
}

%
\maketitle

\begin{abstract}
The steady incompressible viscous flow in the wide gap between spheres rotating rapidly about a common axis at slightly different rates (small Rossby number) has a long and celebrated history. The problem is relevant to the dynamics of geophysical and planetary core flows, for which, in the case of electrically conducting fluids, the possible operation of a dynamo is of considerable interest. A comprehensive asymptotic study, in the small Ekman number limit $E\ll 1$, was undertaken by Stewartson (J.~Fluid Mech.~1966, vol.~26, pp.~131-144). The mainstream flow, exterior to the $E^{1/2}$  Ekman layers on the inner/outer boundaries and the shear layer on the inner sphere tangent cylinder $\CC$, is geostrophic. Stewartson identified a complicated nested layer structure on $\CC$, which comprises relatively thick quasi-geostrophic $E^{2/7}$ (inside $\CC$) and $E^{1/4}$ (outside $\CC$) layers. They embed a thinner \red{ageostrophic} $E^{1/3}$ shear layer (on $\CC$), which merges with the inner sphere Ekman layer to form the $E^{2/5}$ equatorial Ekman layer of axial length $E^{1/5}$. Under appropriate scaling, this $E^{2/5}$--layer problem may be formulated, correct to leading order, independent of $E$. \red{Then} the Ekman boundary layer and ageostrophic shear layer become features of the far-field (as identified by the large value of the scaled axial co-ordinate $z$) solution. We present a numerical solution \red{of the previously unsolved equatorial Ekman layer problem using} a non-local integral boundary condition at finite $z$ to account for the far-field behaviour. Adopting $z^{-1}$ as a small parameter we extend Stewartson's similarity solution for the ageostrophic shear layer to higher orders. This far-field solution agrees well with that obtained from our numerical model.

\end{abstract}


 \section{Introduction\label{Introduction}}

In rapidly rotating geophysical and astrophysical fluid flows, boundary layers play an important role in determining the nature of the mainstream forming the bulk of the flow exterior to them. When a viscous fluid with constant density $\rho^\star$ and kinematic viscosity $\nu$, moves steadily with velocity $\uv^\star$ in a frame rotating with constant angular velocity $\Omegav$, the primary geostrophic force balance is $2\vp{\Omegav}{\uv^\star}=-\bm\nabla^\star \bigl(p^\star/\rho^\star\bigr)$, where $p^\star$ is the pressure, ensuring that the flow is independent of the axial coordinate $z^\star$. In a confined region, length scale $L$, this geostrophy holds almost everywhere except in various layers whose thickness tends to zero in concert with the Ekman number 
\bme
\label{Ek-num}
\be
E\,=\,\bigl(\delta^\star\big/L\bigr)^2\,,\qquad\qquad \delta^\star\,=\,\sqrt{\nu/\Omega}\,,
\ee    
\eme
where $\Omega=|\Omegav|$, which quantifies the relative magnitudes of the viscous and Coriolis forces. In the sense described, motion in the mainstream, i.e., outside the boundary layers, is 2-dimensional but is otherwise an arbitrary function of the coordinate vector $\xv_{\!\perp}^\star$ in the plane perpendicular to the rotation vector $\Omegav$. This geostrophic degeneracy is partially resolved by the geometry of the physical system, because columns of fluid must move without their length changing. In a confined container geostrophic motion can only follow the geostrophic contours \citep[see][]{G68}. However this does not determine the amplitude of this geostrophic velocity, which is finally resolved by consideration of the Ekman jump conditions across the Ekman layers adjacent to the boundaries.

\red{Geostrophic degeneracy and Ekman layers are well known in both geophysical \citep[see, e.g.,][]{P79} and planetary/stellar dynamo \citep[see, e.g.,][]{DS07} applications. In the planetary context, the confined flow inside a shell with concentric inner and outer rigid boundaries of radii $r^\star_i=L$ and $r^\star_o=\alpha L$  ($\alpha>1$) respectively is particularly relevant to the investigation of the dynamics of fluid cores. In that configuration, geostrophic motion is azimuthal independent of the axial coordinate $z^\star$ and lies on circular cylinders $s^\star\!=\;$const., where $s^\star$ is the distance from the rotation axis, but is otherwise an arbitrary function of $s^\star$ (geostrophic degeneracy).} 

\red{The classic \cite{P56} problem (see \S\ref{Proud-prob} below) illustrates many of the fundamental processes involved. It concerns a slightly differentially rotating shell, for which the inner and outer shells rotate with angular velocities $\Omegav^\star_i=\Omegav$ and $\Omegav^\star_o=(1+\epsilon)\Omegav$ respectively ($|\epsilon|\ll 1$). The Ekman layers, which resolve the degeneracy, become singular near the equator and lead to serious mathematical difficulties \citep[considered, e.g., by][]{R07}. The equatorial singularity spawns a free shear layer on the cylinder $\CC:\,s^\star\!=L$ tangent to the inner sphere at its equator with a very complicated nested sublayer structure investigated in detail by \cite{S66} (see \S\ref{Stew-prob} below). Our objective here is to reappraise Stewartson's work and to provide a solution to the equatorial Ekman layer problem.}

\red{The geometry and boundary conditions of the \cite{P56} problem provide a special case (in parameter space)  of spherical Couette flow particularly relevant to the dynamics of planetary fluid cores. Generally such geo- (also astro-) physical applications involve other ingredients including time ($t^\star$) dependant motions (particularly various waves) and body forces which result from buoyancy and, in MHD, Lorentz forces \citep[see, e.g.,][]{G14}. So, whenever the system contains a solid inner core, the various nested Stewartson-layers occurring on and about the tangent cylinder $\CC$ play a central role in fluid core dynamics. In addition to the analytical and many numerical studies of the tangent cylinder shear layers, there has been considerable experimental interest \citep[see, e.g.,][]{Aetal03}. Much of the geodynamo modellers' concern is with finite Reynolds number flows. The issue of the stability of the entire free shear layer encompassing the tangent cylinder has motivated extensive work by \cite{H03,Hetal04} amongst others.}

\red{A topic, that has attracted much interest in the planetary and stellar context, is inertial waves in spherical shells governed by $\partial\uv^\star\big/\partial t^\star+2\vp{\Omegav}{\uv^\star}=-\bm\nabla^\star \bigl(p^\star/\rho^\star\bigr)$.  When normal modes of a particular frequency are sought, the spatial structure of the mode is governed by a hyperbolic (wave) equation with two families of straight line characteristics, each of which are inclined to the geostrophic cylinders at equal and opposite angles. As the frequency of the sought mode decreases, the angle of inclination decreases in concert and vanishes in the steady geostrophic flow limit, when the two families degenerate into a unique one. The problem of solving hyperbolic equations subject to elliptic boundary conditions is not well posed and their solution in a shell geometry leads to serious difficulties \citep[see][]{SR70}. Put simply, a disturbance following a characteristic that meets the boundary is reflected along a member of the other family of characteristics. Therefore a characteristic tangent to the inner sphere has a special significance: those on one side are reflected, while those on the other pass by unimpeded. \cite{RS63} demonstrated that this grazing contact with the inner sphere at mid-latitudes leads to a singularity which excites a boundary layer in a region with radial and lateral extents comparable to that for the steady flow equatorial Ekman layer. The off-equator location of the singularity in the time-dependent case leads to the added complexity of an erupted disturbance along the other family of characteristics \citep[see figure~1 of][for a detailed description of the ensuing shear layer geometry]{K95}. In view of the importance of inertial waves in spherical shells, the problem has been studied extensively both analytically, numerically and experimentally \citep[see][and references therein]{Ketal13,LBetal15}.}

\red{The tangent cylinder also plays a crucial role in non-axisymmetric convection driven by buoyancy \cite[see, e.g.][]{Detal04} and prescribed body forces \cite[see, e.g.,][]{HP93,LH12}, their motivation being to understand the role of a non-axisymmetric Lorentz force in dynamo problems. These applications build on ideas that pertain to our shear layer theme including the notion of a Taylor state \citep[][]{T63}, which demands that the integral of the azimuthal component of the Lorentz force over each geostrophic cylinder $s^\star\!=\;$const.~vanishes. This is an area of considerable ongoing research \citep[for a recent review, see][]{RK13}.}

\red{Though we have focused on the singularities on the inner sphere boundary, we should not overlook the fact that they also occur on the outer sphere boundary. In the case of the steady flows, there is  a major difference between the inner and outer sphere equatorial Ekman layers. In the former the tangent cylinder is inside the fluid and the shear layer lies on it, while in the latter the tangent cylinder is entirely outside so eliminating the possibility of any free shear layer. \cite{P71} extended Stewartson's analysis to the study of the latter outer sphere equatorial Ekman layer and provided some numerical results. The outer sphere equatorial Ekman layer is of particular interest to oceanographers in connection to ocean currents driven by surface wind stresses \citep[see, e.g.][]{G71}. An analytic study of that configuration was undertaken by \cite{D72}.}

\red{Historically, because of its relative simplicity, the flow driven by a rotating disc (rotation axis parallel to $\Omegav$) has received much attention with early experiments performed by \cite{HT67} \citep[see also the split disc configuration discussed by][for recent experimental results on stability]{S57,Vetal15}. \cite{vdV93} has studied numerically the flow near the edge of a disc in an unbounded fluid. His results exhibit remarkable qualitative similarities to our equatorial Ekman layer findings.}

\red{As our present study builds on the results obtained by \cite{P56} and \cite{S66}, we summarise them briefly in the following two subsections.}

\subsection{The Proudman problem \label{Proud-prob}}

\red{We describe the Proudman problem relative to a frame rotating with the angular velocity $\Omegav$ of the inner sphere. We measure distance and angular velocity in units of the inner sphere radius $L$ and the relative outer sphere angular velocity $\epsilon \Omega$} and so write $\xv^\star=L\xvB$ and $\uv^\star=\epsilon L \Omega\uvB$.\red{\footnote{\red{Dimensional variables are distinguished by a{$\,\,\stary\,$}. However, as our primary goal is an equatorial Ekman layer study for which we nondimensionalise on different units without accents, we adopt in our preliminary nondimensionalisation here the unobtrusive breve accent $\breve{\,\,}\,$ to avoid ambiguity later.}}} Then, relative to cylindrical polar coordinates $\sB,\,\phiB,\,\zB$ $\bigl(\rB=\sqrt{\stwoB+\ztwoB}\,\bigr)$ in the \red{rotating frame}, the axisymmetric flow velocity $\vvB\equiv\bigl(\uB,\,\vB,\,\wB)$ may be expressed in terms of the azimuthal angular velocity $\omegaB=\vB/\sB$ and meridional streamfunction $\psiB$ as
\be
\label{velocity}
{\vvB}\,=\biggl(-\,\dfrac1{{\sB}}\pd{\psiB}{\zB},\,{\sB}{\omegaB},\,\dfrac1{\sB}\pd{\psiB}{\sB}\biggr).
\ee
Motion is governed by the \red{linearised ($|\epsilon|\ll 1$)} azimuthal components of the vorticity and momentum equations
\bme
\label{mom-vort}
\be
2\pd{\vB}{\zB}\,=\,E\,\DS^2\biggl(\dfrac{\psiB}{\sB}\biggr),\qquad\qquad
-\,\dfrac2{\sB}\pd{\psiB}{\zB}\,=\,E\,\DS \vB \,
\ee
respectively, where
\be\se
\DS\,=\,\pd{^2\,}{\stwoB}\,+\,\dfrac1{\sB}\pd{\,}{\sB}\,-\,\dfrac1{\stwoB}\,+\,\pd{^2\,}{\ztwoB}\,.
\ee
\eme

The tangent cylinder $\CC$: $\sB=1$ (tangent to the inner sphere at its equator) divides the flow  up into two regions $\DC_\inC$: $\sB< 1$ and $\DC_\outC$: $\sB> 1$. \red{In view of the importance of $\CC$, we find it convenient to define distance from it by
\be
\label{stox}
\xB\,=\,\sB-1\,.
\ee
}
The flow has the symmetries
\be
\label{symmetries}
\psiB\bigl(\sB\,,-\zB\bigr)\,=\,-\,\psiB\bigl(\sB\,,\zB\bigr)\,,\qquad\qquad
\vB\bigl(\sB\,,-\zB\bigr)\,=\,\vB\bigl(\sB\,,\zB\bigr)\,.
\ee
So without loss of generality we may restrict attention to the half-shell $1<\rB<\alpha$, $\zB> 0$, provided that in $\DC_\outC$ we apply the equatorial symmetry conditions 
\be
\label{symmetries-EP}
\psiB\,=\,0\,,\qquad\qquad \pd{\vB}{\zB}\,=\,0\qquad \qquad \mbox{on}\qquad \zB\,=\,0\,.
\ee

The flow domain in the small Ekman number limit, 
\be
\label{E-small}
E\,\,\ll\,1 \,,
\ee
is divided up into various regions. The  essential partition is between the mainstream, where viscous effects can be ignored and the flow is geostrophic, and the various boundary and free shear layers, where viscosity plays a direct role in the dynamics. However, the Ekman boundary layers provide the key to the solutions both in the mainstream and in the free shear layers that reside on the tangent cylinder. They are located adjacent to the inner ($\SC_-$) and outer ($\SC_+$) spherical boundaries with unit normals $\nv_-$ and $\nv_+$ respectively, directed into the fluid region, of thickness
\be
\label{ELwidth}
\deltapm\,=\,\bigl(\delta^\star\!\big/ L\bigr)\bigl(|\Omegav|\big/|\sp{\nv_\pm}{\Omegav}|\bigr)^{1/2}\,,
\ee
i.e.,
\vskip-6mm
\bme
\label{delta-z}
\begin{align}
\deltam\,&=\,\bigl(E\big/\zB_-\bigr)^{1/2},  & \deltap\,&=\,\bigl(E\alpha\big/\zB_+\bigr)^{1/2}\,,\\[-1.0em]
\intertext{where\vskip-2mm}
\zB_-\,&=\,\sqrt{1-\stwoB}\,, & \zB_+\,&=\,\sqrt{\alpha^2-\stwoB}\,.
\end{align}
\eme
\cite{P56} showed that the Ekman jump conditions across them determine
\bme
\label{psi-pm}
\be
\psiB_-  \,=\,\tfrac12 \deltam \sB \vB_-\,,
\qquad\qquad
\psiB_+  \,=\,\tfrac12 \deltap \sB (\sB-\vB_+)\,,
\ee
\eme
where $\psiB_\pm$ and $\vB_\pm$ are the mainstream values taken on leaving the Ekman layers, i.e., $\,\,\,\SC_-$: $(\rB-1)/\deltam\to\infty$; $\,\,\,\SC_+$: $(\alpha -\rB)/\deltap\to\infty$.

The mainstream geostrophic flow, that satisfies $\vB(\sB)=\vB_\pm(\sB)$ and $\psiB(\sB)=\psiB_\pm(\sB)$ (see (\ref{psi-pm}$a$,$b$)), is
\bse
\label{Proud-sol}
\begin{align}
\omegaB\,=\,{\vB}\big/{\sB}\,=\,&\left\{\begin{array}{lll}
\deltap\big/(\deltam\,+\,\deltap)\qquad\qquad&\mbox{in}\quad \DC_{\inC}\,,\qquad\\[0.2em]
1\qquad&\mbox{in}\quad\DC_{\outC}\,,
\end{array}\right.\\[0.2em]
{\psiB}\big/{\stwoB}\,=\,&\left\{\begin{array}{lll}
\tfrac12\,\deltap\deltam\big/(\deltam\,+\,\deltap)\qquad&\mbox{in}\quad \DC_{\inC}\,,\qquad\\[0.2em]
0 \qquad&\mbox{in}\quad\DC_{\outC}
\end{array}\right.
\end{align}
\ese
\citep[][eqs.~(3.17), (3.18)]{P56}. Motion is predominantly azimuthal $(0,\,\vB(\sB),\,0)$. The smaller axial velocity $\wB(\sB)=\sminusoneB\dR\psiB\big/\dR \sB=\OR\bigl(E^{1/2}\vB\bigr)$ is driven by suction (blowing) into (out of) the Ekman boundary layer on the outer $\SC_+$ (inner $\SC_-$) sphere, the equality of which determines the Proudman solution (\ref{Proud-sol}).

On the axis $\sB=0$, where the normals to both the inner and outer sphere are parallel $\big(\nv_-\parallel\nv_+$\,$\Rightarrow$ $\deltapm=E^{1/2}\big)$, the meridional flux  balance is achieved by $\omegaB=\tfrac12$. As $\sB$ increases from zero the direction cosine on the inner boundary decreases faster than on the outer boundary $\bigl(E^{1/2}<\deltap<\deltam\bigr)$ with the consequence that $\omegaB(\sB)$ decreases monotonically to zero at the equator $\sB=1$ (but read on). The value taken by $\deltam$ on the inner sphere close to the equator has serious implications. There, since
\bse
\label{delta-z-sing}
\be
\zB_-\,\approx\,\sqrt{-2\xB} \qquad\quad \bigl(\xB=\sB-1\bigr)\qquad\qquad\mbox{as}\qquad  \sB\,\uparrow \,1
\ee
(see (\ref{stox}) and (\ref{delta-z}$c$)), it follows from (\ref{delta-z}$a$) that $\deltam$ diverges as the tangent cylinder is approached:
\be
\deltam\big/E^{1/2}\,\approx\,(-2\xB)^{-1/4}
\qquad\qquad\qquad\qquad\;\;\mbox{as}\qquad  \sB\,\uparrow \,1\,.
\ee
\ese
Close to the equator, blowing from the Ekman boundary layer on the inner sphere becomes more effective and so the azimuthal geostrophic flow tends to co-rotate with the inner sphere in order to maintain the correct axial mass flux balance: 
\bme
\label{C-Proud}
\be
\omegaB\,\to\,0\,\qquad\qquad \psiB\,\to\,\psiB_{\PR}\qquad\qquad\mbox{as}\qquad  \sB\,\uparrow \,1\,,
\ee
where from (\ref{Ek-num}$a$), (\ref{delta-z}$b$,$d$) and (\ref{Proud-sol}$b$)
\be
\se
\psiB_{\PR}\,=\,\tfrac12\deltap\,=\,\tfrac12 E^{1/2}\alpha^{1/2}\!\big/\bigl(\alpha^2-1\bigr)^{1/4}.
\ee
\eme
Since $\psiB=0$ outside the tangent cylinder $\CC$, $2\pi\psiB_{\PR}$ is the total fluid (Proudman) flux returned from the outer to inner sphere Ekman layers on $\CC$. Curiously  $\dR \omegaB/\dR \sB\to -\infty$ as $\sB\uparrow 1$, but perhaps more significantly $\omegaB$ itself is discontinuous at $\CC$, across which it jumps from $\omegaB=0$ in $\DC_\inC$ to unity throughout $\DC_\outC$, where the fluid co-rotates with the outer sphere. The corresponding $\omegaB$--profile is illustrated in figure~3.7(a) of \cite{DS07}, albeit the figure is sketched essentially for $\epsilon<0$ so that an up-down reflection is needed in order to compare with the present work.

\subsection{The Stewartson problem \label{Stew-prob}}

Though the Proudman solution has an elegant simplicity, the singularities on $\CC$ must be removed by free shear layers \cite[for numerical results at various $E$ see, e.g.,][figure~4 (left panel); again up-down reflection]{DCJ98}, which were largely resolved by \cite{S66}. 

The azimuthal angular velocity $\omegaB$ and its gradient $\dR \omegaB/\dR \sB$ (but notably not $\psiB$) are rendered continuous across $\sB=1$ in quasi-geostrophic (QG) shear layers containing $\CC$, in which the predominant azimuthal velocity remains $\zB$-independent. However, whereas in the mainstream geostrophic degeneracy is resolved by Ekman suction alone, in the QG-region lateral friction, characterised by the term $\DS \vB$ on the right-hand side of the azimuthal momentum equation (\ref{mom-vort}$b$), is also involved. In $\DC_\outC$, the balance of Ekman suction and lateral friction leads to the well known $E^{1/4}$-layer. In $\DC_\inC$, because of the intense Ekman blowing near the equator, the corresponding layer is thinner and referred to as the $E^{2/7}$-layer. The upshot is that the velocity jump is largely accommodated by the thicker $E^{1/4}$-layer in $\DC_\outC$, so that the angular velocity is small,
\bse
\label{DeltaOmega}
\begin{align}
\bigl.\omegaB\bigr|_{\sB=1}\,\equiv\,&\,\omegaBC\,=\,\Delta\Omega\big/\Omega\qquad\qquad\qquad  \bigl(0<\Delta\Omega\ll\Omega\bigr),\\
\intertext{on the tangent cylinder $\CC$. The layers either side of $\CC$ then provide the estimates}
\qquad\biggl.\od{\omegaB}{\sB}\biggr|_{\sB=1}\,\equiv\,&\,\omegaBCP\,
=\left\{\begin{array}{ll}
\OR\bigl(E^{-2/7}\,\omegaBC\bigr) \qquad\,\,\,\,&\mbox{in}\quad \DC_{\inC}\,,\\[0.3em]
\OR\bigl(E^{-1/4}\bigr) \qquad\,\,\,\,&\mbox{in}\quad \DC_{\outC}\end{array}\right. 
\end{align}
for the size of the angular velocity gradient on $\CC$ with the consequence that
\be
\omegaBC\,=\,\OR\bigl(E^{1/28}\bigr)   
\qquad\qquad\Longleftrightarrow\qquad\qquad
\Delta\Omega\,=\,\OR\bigl(E^{1/28}\Omega\bigr).
\ee
\ese
Note that $\epsilon \Delta\Omega$, like $\epsilon \Omega$, is the dimensional angular velocity. As in the case of the Proudman solution, the corresponding $\omegaB$--profile across the shear layer is illustrated in figure~3.8(a) of \cite{DS07} again sketched for $\epsilon<0$.

The solution only becomes ageostrophic (AG, i.e.,~$\zB$-dependent) in a thinner $E^{1/3}$-layer spanning the tangent cylinder $\CC$. The prime purpose of \red{this} layer is to remove discontinuities on $\CC$ that have arisen because of the QG-approximations made so far. Stewartson's starting point was to note that the QG linear shear 
\bme
\label{lin-shear}
\be
\vB_G\,\approx\,\omegaB_G\,=\,\omegaBC\,+\,\omegaBCP\,\xB
\qquad\qquad\mbox{for}\qquad\quad \xB\,=\,\sB-1\,=\,\OR\bigl(E^{1/3}\bigr),
\ee
\eme
obtained from the QG-solution, together with $\psiB=\psiB_G=0$ solves the $E^{1/3}$ shear layer equations (\ref{mom-vort-Ethird}). He then sought an AG-correction $\omegaB_{\!A}$ and considered 
\bme
\label{approx-third}
\be
\omegaB\,=\,\omegaB_G(\sB)\,+\,\omegaB_{\!A}(\sB,\zB)\,,
\qquad\qquad  \psiB\,=\,\psiB_{\!A}(\sB,\zB)
\ee
\citep[][eq.~(6.16)]{S66}\,, where the contributions to $\omegaB-\omegaBC$ have relative sizes
\be
\se
\left.\begin{array}{rl}
\omegaBCP\,\xB\!\!&\!\!=\,\OR\bigl(E^{1/21}\,\omegaBC\bigr),\\[0.4em]
\omegaB_{\!A}\!\!&\!\!=\,\OR\bigl(E^{1/12}\,\omegaBC\bigr)
\end{array}\!\!\right\}
\qquad \mbox{for}\qquad
\left\{\begin{array}{rl}
\!\!\xB\!\!&\!\!=\,\OR\bigl(E^{1/3}\bigr),\\[0.4em]
\!\!\zB\!\!&\!\!=\,\OR\bigl(1\bigr)
\end{array}\right.
\ee
\eme
(\red{for each estimate} use \red{respectively} (\ref{DeltaOmega}$b$) and the result (\ref{Ethird-sol}$a$)). At lowest order the matching of the entire solution (\ref{approx-third}$a$,$b$) with the QG-solution outside, where $|\xB|\gg E^{1/3}$, takes care of itself, i.e., other than rather obvious boundedness conditions on $\omegaB_{\!A}$ and $\psiB_{\!A}$, we may ignore the demands of matching. Essentially, the AG-solution is simply driven by the inner sphere Ekman boundary layer via the condition (\ref{psi-pm}$a$), which in view of \red{(\ref{lin-shear}$a$) and the estimates} (\ref{approx-third}$c$) \red{simply} uses $\omegaB_-=\vB_-/\sB=\omegaBC$ to obtain
\be
\label{psi-pm-third}
\psiB_-\,=\,\tfrac12 \deltam  \,\omegaBC\,,
\ee
where $\deltam\propto (-\xB)^{-1/4}$ is defined by (\ref{delta-z-sing}$b$). This is the only non-zero boundary condition (see (\ref{mom-vort-Ethird-approx}$b$)) and it leads to Stewartson's solution (\ref{Ethird-sol}).

Significantly, whereas the estimate $\bigl|\omegaBC\bigr|\gg\bigl|\omegaBCP\,\xB\bigr|\gg\bigl|\omegaB_{\!A}\bigr|$ (see (\ref{approx-third}$c$)) holds in the bulk of the $E^{1/3}$-layer where $\zB=\OR(1)$, the magnitude of $\omegaB_{\!A}$ increases with decreasing $\zB$ in a sublayer near the equator due to the singular nature of the boundary condition (\ref{psi-pm-third}):
\be
\label{sl-qg--est}
\left.\begin{array}{rl}
\!\!\!\!\!\!\!\omegaBCP\,\xB\!\!&\!\!=\OR\bigl(E^{1/21}\zB^{1/3}\,\omegaBC\bigr),\\[0.4em]
\!\!\!\!\!\!\!\omegaB_{\!A}\!\!&\!\!=\OR\bigl(E^{1/12}\zB^{-5/12}\,\omegaBC\bigr)
\end{array}\!\!\!\right\} 
\qquad \mbox{for}\qquad
\left\{\begin{array}{rl}
\!\!\!\xB\!\!&\!\!=\,\OR\bigl((E\zB)^{1/3}\bigr),\\[0.2em]
\!\!\!\zB\!\!&\!\!\ll\,1\,.
\end{array}\right.
\ee
The relatively thin domain identified is an $E^{1/3}$-type sublayer based on the height $\zB$, rather than the tangent cylinder height $\OR(1)$, in which the solution has similarity form (\ref{Ethird-sol-similarity})\red{, hereafter referred to as the similarity sublayer. Inside it, the ratio 
\be
\label{linear-shear-estimate}
\bigl|\omegaBCP\,\xB\bigr|\,\bigl/\,{|\omegaB_{\!A}|}\,=\,\OR\bigl(\zB^{3/4}E^{1/28}\bigr)
\ee
(see (\ref{sl-qg--est})) decreases with $\zB$ becoming $\OR(1)$ at $\zB=\OR(E^{1/21})$.} 

\red{Whereas the similarity sublayer width $\OR\bigl((E\zB)^{1/3}\bigr)$ decreases in concert with $\zB$, the Ekman layer width $\deltam = (E/\zB)^{1/2}$ (see (\ref{delta-z}$a$)) increases. The solutions merge on the tangent cylinder when 
\be
\label{EEL-dimensions}
\left.\begin{array}{rl}
\omegaBCP\,\xB\!\!&\!\!=\,\OR\bigl(E^{4/35}\,\omegaBC\bigr),\\[0.4em]
\omegaB_{\!A}\!\!&\!\!=\,\OR\bigl(\omegaBC\bigr)
\end{array}\!\!\right\} 
\qquad \mbox{for}\qquad
\left\{\begin{array}{rl}
\!\!\xB\!\!&\!\!=\,\OR\bigl(E^{2/5}\bigr),\\[0.3em]
\!\!\zB\!\!&\!\!=\OR\bigl(E^{1/5}\bigr),
\end{array}\right.
\ee
namely the dimensions of the equatorial Ekman layer.}

\red{Significantly the ratio $\bigl|\omegaBCP\,\xB\bigr|\,\bigl/\,{|\omegaB_{\!A}|}$  (see (\ref{linear-shear-estimate})) continues to decrease within the similarity sublayer region over the range $\OR\bigl(E^{1/21}\bigr)\gg\zB\gg\OR\big(E^{1/5}\bigr)$, on which (\ref{sl-qg--est}) provides the estimate $\bigl|\omegaBC\bigr|\gg\bigl|\omegaB_{\!A}\bigr|\gg\bigl|\omegaBCP\,\xB\bigr|$. On reaching the the equatorial Ekman layer the estimate is modified to $\bigl|\omegaB_{\!A}\bigr|=\OR\bigl(\omegaBC\bigr)\gg\bigl|\omegaBCP\,\xB\bigr|$ and $\bigl|\omegaBCP\,\xB\bigr|\,\bigl/\,{|\omegaB_{\!A}|=\OR\bigl(E^{4/35}\bigr)}$  (see (\ref{EEL-dimensions})). There, the small size of the shear $\omegaBCP\,\xB$ permits us to neglect it in our leading order formulation of the equatorial Ekman layer problem in \S\ref{formulation}.}

\begin{figure}
\centerline{}
\vskip 5mm
\centerline{\hskip10mm\includegraphics[width=0.8\textwidth,clip=true,trim=5 100 190 5]{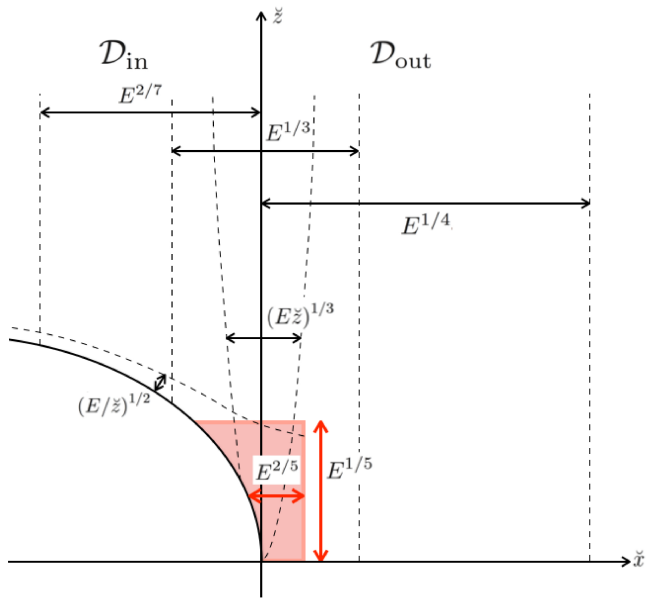}}
\caption{\label{schematic} \red{A schematic drawing identifying the the various nested boundary layers close to the inner sphere equator.} It is not to scale, but the order of magnitude of the layer thicknesses are indicated. Further the $\zB$-scale has been shrunk relative to the $\xB$-scale, which itself is short and whence the parabolic shape of the sphere boundary.}
\end{figure}

\red{Figure~1 of \cite{S66} provides a (quote) ``Schematic drawing (not to scale) of the intersection region of the Ekman layer near the inner sphere and the shear layer near $\CC\,$''.  We reproduce its content in our figure~\ref{schematic} ($E^{-1}$ is his $R$) but add the $(E\zB)^{1/3}$ similarity sublayer to emphasise how it is spawned by the equatorial Ekman layer. That detail alone is identified in figure~3.9 of \cite{DS07}.}

%

\subsection{Outline \label{Outline}}

The outline of our paper is as follows. 

In \S\ref{S-problem}  we carefully summarise and review the nature of Stewartson's nested free boundary layers on the tangent cylinder \red{identified in figure~\ref{schematic}}. That includes a description of \red{the QG-flow in} the $E^{1/4}$-layer (\S\ref{DoutSol}) together with new $E^{2/7}$-layer results (\S\S\ref{DinSol}, \ref{DNsol} and Appendix~\ref{Bessel}). The detailed survey of the AG-sublayer in \S\ref{Ag-shear-layer} leads\red{, via the similarity sublayer solution (\S\ref{Stew-sim-sol}), on} to the equatorial Ekman layer problem (\S\ref{EE-layer}) as \red{proposed} by \cite{S66} in his final \S7 entitled ``The terminal form of the Ekman layer on the inner sphere''. In \S\ref{formulation} we formulate the $E^{2/5}$-equatorial Ekman layer in terms of local units and set up the governing equations (\S\ref{gov-eq-bdry-condit}) relative to the frame rotating with angular velocity \blue{$\Omegav+\epsilon\Delta\Omegav$}, rather than \blue{$\Omegav$}, where \blue{$\Delta\Omegav=\,\omegaBC\,\Omegav$}. A non-local mainstream (including the \red{similarity sublayer} but not the Ekman layer on $\SC_-$)\red{\footnote{\red{``Mainstream'' is a natural description, which we adopt in the equatorial Ekman layer context, but  in truth is a misnomer as the region lies entirely within $E^{1/3}$-layer}}} top boundary condition at \red{$\zB=E^{1/5}H$, where $H$ is a moderately large $O(1)$ constant,} is developed in \S\ref{int-cond}. \red{It} proves very useful in the implementation of the numerical model described in \S\ref{num-model}. In \S\ref{sl-ff} and Appendix~\ref{appA} we extend Stewartson's far-field similarity solution \red{(see \S\ref{Stew-sim-sol})} to higher orders. The amplitude of each higher order correction is determined to the linked higher order corrections to the Ekman layer on $\SC_-$ considered in \S\ref{El-s} with results for each order reported in \S\S\ref{El-s0}, \ref{El-s1} and \ref{El-s2} respectively. In \S\ref{Numerical-results} we discuss our numerical results. We \red{stress} the connection of our numerics to the \S\ref{sl-ff} far-field similarity solutions (\S\ref{far-field}) and assess the importance of other \cite{MS69} similarity forms (\S\ref{MS-sols}). We also draw attention to the remarkable topological equivalence of the equatorial Ekman flow to that identified by \cite{vdV93} for his aforementioned rotating disc problem in \S\ref{rel-stud}. We finalise with a few concluding remarks in \S\ref{Conclusions}.

\red{To place our results in perspective, we must emphasise that Stewartson's expansions involve rather bizarre powers of $E$ such as $\omegaBC\,=\OR\bigl(E^{1/28}\bigr)$ (see (\ref{DeltaOmega}$c$)), which means that their usefulness is limited to extremely small $E$. The Ekman number in the case of the Earth's core, based on current estimates for the kinematic viscosities $\nu$ is about $10^{-14}$ for which $E^{1/28}\sim 1/3$ is hardly a small number! If  a turbulent value of $\nu$ is taken that estimate may rise as high as $E\sim 10^{-8}$, a value that recent geodynamo simulations are close to attaining \citep[see][for full DNS of planetary core flows at low Ekman number, $E=5\times10^{-7}$]{SR09}. In view of that caveat, we now expand on our objectives.}

\red{Our primary goal is the combined analytic and numerical solution of the equatorial Ekman layer problem (\S\S\ref{formulation}--\ref{El-s} and \S\ref{far-field}). For that, much of the comprehensive survey of all the free shear layers in \S\ref{S-problem} could be sensibly bypassed on a first read. However, the relevance and limitations of our local equatorial Ekman layer study to the results from the Direct Numerical Simulation (DNS) of the equations (\ref{mom-vort}$a$,$b$) governing motion in the entire spherical shell (\S\ref{more-DNS}) at small but finite $E$ (see, e.g., new results illustrated in figure~\ref{full_shell_num}) can only by understood and appreciated through a proper understanding of the \S\ref{S-problem} survey of the nested sublayers on the tangent cylinder.}

\red{The most important part of the shear layer flow is the dominant QG-contribution $\omegaB_G(\sB)$ (see (\ref{approx-third}$a$)) fixed by the solution of ordinary differential equations (ODE's), that govern it within the $E^{1/4}$ and $E^{2/7}$-layers. It is their length scale ratio that determines the tangent cylinder value $\omegaB_G(1)=\omegaBC\,=\OR\bigl(E^{1/28}\bigr)$, which fixes our scaling (non-dimensionalisation) of the equatorial Ekman layer problem based on} \blue{the angular velocity $\epsilon\Delta\Omega=\,\epsilon\omegaBC\,\Omega$, rather than $\epsilon\Omega$}. \red{Due to its importance, a secondary accomplishment has been our new $E^{2/7}$-layer results. A preliminary series solution, taken to three orders of magnitude in powers of $E^{1/28}$ (\S\ref{DinSol}), determines an expression for $\omegaBC$ correct to the same order of accuracy (see (\ref{sl-cont-0}$b$), (\ref{sl-sim-consts}) and (\ref{sl-deriv0}$a,b$)). Unfortunately the series converges slowly with decreasing $E$ achieving only two significant accuracy by $E$ as small as $10^{-14}$ (see table~\ref{table-1}). We bypass this asymptotic difficulty by directly solving the governing ODE numerically at various small fixed $E$ including the Earth-like value $10^{-14}$ (\S\ref{DNsol}). The \red{Direct Numerical (DN}) solution portrayed in figure~\ref{QGin} at small $E$ even as large as $10^{-7}$ is perfectly reliable because it gives good agreement with the DNS results for both $E=10^{-5}$ and $10^{-7}$ (see (\ref{DN-DNS})).}

\red{Our comparison in \S\ref{more-DNS} of our equatorial Ekman layer solution with full shell DNS at $E=10^{-7}$ sheds more light on the small $E$ issue. The meridional streamlines for each, illustrated respectively in figures~\ref{contours}($b$) and \ref{full_shell_num-blow_up}($b$), compare well. However, though inside the equatorial Ekman layer tolerable agreement between the contours of constant $\vB$ is visible (see figures~\ref{contours}($a$) and \ref{full_shell_num-blow_up}($a$)), as either $E^{-2/5}{\breve x}$ or $E^{-1/5}{\breve z}$ increases detailed comparison becomes less encouraging. In the light of our experience with the $E^{2/7}$-layer, we may reasonably expect good agreement when $E$ is really small taking Earth-like values of order $10^{-14}$ but that is well outside today's numerically accessable regime. The reason for the weak agreement at $E=10^{-7}$ may be traced to the fact that the QG-shear $\omegaBCP\,\xB$ in (\ref{lin-shear}$a$) is asymptotically small and so does not appear in our equatorial Ekman layer formulation. In reality the QG-shear $\omegaBCP\,\xB$ is not that small (see (\ref{EEL-dimensions})) but one could rectify its omission by reinstating} \red{this term as part of the boundary conditions. Then $E$ would appear explicitly in the problem, which would need to be solved at various small fixed $E$ just as we do for the $E^{2/7}$-layer. That ambitious project is outside the scope of our remit.}


 \section{The Stewartson problem\label{S-problem}}

Here we outline in more detail the nature of the QG and AG shear layers.

\subsection{$E^{2/7}$-- and $E^{1/4}$--quasi-geostrophic (QG) shear layers\label{Qg-shear-layer}}

Away from the Ekman layers, QG-flow exists in the region outside the AG shear layer, which contains the inner sphere tangent cylinder $\CC$. Here the flow velocity components $\vvB_{\!\perp}\equiv\bigl(\uB,\,\vB,\,0\bigr)$, in the plane perpendicular to $\zB$, are dependent on $\sB$ alone. In view of (\ref{velocity}) this implies that $\psiB$ is linear in $\zB$ and given by
\be
\label{psi-qg}
\psiB\,=\,\left\{\begin{array}{lll}
\bigl[\psiB_+\bigl(\zB-\zB_-\bigr)+\,\psiB_-\bigl(\zB_+-\zB\bigr)\bigr]\big/\bigl(\zB_+\,-\,\zB_-\bigr)\qquad&\mbox{in}\quad \DC_{\inC}\,,\\[0.4em]
\psiB_+\zB\big/\zB_+&\mbox{in}\quad \DC_{\outC}\,,
\end{array}\right.
\ee
where $\zB_\pm$ are defined \red{by (\ref{delta-z}$c$,$d$)}.

Now the azimuthal component of the momentum equation
\be
\label{azim-mom}
2 \uB\,\Bigl(=\,E\,\DS \vB\Bigr)\,=\,\dfrac{E}{\stwoB}\od{\,}{\sB}\biggl(\sthreeB\od{\,}{\sB}\biggl(\dfrac{\vB}{\sB}\biggr)\biggr)
\ee
(see (\ref{mom-vort}$b$)) integrated with respect to $\zB$ determines
\bse
\label{int-azim-mom}
\be
\dfrac{E}{\sB}\od{\,}{\sB}\biggl(\sthreeB\od{\,}{\sB}\biggl(\dfrac{\vB}{\sB}\biggr)\biggr)=\,\left\{\begin{array}{lll}
2\bigl(\psiB_-\,-\,\psiB_+\bigr)\big/\bigl(\zB_+\,-\,\zB_-\bigr)
\qquad&\mbox{in}\quad \DC_{\inC}\,,\\[0.4em]
-\,2\psiB_+\big/\zB_+\qquad&\mbox{in}\quad \DC_{\outC}\,.
\end{array}\right.
\ee
Substitution of $\psiB_\pm$ defined by (\ref{psi-pm}) into the right-hand side of (\ref{int-azim-mom}$a$) and noting that $\vB_\pm=\vB$ gives
\begin{align}
\dfrac{E}{\stwoB}\od{\,}{\sB}\biggl(\sthreeB\od{\,}{\sB}\biggl(\dfrac{\vB}{\sB}\biggr)\biggr)
=\left\{\begin{array}{lll}
\bigl[(\deltam\,+\,\deltap)\vB\,-\,\deltap \sB\bigl]\big/\bigl(\zB_+\,-\,\zB_-\bigr)
\qquad&\mbox{in}\quad \DC_{\inC}\,,\\[0.4em]
\deltap(\vB\,-\,\sB)\big/\zB_+
\qquad&\mbox{in}\quad \DC_{\outC}\,,
\end{array}\right.\label{sl-eq}
\end{align}
\ese
to be solved subject to $ \vB$ and $\dR \vB/\dR \sB$ continuous at $\sB=1$. 

In the mainstream, the left-hand side of (\ref{int-azim-mom}$b$) is smaller than its right-hand side by a factor $E^{1/2}$ (recall that $\deltapm=O(E^{1/2})$; \red{see (\ref{delta-z}$a$,$b$)}) so recovering the Proudman solution (\ref{Proud-sol}). The discontinuities of $\vB$ and $\dR \vB/\dR \sB$ at $\sB=1$ described by (\ref{Proud-sol}) are smoothed out across a thin QG shear layer as indicated in \S\ref{Stew-prob}. A key feature of these QG-layers is the remaining weak singular behaviour of $\vB=\sB\omegaB$:
\be
\label{signg-beh} 
\omegaB\,-\,\omegaB_G \,\approx \left\{\begin{array}{ll}
\dfrac{16}{21}\,\dfrac{\omegaBC(-\xB)^{7/4}}{E^{1/2}2^{1/4}\zB_+} \qquad&\mbox{in}\quad \DC_{\inC}\,,\\[0.8em]
0 \qquad&\mbox{in}\quad \DC_{\outC}
\end{array}\right\}\qquad\mbox{as} \qquad \sB\,\to \,1\,,
\ee
where $\omegaB_G=\omegaBC+\omegaBCP\,\xB$ (see (\ref{lin-shear}$a$)), that we identify in (\ref{sl-near-0}). The discontinuity of the  second derivative $\dR^2 \omegaB\big/\dR \stwoB$ at $\sB=1$, implied by (\ref{signg-beh}), is smoothed out in a thinner AG $E^{1/3}$-sublayer considered in \S\ref{Ag-shear-layer}. There matching of $\omegaB_{\!A}$ to $\omegaB-\omegaB_G$ defined by (\ref{signg-beh}), when $\xB=\sB-1=\OR\bigl(E^{1/3}\bigr)$, provides the estimate $\omegaB_{\!A}=\OR\bigl(E^{1/12}\omegaBC\bigr)$ invoked in (\ref{approx-third}$c$) and predicted by (\ref{Ethird-sol}$a$).

\subsubsection{The $\DC_{\outC}$ solution \label{DoutSol}}

Outside the tangent cylinder, $\DC_{\outC}$: $\xB=\sB-1>0$, we make the approximations
\bme
\be
\sB\approx 1, \qquad\qquad  \zB_+\approx \sqrt{\alpha^2-1}
\ee 
\eme
to obtain the leading order solution 
\bse
\label{inner-sl-v}
\be
\omegaB\,\approx\,\vB\,=\,1\,- \,(1-\omegaBC)\,\exp\!\big(-\,\xB\big/\delta_\outC\bigr),
\ee
where
\be
\delta_\outC^2\,=\,E\zB_+\big/\deltap\,=\,\bigl(E \zB_-^3\big/\alpha\bigr)^{1/2}
\ee
\ese
(see (\ref{delta-z}$b$)), of (\ref{sl-eq}) subject to the boundary conditions $\omegaB\to 0$ as $\xB\to\infty$ and $\omegaB=\omegaBC$ at $\xB=0$. There the value of $\omegaBCP=\dR\omegaB/\dR \sB$ determined by (\ref{inner-sl-v}$a$) ($\xB=\sB-1$) is
\be
\label{dDeltaOmega}
\omegaBCP\,=\,(1-\omegaBC)\big/\delta_\outC\,=\,\OR\bigl(E^{-1/4}\bigr).
\ee
With the help of (\ref{psi-pm}$b$) and (\ref{psi-qg}), the formula (\ref{inner-sl-v}$a$) for $\omegaB$ determines
\be
\label{inner-sl-psi}
\psiB\,=\,\tfrac12\bigl(E\big/ \delta_\outC^2\bigr)(1-\omegaBC)\,\zB\,\exp\!\big(-\,\xB\big/\delta_\outC\bigr)\,=\,\OR\bigl(E^{1/2}\bigr)
\ee
\citep[see][eqs.~(5.3$a,b$)]{S66}.

\subsubsection{The $\DC_{\inC}$ solution and $\omegaBC$ \label{DinSol}}

The solution inside the tangent cylinder, $\DC_{\inC}$: $\xB=\sB-1<0$, is rather more complicated because of the singular behaviour of $\zB_-$ and $\deltam$ described by (\ref{delta-z-sing}$a$,$b$). To appreciate its complexity we write out the leading order form of (\ref{sl-eq}) explicitly:
\bme
\label{sl-eq-inside}
\be
\delta_\outC^2\,\od{^2\omegaB}{\xtwoB}\,-\,\dfrac{E^{1/2}}{\deltap}(-2\xB)^{-1/4}\omegaB\,=\,-\bigl(1\,-\,\omegaB \bigr)\,,
\qquad\quad
\dfrac{E^{1/2}}{\deltap}\,=\biggl(\frac{\zB_+}{\alpha}\biggr)^{\!1/2}
\ee
\eme
\citep[cf.][eq.~(5.5)]{S66}. On balancing the two terms on the left-hand side of (\ref{sl-eq-inside}$a$) and noting the $(-\xB)^{-1/4}$ singularity in the coefficient of $\omegaB$, we may identify the $\DC_{\inC}$ length scale $E^{2/7}$ (short compared to the $\DC_{\outC}$ length scale $E^{1/4}$) and so introduce the stretched coordinate
\bme
\label{sl-coord-inside}
\be
\xs\,=\,\xB\big/\delta_\inC\,,\qquad\qquad \delta_\inC\,=\,\bigl(2E^2\zB_+^4\bigr)^{1/7}\,.
\ee
\eme
We note that the ratio of the inner ($\delta_\inC$) to outer ($\delta_\outC$) shear layer length scales
\bme
\label{sl-param-inside}
\be
\se
{\delta_\inC}\big/{\delta_\outC}\,=\,E^{1/28}\aleph
\ee
is small, where
\be
\dfrac{\deltap}{E^{1/2}}\bigl(2\delta_\inC\bigr)^{1/4}\,=\,E^{1/14}\aleph^2\,=\,\biggl(\dfrac{\delta_\inC}{\delta_\outC}\biggr)^{\!2}\,,\qquad\qquad
\aleph\,=\,\dfrac{2^{1/7}\alpha^{1/4}}{\bigl(\alpha^2-1\bigr)^{5/56}}\,.
\ee
\eme

The power series solution of (\ref{sl-eq-inside}$a$), that meets the $\xs=0$ boundary conditions, begins
\be
\label{sl-near-0}
\omegaB\,=\,\omegaBC\,+\,\delta_\inC\,\omegaBCP\, \xs\,+\,({16}/{21})\,\omegaBC\, (-\xs)^{7/4}\,+\,\cdots\,.
\ee
Matching with the mainstream Proudman solution (\ref{Proud-sol}a), identified by neglecting the term $\delta_\outC^2\dR^2 \omegaB\big/\dR \xtwoB$ in (\ref{sl-eq-inside}$a$), requires
\bse
\label{sl-Proud}
\be
\omegaB\,\approx\, E^{1/14}\aleph^2 \GC_\infty(\xs) \qquad\qquad \mbox{as} \qquad \xs\,\to\,-\,\infty\,,
\ee
where
\be
\GC_\infty(\xs)\,=\,(-\xs)^{1/4}\big/\bigl[1+E^{1/14}\aleph^2(-\xs)^{1/4}\bigr].
\ee
\ese

Guided by (\ref{sl-near-0}) and (\ref{sl-Proud}) the solution of (\ref{sl-eq-inside}$a$) may be expressed as the sum 
\be
\label{sl-FG}
\omegaB\,=\,\omegaBC\,\FC(\xs)\,+\,E^{1/14}\aleph^2\,\GC(\xs)
\ee
\citep[cf.][eq.~(5.7)]{S66}. Under this partition, $\FC(\xs)$ is chosen to solve
\bme
\label{sl-F-eq}
\be
\se
\FC^{\,\prime\prime}-\bigl[(-\xs)^{-1/4}+E^{1/14}\aleph^2\bigr]\FC\,=\,0
\ee
(here, the $^\prime$ denotes the $\xs$-derivative) subject to
\be
\FC(0)\,=\,1\,,\qquad\qquad\quad  \FC\,\approx \,0 \qquad \mbox{for}\qquad 1\,\ll\, -\,\xs\,<\,\OR\bigl(E^{-2/7}\bigr),
\ee
\eme
while $\GC(\xs)$ solves
\bme
\label{sl-G-eq}
\be
\se
\GC^{\,\prime\prime}-\bigl[(-\xs)^{-1/4}+E^{1/14}\aleph^2\bigr]\GC\,=\,-\,1
\ee
subject to
\be
\GC(0)\,=\,0\,,\qquad\qquad\quad \GC\,\approx\, \GC_\infty(\xs)\qquad \mbox{for}\qquad 1\,\ll\, -\,\xs\,<\,\OR\bigl(E^{-2/7}\bigr)
\ee
\eme
\citep[cf.][eqs.~(5.8), (5.9)]{S66}. Finally, using (\ref{dDeltaOmega}) and (\ref{sl-param-inside}$a$), continuity of $\dR\omegaB\big/\dR \sB$ requires
\bse
\label{sl-cont-0}
\be
\omegaBC\,\FC^{\,\prime}(0)\,+\,E^{1/14}\aleph^2\,\GC^{\,\prime}(0)\,=\,\delta_\inC\omegaBCP\,=\,E^{1/28}\aleph\bigl(1-\omegaBC\bigr)
\ee
\citep[cf.][eqs.~(6.13), (6.14)]{S66} implying that
\be
\omegaBC\,=\,E^{1/28}\aleph\,\,\dfrac{1\,-\,E^{1/28}\aleph\,\GC^{\,\prime}(0)}{\FC^{\,\prime}(0)\,+\,E^{1/28}\aleph}\,.
\ee
\ese

The complete asymptotic solution of the combined $E^{1/4}$ and $E^{2/7}$ QG shear layers involves expansions in powers of the parameter $E^{1/28}$ identified by (\ref{sl-param-inside}a). Accordingly in (\ref{sl-FG}), we introduce the expansions 
\bse
\label{sl-series}
\begin{align}
\FC(\xs)\,=\,&\,\FC_0(\xs)\,+\,E^{1/14}\aleph^2\FC_1(\xs)\,+\,\cdots\,, \\
\GC(\xs)\,=\,&\,\GC_0(\xs)\,+\,E^{1/14}\aleph^2\GC_1(\xs)\,+\,\cdots\,.
\end{align}
\ese

The dominant contribution $\FC_0(\xs)$ to (\ref{sl-FG}), which solves $\FC_0^{\,\prime\prime}-(-\xs)^{-1/4}\FC_0=0$ and decays to zero as $\xs\to-\infty$, may be expressed in terms of the modified Bessel function:
\bme
\label{sl-sim-sol}
\begin{align}
\FC_0(\xs)\,=\,&\,\Upsilon (-\xs)^{1/2}\,\KR_{4/7}(\sigma)&&\mbox{with} &
\FC_0^{\,\prime}(x)\,=\,&\,\Upsilon (-\xs)^{3/8}\,\KR_{3/7}(\sigma)\,,\\
\intertext{where}
\sigma\,=\,&\,(8/7)\,(-\xs)^{7/8} &&\mbox{and} &
\Upsilon\,=\,&\,{2(4/7)^{4/7}}\Big/{\Gamma(4/7)}
\end{align}
\eme
ensures that $\FC_0(0)=1$ (use (\ref{I-nr-0}$b$)) as well as fixing
\be
\label{sl-sim-consts}
{\bigl.\FC_0\bigr.}^{\!\prime}(0)\,=\,{(4/7)^{1/7}\Gamma(3/7)}\big/{\Gamma(4/7)}
\ee
\citep[see][eq.~(5.11)]{S66} sufficient to determine the zeroth order approximation
\be
\label{Delt-Om-0}
\omegaBCzero\,=\,{E^{1/28}\aleph}\Big/{{\bigl.\FC_0\bigr.}^{\!\prime}(0)}
\ee
to (\ref{sl-cont-0}$b$) \citep[cf.][eq.~(6.15)]{S66}.

In Appendix~\ref{Bessel} we extend Stewartson's zeroth order results and construct the next order contributions $\GC_0(\xs)$ and $\FC_1(\xs)$, from which we obtain the $\xs=0$ values
\bme
\label{sl-deriv0}
\be
-\,{\bigl.\GC_0\bigr.}^{\!\prime}(0)\,=\,(7/4)^{1/7}\Gamma(8/7)\,,\qquad\quad
{\bigl.\FC_1\bigr.}^{\!\prime}(0)\,=\,(7/4)^{1/7}\,\dfrac{\Gamma(12/7)\,\bigl[\Gamma(8/7)\bigr]^2}{\Gamma(16/7)\Gamma(4/7)}
\ee
\eme
needed to determine $\omegaBC$ from (\ref{sl-cont-0}$b$) correct to $\OR\bigr(E^{3/28}\bigr)$ (see also (\ref{truncation})). Note too that even the smallest contribution $E^{1/14}\aleph^2\omegaBC\,\FC_1(\xs)$ to $\omegaB$ in (\ref{sl-FG}) is just larger than the AG-contribution $\omegaB_{\!A}=\OR\bigl(E^{1/12}\omegaBC\bigr)$ (see (\ref{approx-third}$c$)) that we will consider in \S\ref{Ag-shear-layer}.

Interpretation of the leading order solution $\omegaB=\omegaBCzero \FC_0(\xs)$ requires care. For though it dominates for $\xs=\OR(1)$, it decays exponentially for large $-\,\xs$ and so does not match up with the Proudman solution (\ref{sl-Proud}$a$). The needed composite solution valid in the overlap domain $1\ll -\,\xs\ll E^{-2/7}$, determined by the large $\sigma$ asymptotic form of $\KR_{4/7}(\sigma)$ and leading order approximation $\GC_\infty=(-\xs)^{1/4}$ of (\ref{sl-Proud}$b$), is
\be
\label{composite}
\omegaB\,=\,E^{1/28}\aleph\biggl[\dfrac{\sqrt\pi\,(-\xs)^{1/16}}{(4/7)^{1/14}\Gamma(3/7)}\,\exp\Bigl(-\dfrac87\, (-\xs)^{7/8}\Bigr)+\,E^{1/28}\aleph\,(-\xs)^{1/4}\biggr].
\ee
It shows that, though $\omegaBCzero=\OR\bigl(E^{1/28}\bigr)$ (see (\ref{Delt-Om-0})), the value of $\omegaB$ continues to decrease in size with increasing $-\xs$ by an order of magnitude attaining its minimum of $\OR\bigl(E^{1/14}\bigr)$ at large $-\xs=-\xs_M$ \citep[say, and see figure~\ref{QGin}, also][figure~4, left panel]{DCJ98}. As the value of $\psiB$ on the outer boundary is given correct to leading order by $\psiB_+\approx\psiB_{\PR}-\tfrac12 \deltap\omegaB$ (see (\ref{psi-pm}$b$) and (\ref{C-Proud}$c$)), it follows that the maximum of $\psiB_+$ is also located at $-\xs\approx -\xs_M\gg 1$. 

\subsubsection{The direct numerical (DN-)solution: $\alpha^{-1}=0.35$ \label{DNsol}}

\begin{figure}
\centerline{}
\centerline{\includegraphics[width=0.6\textwidth,clip=true]{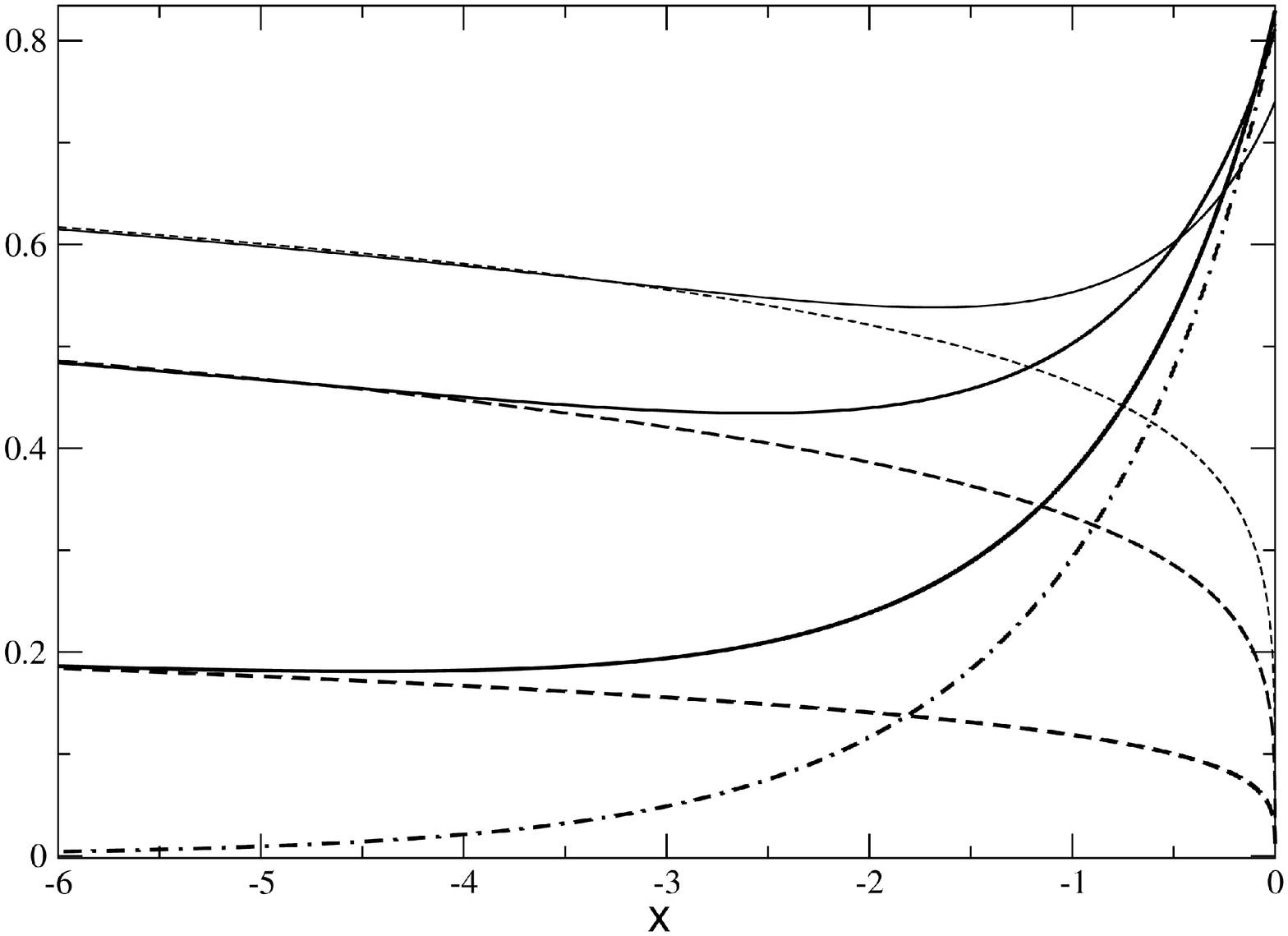}}
\caption{\label{QGin} The $E^{2/7}$-layer DN-solutions (solid lines) for $\omegaB\big/\bigl(E^{1/28}\aleph\bigr)$ (see (\ref{sl-FG}); $\alpha^{-1}=0.35$) of the problem (\ref{sl-F-eq})--(\ref{sl-cont-0}) plotted versus $\xs$ for $E=10^{-7}$ (upper), $10^{-14}$ (middle),  $10^{-28}$ (lower); the $E$-values listed in table~\ref{table-1}. The Stewartson solution $\FC_0(\xs)\big/{\bigl.\FC_0\bigr.}^{\!\prime}(0)$ (see also (\ref{sl-sim-consts}), (\ref{Delt-Om-0}))  and the Proudman solutions  $E^{1/28}\aleph \GC_\infty(\xs)$ (see (\ref{sl-Proud})) are identified  by the dash-dotted and dashed curves respectively.}
\end{figure}

The complete resolution of the QG shear layers relies on the solution of the problem (\ref{sl-FG})--(\ref{sl-cont-0}). The expansions (\ref{sl-series}$a$,$b$) of $\FC(\xs)$ and $\GC(\xs)$ in powers of $E^{1/28}$ limit their applicability to extremely small $E$. Nevertheless, the direct numerical (DN-)solution at fixed $E$ is not impeded by this consideration. Such DN-solutions for the oft-used value
\bme
\label{apha-val}
\be
\alpha^{-1}\,=\,0.35\qquad\qquad\quad\Longrightarrow\qquad\qquad\quad \aleph\,\doteqdot\, 1.2040
\ee
\eme
(see (\ref{sl-param-inside}$c$)) are plotted in figure~\ref{QGin} for $E=10^{-7}$, $10^{-14}$, $10^{-28}$, while the corresponding DN-values of $\omegaBC$ (see (\ref{sl-cont-0}$b$)) are listed in table~\ref{table-1}. They are compared with the asymptotic values of (\ref{sl-cont-0}$b$) with $\bigl[{\bigl.\FC\bigr.}^{\,\prime}\!(0),\,{\bigl.\GC\bigr.}^{\,\prime}\!(0)\bigr]$ truncated at various levels:
 \be
\label{truncation}
\bigl[{\bigl.\FC\bigr.}^{\,\prime}\!(0),\,{\bigl.\GC\bigr.}^{\,\prime}\!(0)\bigr]=\left\{\begin{array}{ll}
\bigl[{\bigl.\FC_0\bigr.}^{\!\prime}(0),\,0\bigr],&  \OR\bigl(\aleph E^{1/28}\bigr),\\[0.2em]
\bigl[{\bigl.\FC_0\bigr.}^{\!\prime}(0),\,{\bigl.\GC_0\bigr.}^{\!\prime}(0)\bigr],&  \OR\bigl(\aleph E^{1/14}\bigr),\\[0.2em]
\bigl[{\bigl.\FC_0\bigr.}^{\!\prime}(0)\,+\,E^{1/14}\aleph^2{\bigl.\FC_0\bigr.}^{\!\prime}(0),\, {\bigl.\GC_0\bigr.}^{\!\prime}(0)\bigr],\qquad&  \OR\bigl(\aleph E^{3/28}\bigr)
\end{array}\right.
\ee
giving $\omegaBC$ accurate to the orders of magnitude indicated. Though the comparison at $E=10^{-7}$ is fair, at $E=10^{-14}$ it is excellent. Certainly the values of $\omegaB\big/\bigl(E^{1/28}\aleph\bigr)$ on the $\xs=0$ axis of figure~\ref{QGin} (namely, the tangent cylinder $\CC$) for $E=10^{-14}$  and $10^{-28}$ are indistinguishable from the lowest order value $\omegaBCzero\big/\bigl(E^{1/28}\aleph\bigr)=1\big/ {\bigl.\FC_0\bigr.}^{\!\prime}\!(0)$ (see (\ref{Delt-Om-0})). We also show the limiting forms of the Proudman solutions $E^{1/28}\aleph \GC_\infty(\xs)$ (see (\ref{sl-Proud})), because they provide the $-\,\xs \gg 1$ asymptotes to the numerical solutions. Evidently on each finite $E$ curve a minimum at negative $\xs_M$ (dependent on $E$) can be identified as predicted by the discussion in \S\ref{DinSol} following (\ref{composite}).

\begin{table}
{
 \renewcommand*{\arraystretch}{1.3}
 \setlength{\tabcolsep}{12pt}
   \begin{center}
     \begin{tabular}{|l|l|l|l|l|l|}    
\cline{3-5}
\multicolumn{2}{c}{}&\multicolumn{3}{|c|}{Asymptotics correct to order}&\multicolumn{1}{c}{}\\
\cline{1-2}\cline{6-6}
$\,\,\,E$& {$\!\aleph E^{1/28}$} &   {$\OR\bigl(E^{1/28}\bigr)$}&{$\OR\bigl(E^{1/14}\bigr)$}&{$\OR\bigl(E^{3/28}\bigr)$}&$\,\,\,\,$DN\\
\cline{1-6}
 {$10^{-5}$}  &  $0.7981$ &  $\,\,\,0.6517$ & $\,\,\,0.7137$ & $\,\,\,0.6201$ & $0.5628$ \\
 {$10^{-7}$}  &  $0.6771$ &  $\,\,\,0.5529$ & $\,\,\,0.6003$ & $\,\,\,0.5381$ & $0.5012$ \\
 {$10^{-14}$} &  $0.3807$ &  $\,\,\,0.3109$ & $\,\,\,0.3287$ & $\,\,\,0.3150$ & $0.3088$ \\
 {$10^{-28}$} &  $0.1204$ &  $\,\,\,0.0983$ & $\,\,\,0.1004$ & $\,\,\,0.0993$ & $0.0998$ \\
\cline{1-6}
     \end{tabular}
   \end{center}
\caption{The values of  
$\omegaBC$ for $\alpha^{-1}=0.35$ at various values of $E$ (see (\ref{apha-val}$b$) for $\aleph E^{1/28}$ in column 2). Asymptotic results truncated at the three levels indicated in (\ref{truncation}) are listed in middle columns 3--5 respectively\red{, the first of which, column 3, gives the values of $\omegaBCzero$ defined by (\ref{Delt-Om-0})}. The DN-results are listed in the last column 6.}
\label{table-1}
}
\end{table} 
%

\subsection{$E^{1/3}$-ageostrophic(AG) shear layer\label{Ag-shear-layer}}

The relatively thick QG shear layers have ensured continuity of the geostrophic flow close to the tangent cylinder $\CC$, where $\omegaB\,\approx\,\omegaB_G(\sB)$ (see (\ref{lin-shear}$a$)). Nevertheless, since both $\psiB$, $\partial^2\omegaB/\partial \stwoB$ and their derivatives remain discontinuous at $\sB=1$, it is the role of the AG shear layer to smooth them out over the $\sB$-directed length scale $E^{1/3}$. As anticipated in (\ref{approx-third}$a$,$b$), we set $\omegaB=\omegaB_G+\omegaB_{\!A}$, $\psiB=\psiB_{\!A}$ and solve for $\omegaB_{\!A}$, $\psiB_{\!A}$. In view of the estimates (\ref{approx-third}$c$), it is evident that the leading order approximation to (\ref{lin-shear}$a$) is 
\be
\label{omegaGapprox}
\omegaB_G\,\approx\,\omegaBC\,.
\ee

\subsubsection{The Fourier transform solution: $\xB=\OR(E^{1/3})$,\,\, $\zB=\OR(1)$\label{E13-full}}

On the short radial length scale $\xB=\OR\bigl(E^{1/3}\bigr)$, the governing equations (\ref{mom-vort}$a$,$b$) reduce to
\bme
\label{mom-vort-Ethird}
\be
2\pd{\vB_A}{\zB}\,=\,E\pd{^4\psiB_{\!A}}{\xfourB}\,,
\qquad\qquad\qquad
-\,2\pd{\psiB_{\!A}}{\zB}\,=\,E\pd{^2\vB_A}{\xtwoB}\,.
\ee
\eme
\red{Correct to leading order} the boundary conditions (\ref{psi-pm}) determine
\bme
\label{mom-vort-Ethird-approx}
\be
\psiB_{A+}\,=\,\tfrac12 \deltap\,\approx\,\psiB_{\PR}\,,
\qquad\qquad
\psiB_{A-}\,=\left\{\!\!\begin{array}{lll}
\tfrac12 \deltam \,\omegaBC \qquad&\mbox{in}\quad\DC_{\inC}\,,\\[0.6em]
0\qquad&\mbox{in}\quad\DC_{\outC}\end{array}\right.
\ee
\eme
(see also (\ref{C-Proud}$c$), (\ref{psi-pm-third})). On use of the estimates (\ref{delta-z-sing}$b$) and (\ref{DeltaOmega}$c$), we obtain
\be
\label{neglect-psiA}
(\deltam/E^{1/2})\,\omegaBC\,=\,\OR\bigl(E^{-1/12}\,\omegaBC\bigr)=\,\OR\bigl(E^{-1/21}\bigr)\gg\,(\deltap/E^{1/2})\,=\,\OR(1)\,.
\ee
So despite the small size of $\omegaBC$, the value of $\psiB_{\!A}$ on the bottom boundary in $\DC_\inC$ exceeds $\psiB_{\PR}$ on the top by a moderate factor $\OR\bigl(E^{-1/21}\bigr)$. With $\psiB_{\PR}$ neglected, we are left with
\bse
\label{Ethird-bc}
\begin{align}
\psiB_{\!A}&=0        & \mbox{at} & &   \zB&=\sqrt{\alpha^2-1}\,,\\
\psiB_{\!A}&=\left\{\!\!\begin{array}{ll}
\tfrac12
\biggl[\dfrac{E^{1/3}}{(-2\xB)}\biggr]^{\!1/4}E^{5/12}\,\omegaBC\,\,\,\,& \mbox{in}\,\,\,\DC_{\inC}\,,\\[0.8em]
0\quad& \mbox{in}\,\,\,\DC_{\outC}
\end{array}\!\!\right\} &\mbox{at} &&  \zB&=\,0\,.
\end{align}
\ese
Here, since the bottom boundary is located at $\zB_-\approx\sqrt{-2\xB}=\OR(E^{1/6})$, the boundary condition (\ref{Ethird-bc}$b$) in $\DC_{\inC}$ may be applied, correct to leading order, at $\zB=0$ as stated. Together with natural boundedness requirements as $E^{-1/3}\xB\to \pm \infty$, the boundary conditions (\ref{Ethird-bc}) are sufficient to determine the Fourier transform solution $(\vB_A\approx\omegaB_{\!A})$
\bse
\label{Ethird-sol}
\begin{align}
\omegaB_{\!A}\,&=\,-\,\dfrac{E^{1/12}\,\omegaBC}{2^{7/4}\Gamma\bigl(1/4\bigr)}\int_{-\infty}^\infty\,
\dfrac{{\bigl.\kT\bigr.}^{\,1/4}\cosh\Bigl[\tfrac12 {\bigl.\kT\bigr.}^{\,3}\bigl(\zT_+-\zT\,\bigr)\Bigr]}{\sinh\Bigl[\tfrac12 {\bigl.\kT\bigr.}^{\,3}\zT_+\Bigr]}
\exp\Bigl[\iR\Bigl(\kT\xT\,+\,\dfrac{3\pi}{8}\Bigr)\Bigr]\,\dR \kT\,,\\[0.6em]
\!\!\!
\psiB_{\!A}\,&=\,\dfrac{E^{5/12}\,\omegaBC}{2^{7/4}\Gamma\bigl(1/4\bigr)}\,\int_{-\infty}^\infty\,\dfrac{\sinh\Bigl[\tfrac12 {\bigl.\kT\bigr.}^{\,3}\bigl(\zT_+-\zT\,\bigr)\Bigr]}{{\bigl.\kT\bigr.}^{\,3/4}\sinh\Bigl[\tfrac12 {\bigl.\kT\bigr.}^{\,3}\zT_+\Bigr]}
\exp\Bigl[\iR\Bigl(\kT\xT\,+\,\dfrac{3\pi}{8}\Bigr)\Bigr]\,\dR \kT
\end{align}
\ese
\citep[see][eq.~(6.22$a,b$)]{S66}, where
\bme
\label{breve-sz}
\te
\be
\xT\,=\,E^{-1/3}\xB\,, \qquad\qquad\zT\,=\,\zB\,, \qquad\qquad   \zT_+\,=\,\sqrt{\alpha^2-1}
\ee
\eme
and ${\bigl.\kT\bigr.}^{\,1/4}=\bigl|\kT\bigr|^{1/4}\eR^{\iR\pi/4}$ and  ${\bigl.\kT\bigr.}^{\,-3/4}=\bigl|\kT\bigr|^{-3/4}\eR^{-\iR 3\pi/4}$ for $\kT<0$.

Significantly the result (\ref{Ethird-sol}$a$) shows that $\omegaB_{\!A}=\OR(E^{1/12}\omegaB_G)$, where $\omegaB_G=\omegaBC$ (see (\ref{omegaGapprox})). The small relative size of $\omegaB_{\!A}$ to $\omegaB_G$, albeit by a factor $\OR(E^{1/12})$, ensures that any Ekman suction produced by $\omegaB_{\!A}$ is small compared to that assumed in our applied bottom boundary condition (\ref{Ethird-bc}$b$) by an $\OR(E^{1/12})$ factor. This estimate is sufficient to confirm the consistency of the approximations made.

Finally, we consider the $\zB$-average
\be
\label{omega-av}
\bigl\langle\omegaB_{\!A}\bigr\rangle\,=\,\dfrac{1}{\zT_+}\int_0^{\zT_+}\omegaB_{\!A}\,\dR \zT\,=\,
\left\{\!\!\begin{array}{ll}
\dfrac{E^{1/12}\,\omegaBC}{2^{1/4}\zT_+}\, \dfrac{16}{21}\,(-\xT\,)^{7/4}\,\,\,\,& \mbox{in}\,\,\,\DC_{\inC}\,,\\[0.8em]
0\quad& \mbox{in}\,\,\,\DC_{\outC}\,,
\end{array}\!\!\right.
\ee
which may be determined from (\ref{Ethird-sol}$a$) or more easily by direct integration of the $\zB$-average of (\ref{mom-vort-Ethird}$b$).

\red{The combination $\omegaB_G+\bigl\langle\omegaB_{\!A}\bigr\rangle= \omegaBC\,+\,\omegaBCP\,\xB+\bigl\langle\omegaB_{\!A}\bigr\rangle$ (see (\ref{lin-shear}$a$)) for $\DC_{\inC}$, determined by (\ref{omega-av}), is simply the QG asymptotic behaviour in the $E^{2/7}$-layer identified in (\ref{signg-beh}) and established by (\ref{sl-near-0}). So if we denote the entire QG-solution of (2.3$b$), obtained by combining the Proudman mainstream solution (\ref{Proud-sol}$a$) and the $E^{2/7}$- and $E^{1/4}$-QG shear layer solutions (\ref{inner-sl-v}$a$) and (\ref{sl-FG}), as $\omegaB_{PQG}(\sB)$, it means that the $\zB$-average of $\omegaB$ is given correct to leading order by 
\bse
\label{mean-fluct}
\be
\bigl\langle\omegaB\bigr\rangle\,\approx\,\omegaB_{PQG}(\sB)
\ee
everywhere outside the Ekman layers (including the equatorial Ekman layers); a very useful result. It has the important non-trivial consequence that
\be
\omegaB\,\approx\,\omegaB_{PQG}\,+\,\bigl(\omegaB_{\!A}\,-\,\bigl\langle\omegaB_{\!A}\bigr\rangle\bigr).
\ee
\ese
everywhere outside Ekman layers too. Here the singularities of $\bigl\langle\omegaB_{\!A}\bigr\rangle$ and $\omegaB_{PQG}$ on $\CC$ cancel each other out, while $\omegaB_{\!A}$ defined by (\ref{Ethird-sol}$a$) is continuous. We conclude that, whereas $\bigl\langle\omegaB\bigr\rangle$ is singular on $\CC$,\ $\omegaB$ itself is smooth across $\CC$ except for singular behaviour in the immediate neighbourhood of the equator, $(\xT,\,\zT)=(0,0)$ (see \S\S\ref{Stew-sim-sol}, \ref{EE-layer}).} 

\subsubsection{The similarity \red{sublayer}: $\xB=\OR\bigl(E^{1/3}\zthirdB\bigr)$, $\,\,E^{1/5}\ll \zB\ll 1$\label{Stew-sim-sol}}

Close to the equator, where
\bme
\be
\xT=\OR\bigl(\zT^{\,1/3}\bigr)\qquad\quad\mbox{when} \qquad\quad\zT\ll 1\,,
\ee
\eme
the solutions (\ref{Ethird-sol}$a$,$b$) may be approximated as
\bse
\label{Ethird-sol-similarity}
\begin{align}
\omegaB_{\!A}\,=\,&\,-\, E^{1/12}\,\omegaBC\,\zT^{\:-5/12}\,V_0(\Phi)\,,\\[0.3em]
\psiB_{\!A}\,=\,&\,-\,E^{5/12}\,\omegaBC\,\zT^{\:-1/12}\,\Psi_0(\Phi)\,,
\end{align}
in which
\begin{align}
V_0(\Phi)\,&=\,\dfrac{2^{-3/4}}{\Gamma\bigl(1/4\bigr)}\,\int_0^\infty
\varpi^{1/4}\cos\Bigl(\varpi\Phi\,+\,\dfrac{3\pi}{8}\Bigr)\exp\bigl(-\tfrac12 \varpi^3\bigr)\,\dR \varpi\,,\\[0.4em]
\!\!\!
\Psi_0(\Phi)\,&=\,-\,\dfrac{2^{-3/4}}{\Gamma\bigl(1/4\bigr)}\,\int_0^\infty
\varpi^{-3/4}\cos\Bigl(\varpi\Phi\,+\,\dfrac{3\pi}{8}\Bigr)\exp\bigl(-\tfrac12 \varpi^3\bigr)\,\dR \varpi
\end{align}
\ese
\citep[cf.][eqs.~(6.24), (6.25$a,b$)]{S66}, where
\bme
\label{Ethird-sol-similarity-variable}
\be
\varpi\,=\,\kT\zT^{\,1/3}\,,\qquad\qquad\qquad  \Phi\,=\,\xT\big/\zT^{\,1/3}\,.
\ee
\eme

\subsubsection{$E^{2/5}$--equatorial Ekman layer:$\,\,\xB=\OR\bigl(E^{2/5}\bigr)$,\,\,  $\zB=\OR\bigl(E^{1/5}\bigr)$\label{EE-layer}}

The similarity forms (\ref{Ethird-sol-similarity}$a$,$b$) indicate that the solution becomes singular as the equator $(\xB,\, \zB)=(0,\,0)$ is approached. For small $\zT$, we may identify the following lengths \red{(see figure~\ref{schematic})}: the Ekman boundary layer thickness $\deltam=E^{1/2}\,\zT^{\:-1/2}$; the distance $\deltas=\tfrac12 \zT^{\,2}$ of the tangent cylinder to the boundary; and the \red{similarity sublayer} width \blue{$\deltaA=E^{1/3}\,\zT^{\,1/3}$ (for $\Phi=1$, say),}
of the solution (\ref{Ethird-sol-similarity}) triggered at the equator. As $\zB=\zT\,\bigl(\gg E^{1/5}\bigr)$ decreases, the Ekman layer width $\deltam$ increases and remains clear of the \red{similarity sublayer} ($\deltas-\deltam\gg \deltaA$) until $\zB=\OR\bigl(E^{1/5}\bigr)$, when the three lengths become comparable: $\OR\bigl(\deltam\bigr)=\OR\bigl(\deltas\bigr)=\OR\bigl(\deltaA\bigr)=\OR\bigl(E^{2/5}\bigr)$. This length scale coincidence together with $\zB=\OR\bigl(E^{1/5}\bigr)$ identifies the dimensions of the equatorial Ekman layer \red{(highlighted in figure~\ref{schematic}),} inside which $\omegaB_{\!A}=\OR(\omegaBC)$ (see (\ref{EEL-dimensions})); outside $\omegaB_{\!A}\ll \omegaBC$ (see (\ref{sl-qg--est})). The main thrust of our paper is the formulation and solution of the equatorial Ekman layer problem.

\subsubsection{Full shell DNS-results: $\alpha^{-1}=0.35$, $E=10^{-5}$ and $10^{-7}$\label{num-res}}

\begin{figure}
  \centerline{\raisebox{26mm}{\small (a)}\includegraphics[width=0.47\textwidth,clip=true]{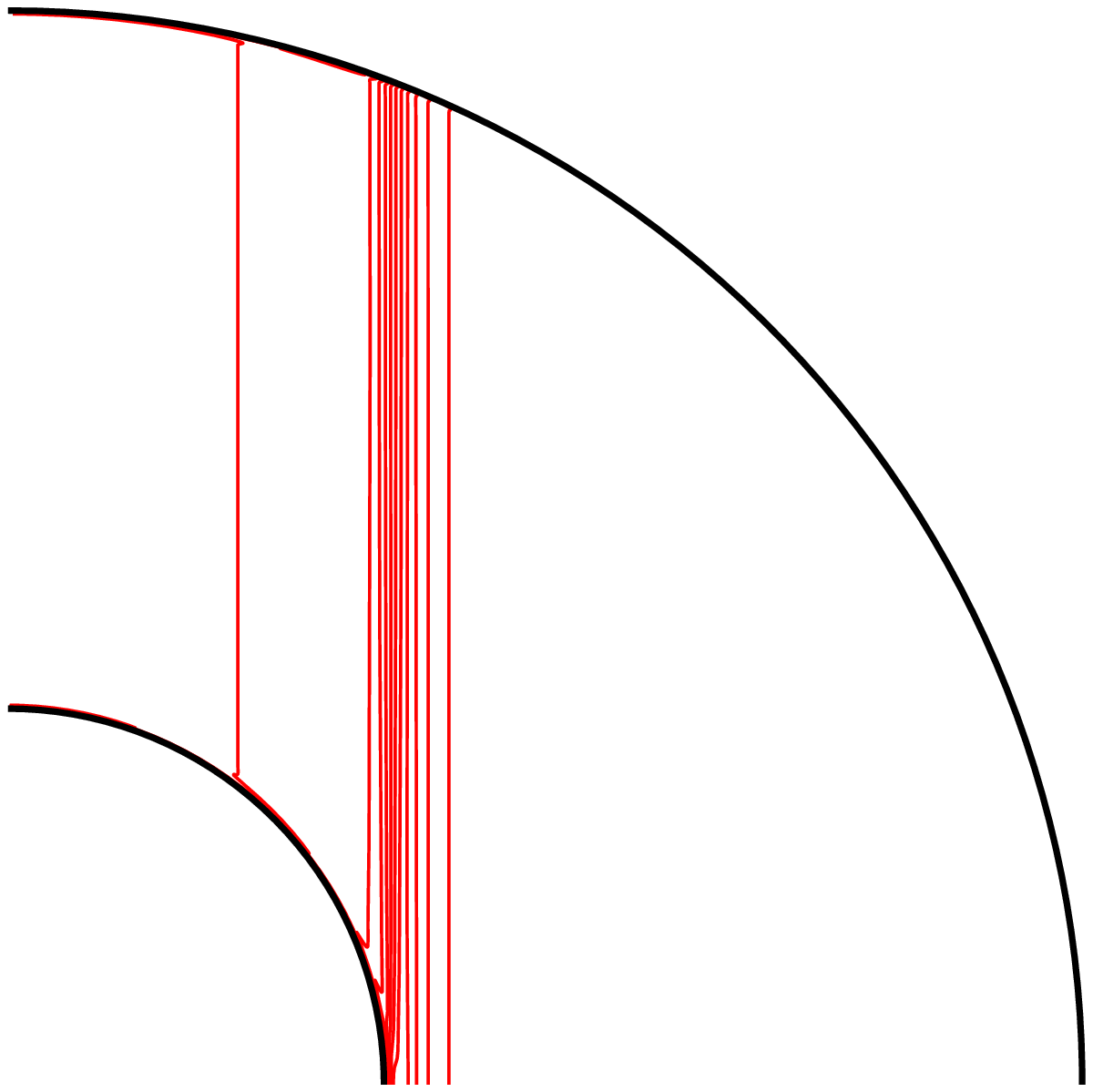}\,
              \raisebox{26mm}{\small (b)}\includegraphics[width=0.47\textwidth,clip=true]{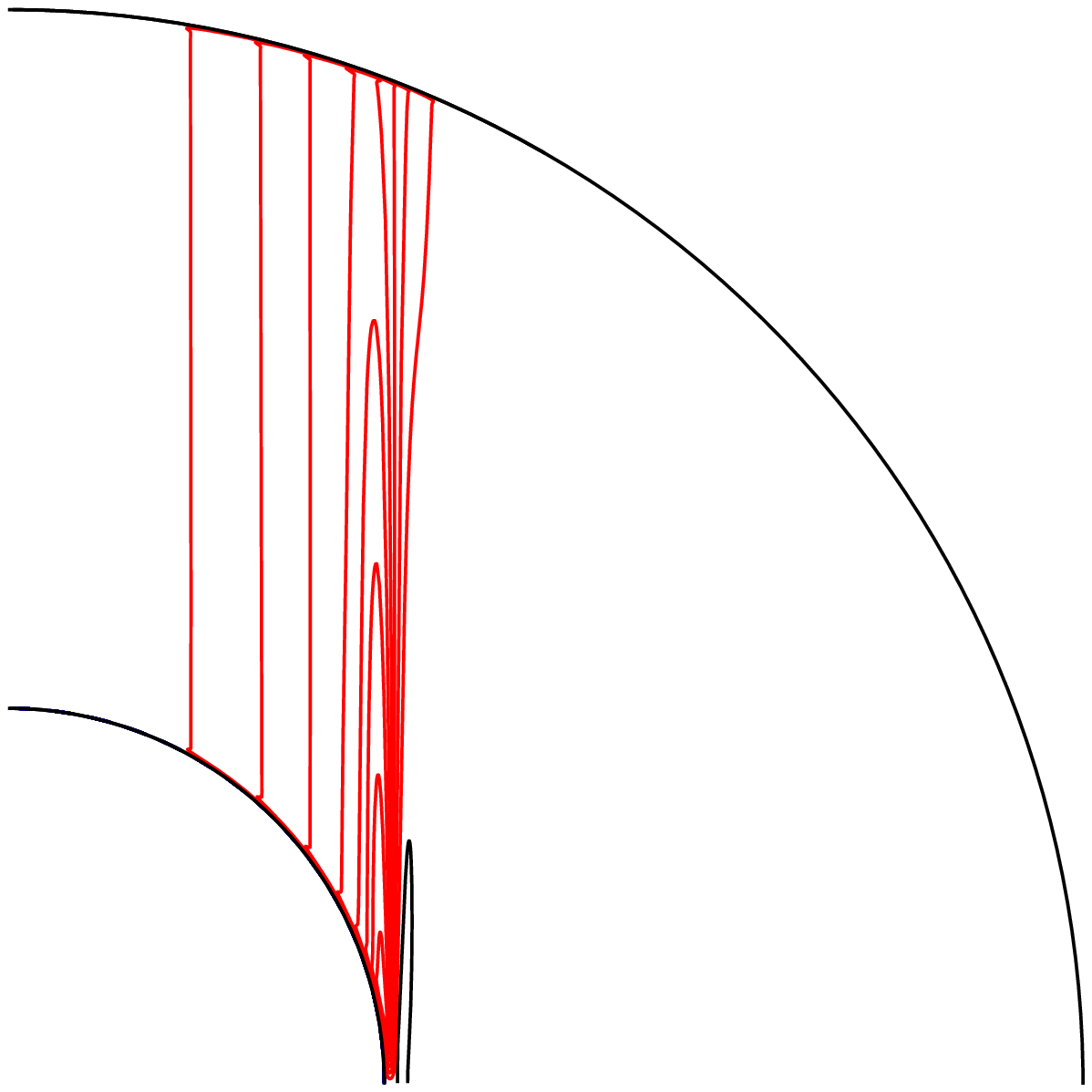}}
  \caption{\label{full_shell_num}Contours of ($a$) the azimuthal angular velocity $\omegaB$ and ($b$) the streamfunction $\psiB$ obtained from the full shell DNS for the case $\alpha^{-1}=0.35$, $E=10^{-5}$. A weak reversed flow eddy is identified by the single black $\psiB$-contour just outside the tangent cylinder.} 
\end{figure}

DNS-results for the full shell equations (\ref{mom-vort}), again at $\alpha^{-1}=0.35$, are illustrated in figures~\ref{full_shell_num}($a$,$b$) for the particular case  $E=10^{-5}$. 

\red{Well inside the tangent cylinder $\CC$, the Proudman solution is illustrated by the contours parallel to the rotation axis both for $\omegaB=\omegaB(\sB)$ and $\psiB=\psiB(\sB)$. For $\omegaB$, this feature continues into the QG-layers either side of $\CC$, where  $\omegaB\approx \omegaB_{PQG}(\sB)$ (defined above (\ref{mean-fluct}$a$)). There $\psiB$ is linear in $\zB$ (see (\ref{psi-qg})) with the streamline structure in figure~\ref{full_shell_num}($b$) consistent with the QG-solution described in \S\ref{DinSol}. Just outside the tangent cylinder, $\DC_{\outC}$, the small tipping of the $\psiB$-contours near the outer sphere is implied by the $E^{1/4}$-layer solution (\ref{inner-sl-psi}). Inside, $\DC_{\inC}$, closed $\psiB$-contours identify a clockwise eddy \citep[cf., figure~5, middle panel of][where a very similar eddy structure is visible albeit in an alternative parameter range]{WH08}. The analysis at the end of \S\ref{DinSol} reveals that the outermost closed contour touches the outer sphere at $\sB=\sB_M=1+\delta_{\inC}\xs_M$ (see explanation below (\ref{composite})), which since $-\xs_M\gg 1$ is at the edge of the $E^{2/7}$-layer. It means that almost all of $E^{2/7}$-layer embraces return flow from the outer Ekman layer \citep[a striking feature anticipated remarkably by][at the end of his \S5]{S66}.}

\red{Interestingly, virtually none of the structure associated with the $E^{1/3}$-AG shear layer is distinguishable on either figure~\ref{full_shell_num}($a$) or ($b$). Nevertheless, we should note that the more intense part of the $E^{2/7}$-layer clockwise eddy (close to the equator of the inner sphere) is located within the smaller $(E\zB)^{1/3}$-Stewartson similarity sublayer, which also contains a weaker counter-clockwise eddy outside, in $\DC_{\outC}$. As we will see (figure~\ref{full_shell_num-blow_up}($b$)) the entire double eddy structure continues into the  equatorial Ekman layer.} 

\red{Asymptotically, the $\zB$-dependence of $\omegaB(\sB,\zB)\approx\omegaB_{PQG}(\sB)+\bigl(\omegaB_{\!A}(\sB,\zB)-\bigl\langle\omegaB_{\!A}(\sB)\bigr\rangle\bigr)$ predicted by (\ref{mean-fluct}$b$) in the $E^{1/3}$-layer is small except in and near the equatorial Ekman layer. However, the value $E=10^{-5}$ used to obtain the DNS-results illustrated in figure~\ref{full_shell_num}($b$) is not small enough to reach that asymptotic regime. So, though not visible in the figure, there is significant $\zB$-dependence of $\omegaB$ on the tangent cylinder $\CC$. Nevertheless, as explained at the end of \S\ref{E13-full} with particular reference to (\ref{mean-fluct}$a$), the $\zB$-average $\bigl\langle\omegaB\bigr\rangle$ of $\omegaB$ is approximated to a high order of accuracy by $\omegaB_{PQG}(\sB)$ throughout the shear layers. The comparison on $\CC$ of 
\be
\begin{array}{c}\bigl\langle\omegaB\bigr\rangle_\CC\\
\mbox{DNS}\end{array}=
\biggl\{\begin{array}{l}   0.5609\\[0.3em]  0.4965\end{array}\biggr.   \qquad\mbox{with}\qquad
\begin{array}{c}\omegaB_\CC\\
\mbox{DN}\end{array}=
\biggl\{\begin{array}{l}   0.5628\\[0.3em]  0.5012\end{array}\biggr.   \qquad\mbox{at}\quad
E=
\biggl\{\begin{array}{l}   10^{-5}\\[0.3em]  10^{-7}\end{array}
\label{DN-DNS}
\ee
(DN-values are given in the last column of table~\ref{table-1}) is excellent.}


\section{Formulation of the equatorial Ekman layer problem\label{formulation}}

Our initial objective is to formulate the equatorial Ekman layer problem identified in \S\ref{EE-layer} in dimensionless variables that do not involve the Ekman number $E$. To that end, we take the equator as our origin and adopt local dimensional rectangular Cartesian co-ordinates $(x^\star,\,y^\star,\,z^\star)=L(\xB,\,\yB,\,\zB)$, where $x^\star=s^\star-L$ and the $y^\star$-direction is the azimuth. So locally for $|x^\star|\ll L$, the sphere boundary is  $2Lx^\star+z^{\star 2}=0$ (see (\ref{delta-z-sing}$a$)). To capture the location of the equatorial Ekman layer, we non-dimensionalise distance by writing
\bme
\label{dim-dist}
\be\te
x^\star\,=\, \ell x\,, \qquad\qquad z^\star\,=\,\sqrt{\ell L}\,z\,, \qquad\qquad \ell\,=\,E^{2/5}L
\ee
so that the $E^{1/3}$-layer coordinates (\ref{breve-sz}$a$,$b$) and \red{$(E\,\zB)^{1/3}$ similarity} sublayer coordinate (\ref{Ethird-sol-similarity-variable}b) become
\be\te
\xT\,=\,E^{1/15}x\,, \qquad\qquad   \zT\,=\,E^{1/5}z\,, \qquad\qquad \Phi\,=\,x/z^{1/3} 
\ee
\eme
\red{respectively.} The sphere boundary $\SC$ (formerly $\SC_-$) is
\be
\label{dim-bdry}
\SC\,: \qquad   2x\,+\,z^2\,=\,0\,.
\ee
Note that the radial Ekman layer coordinate $r^\star-L\approx x^\star+\tfrac12 x^{\star 2}$ non-dimensionalised on its thickness $L \deltam$ (see (\ref{delta-z}$a$) with $\zB_-\approx \zB$) is
\bme
\label{EL-coord}
\be
\zeta\,=\,(\rB-1)/\deltam\, =\,z^{1/2}y\qquad\qquad\mbox{with}\qquad\qquad y\, =\,x\,+\,\tfrac12 z^2\,,
\ee
\eme
which is relevant for $z\gg 1$. A geostrophic cylinder $x\;$const.~through $\bigl({\bar y}, \,{\bar z}\bigr)$ is then
\be
\label{geo-cyl}
\CC\bigl({\bar y}, \,{\bar z}\bigr)\,: \qquad   y\,=\,{\bar y}\,+\,\tfrac12 \bigl(z^2\,-\,{\bar z}^2\bigr).
\ee

\subsection{The governing equations and boundary conditions\label{gov-eq-bdry-condit}}

The far-field boundary condition on $\omegaB$, as either $x\to \infty$ or $z\to \infty$, is 
\be
\label{om-inf}
{\omegaB}\big/{\omegaBC}\,\to\,{\omegaB_G}\big/{\omegaBC}\,=\,1\,+\,E^{2/5}\bigl({\omegaBCP}\big/{\omegaBC}\bigr)\,x\,=\,1\,+\,\OR\bigl(E^{4/35}x\bigr)
\ee
(see (\ref{lin-shear}$a$) and (\ref{DeltaOmega}$b$): ${\omegaBCP}\big/{\omegaBC}=\OR(E^{-2/7})$\red{, but c.f.~(\ref{EEL-dimensions})}). As in \S\ref{Ag-shear-layer}, we will neglect the linear shear flow correction and simply apply $\omegaB\to\omegaBC$ (see (\ref{omegaGapprox})). The error estimate is included in (\ref{om-inf}) only to make clear the level of accuracy possible inside the equatorial Ekman layer at finite $E$. We dimensionalise velocity in the form
\bse
\label{dim-vel}
\be
\bigl(u^\star\,,v^\star\,,w^\star\bigr)=\,\epsilon L\Omega(\uB,\,\vB,\,\wB)
=\,-\,\epsilon L\,\Delta\Omega\,\bigl(E^{1/5}u\,,\,\,v-1\,,\,w\bigr)
\ee
based on the angular velocity \blue{$\epsilon \Delta\Omegav=\epsilon\Omegav\,\omegaBC$} rather than \blue{$\epsilon \Omegav$} (see (\ref{DeltaOmega}$a$)); by introducing $-\epsilon$ we have essentially reversed the sign of the velocity. However, the ploy has the merit that $v=1$ on $\SC$ and $v\to 0$ far from  $\SC$. By implication our new rest frame $\vv={\mathbf 0}$ rotates with the angular velocity $\Omegav+\epsilon\Delta \Omegav$ of the dominant flow in the far-field. The corresponding stream function $\psiB$ becomes 
\be
\psiB\,=\,-\, E^{2/5}\,\omegaBC\,\psi\,,
\ee
\ese
so that from (\ref{velocity}) we have
\bme
\label{velocity-new}
\be
u\,=\,-\,\pd{\psi}{z}\,,\qquad\qquad
w\,=\,\pd{\psi}{x}\,.
\ee
\eme
Hence the $E^{1/3}$-layer azimuthal vorticity and momentum equations (\ref{mom-vort-Ethird}$a$,$b$) become 
\bme
\label{gov-eq}
\be
2\pd{v}{z}\,=\,\pd{^4\psi}{x^4}\,,\qquad\qquad\qquad
2\pd{\psi}{z}\,=\,-\,\pd{^2v}{x^2}\,.
\ee
\eme
On regarding $v$, $w$ and $\psi$ as functions of $(y,\,z)$ (see (\ref{EL-coord}$b$)) rather than $(x,\,z)$, we may recast (\ref{gov-eq}) as
\bme
\label{gov-eq-y}
\be
2\biggl(\pd{v}{z}\,+\,z\pd{v}{y}\biggr)=\,\pd{^4\psi}{y^4}\,,\qquad\qquad
2\biggl(\pd{\psi}{z}\,+\,z\pd{\psi}{y}\biggr)=\,-\,\pd{^2v}{y^2}
\ee
\eme
\citep[c.f.][eq.~(4.3), who was concerned with the outer sphere Ekman layer]{P71}. This has the advantage that the fluid domain becomes simply $y>0$.

The system (\ref{gov-eq}) (or equivalently (\ref{gov-eq-y})) is to be solved subject to 
\bse
\label{bcs}
\begin{align}
\partial v/\partial z\,&=\,0\,,&&&     \psi\,&=\, 0  \qquad \mbox{on} &   z\,&=\, 0\,,& x\,&>\,0\,,\quad\\
v\,&=\,1\,,   & w\,&=\, 0\,,  &   \psi\,&=\,0 \qquad \mbox{on}  &     z\,&=\,\sqrt {-2x}\,,& x\,&<\,0\nonumber\\
&  & & & &\!\!\!\mbox{or simply}  &    y\,&=\,0\,,& &\\ 
v\,&\to\, 0\,, &&& \psi\,&\to\,0   \qquad\mbox{as}  & x\,&\to\,\infty&\!\!\!\!\!\!\!\!\!\!\!\!\mbox{or }\quad y\,&\to\, \infty\,,\\
v\,&\sim\, v_0^{bl}(\zeta)\,,  & w\,&\sim\, w_0^{bl}(\zeta)\,,  &\psi\,&\to\,0  \qquad \mbox{as} &   z\,&\to \, \infty\,,& \zeta\,&>\,0\,,
 \end{align}    
\ese
where
\bme
\label{z-to-infinity}
\be
v_0^{bl}\,=\,\eR^{-\zeta}\,\cos\zeta\,,  \qquad\qquad\qquad
w_0^{bl}\,=\,-\,\eR^{-\zeta}\,\sin\zeta\,,
\ee
\eme
in which $\zeta$ is the Ekman layer coordinate (\ref{EL-coord}$a$). The boundary conditions (\ref{bcs}) are homogeneous except for the requirement $v=1$ (\ref{bcs}$b$)$_1$ on the sphere boundary $\SC$ and the persistence of the associated Ekman boundary layer as $z\to\infty$ (\ref{bcs}$d$)$_{1,2}$. Also (\ref{bcs}$a$)$_{1,2}$ follow from the symmetry conditions (\ref{symmetries-EP}) across the equatorial plane in $\DC_\outC$. These are essentially the boundary conditions proposed in \S7 of \cite{S66} for the terminal $E^{2/5}$ Stewartson layer. There he also raises the possibility that (\ref{bcs}$d$) is not enough and that matching, as $z\to\infty$, with the similarity forms (\ref{Ethird-sol-similarity}$a$,$b$), namely
\bme
\label{EEL-similarity}
\be
v\,=\,z^{-5/12}V_0(\Phi)\,,\qquad\qquad\qquad  \psi\,=\,z^{-1/12}\Psi_0(\Phi)
\ee
\eme
in our dimensionless notation, may be needed \citep[but see also][eq.~(6.29)]{S66}. Our view is that (\ref{EEL-similarity}) follows consistently from the solution of the problem posed by (\ref{gov-eq}$a$,$b$) and (\ref{bcs}$a$-$d$).

On introduction of the complex variable
\be
\label{comp-w}
\wS\,=\,v(x,z)\,+\,\iR w(x,z)\,,
\ee
and differentiation of (\ref{gov-eq}$b$) with respect to $x$, the system of equations (\ref{gov-eq}$a$,$b$) may be expressed compactly, noting (\ref{velocity-new}$b$), as
\be
\label{velocity-new-complex}
\pd{^3\wS}{x^3}\,=\,2\iR\pd{\wS}{z}
\ee
\citep[see, e.g.,][]{MS69}, while the boundary conditions (\ref{bcs}) become
\bse
\label{bcs-complex}
\begin{align}
\left.\begin{array}{c} \Im\{\wS\}\\[0.2em]  \Re\{\partial \wS/\partial z\} \end{array}\right\}
\,&=\,0\,,&&&    &   \qquad \mbox{on} &   z\,&=\,0\,,& x\,&>\,0\,,\qquad\\
\wS\,&=\,1\,,   & &  &   & \qquad \mbox{on}  &    y\,&=\, 0\,,&&\\
\wS\,&\to\, 0\,, &&& & \qquad\mbox{as}  & x\,&\to\,\infty\,,&&\\
\wS\,&\sim\, \wS_0^{bl}(\zeta)\,,  &  &  &&  \qquad \mbox{as} &   z\,&\to \,\infty\,,& \zeta\,&>\,0\,,
\end{align}  
where
\vskip -4mm
\be
\wS_0^{bl}(\zeta)\,=\,\exp\bigl[-(1+\iR)\zeta\bigr]
\ee
$\bigl(\zeta=z^{1/2}y,\,\,y=x+\tfrac12 z^2,\,\,z>0\bigr)$, 
together with
\be
\Re\biggl\{\int_{-z^2/2}^\infty \wS\,\dR x \biggr\}= 0\,
\ee
\ese
which implements the zero axial flux condition: $\psi=0$ on $\SC$ with $\psi\to 0$ far from $\SC$.

\subsection{A \red{far-field} (including the \red{similarity sublayer}) boundary condition at large $z$\label{int-cond}}

Anticipating the numerical solution on a finite domain, our objective here is to replace the \red{far-field} boundary conditions $v\to 0$, $\psi\to 0$ on $-\infty<x<\infty$ as $z\to\infty$ by a boundary condition at finite $z=H\,(\gg 1)$. To that end, we restrict attention to the far-field $z\ge H\,(=\mbox{const.})$ on which we consider the Fourier transforms ${\hat \psi}(k,z)$, ${\hat v}(k,z)$:
\be
\label{FT}
\bigl[\psi,\,v\bigr](x,\,z)\,=\,\int_{-\infty}^\infty \bigl[{\hat \psi},\,{\hat v}\bigr](k,\,z)\exp(\iR kx)\,\dR k\,.
\ee

 The Fourier transform of (\ref{gov-eq}$a$,$b$) determines
\bme
\label{gov-eq-FT}
\be
2\pd{\hat v}z\,=\,k^4\hat\psi\,,\qquad\qquad\qquad
2\pd{\hat \psi}z\,=\,k^2\hat v\,.
\ee
\eme
The solution non-divergent as $z\to\infty$ is
\be
\label{int-constraint}
\hat v(k,z)\,=\,a(k) \exp\bigl(-\tfrac12 |k|^3z\bigr),
\ee
which together with (\ref{gov-eq-FT}$a$) leads to the relation
\be
\label{FT-solution}
\widehat{\pd{^2\psi}{x^2}}\,=\,-k^2{\hat \psi}\,=\,-\,\dfrac{2}{k^2}\,\pd{\hat v}{z}\,=\,|k|\,\hat v\,.
\ee
At $z=H$, it gives
\vskip -4mm
\be
\label{top-bc-FT}
\hat v(k,H)\,=\,\dfrac{1}{|k|}\widehat{\pd{^2\psi}{x^2}}(k,H)\,,
\ee
whose inverse determines the convolution integral
\be
\label{top-bc-0-new}
v(x,H)\,=\,-\,\dfrac{1}{\pi}\int_{-\infty}^\infty \ln|x-x'|\,\pd{^2\psi}{x^2}(x',H)\,\dR x'\,
=\,-\,\dfrac{1}{\pi}\,\dashint_{-\infty}^\infty \dfrac{1}{x-x'}\,\pd{\psi}{x}(x',H)\,\dR x'
\ee
($\dashint$ denotes the principle part). It will provide the basis of our boundary condition (\ref{symbc}$e$)\red{, which is implemented (see Appendix~\ref{iteration})  in the solution of our numerical model}.

Of course, the Stewartson similarity solution (\ref{EEL-similarity}), which determines
\bme
\label{EEL-similarity-H}
\be
v(x,H)\,=\,H^{-5/12}V_0\bigl(x/H^{1/3}\bigr),\qquad\qquad \psi(x,H)\,=\,H^{-1/12}\Psi_0\bigl(x/H^{1/3}\bigr),
\ee
\eme
satisfies (\ref{top-bc-0-new}), but significantly is only an approximation to the true \red{far-field} solution at finite $z=H$.  

\subsection{The numerical model \label{num-model}}

\begin{figure}
  \centerline{\includegraphics[width=0.8\textwidth,clip=true,trim=15 300 20 150]{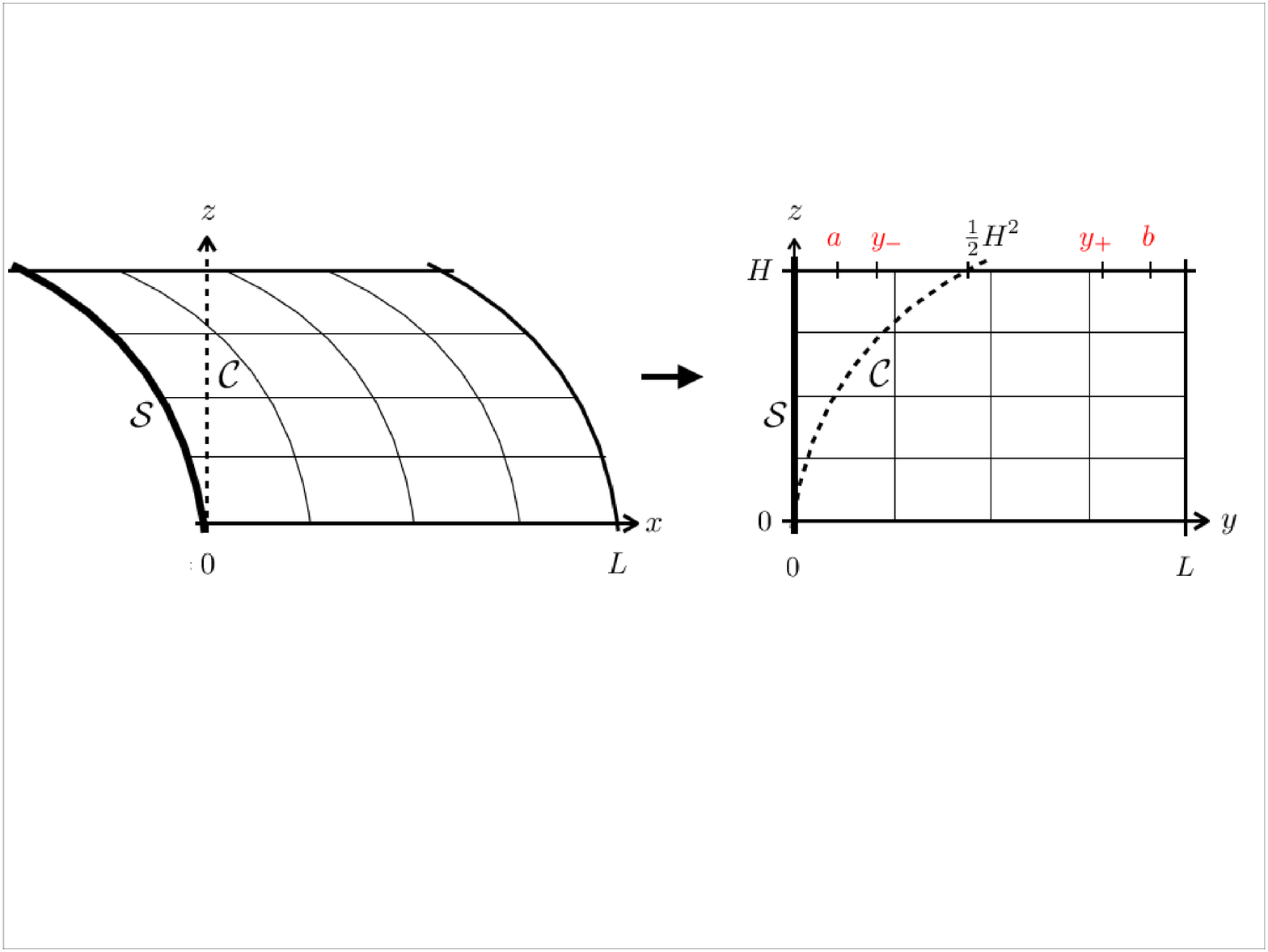}}
  \caption{\label{cdefo}\small{The map from the $x$-$z$ plane to the numerical box $[0,L]\times [0,H]$ in the $y$-$z$ plane. The tangent cylinder $\CC$ (dashed line) intersects the top boundary $z=H$ at $y=\tfrac12 H^2$, as indicated. The other locations $y=a$, $y_\pm$, $b$ \red{(labelled red in the print version) are} used in \red{our} implementation, described in Appendix~\ref{iteration}, of the top boundary condition (\ref{symbc}$e$) \red{based on (\ref{top-bc-0-new})}.}}
\end{figure}

In order to capture the equatorial structure of the Ekman layer and the emergence for $z\gg 1$ of Stewartson's self-similar solution (\ref{EEL-similarity}), we consider a 2-dimensional numerical domain of finite extent, and focus attention on a region bounded below by the equatorial plane $z=0$ and on its left-hand side by the parabolic contour $\SC$: $x=-\tfrac12 z^2$ (see (\ref{dim-bdry})). By adopting the alternative $(y,z)$--coordinates, recall that $x(y,z)\,=\,y\,-\,\tfrac12 z^2$ (see (\ref{EL-coord}$b$)), $\SC$ becomes the left-hand ``vertical'' boundary of our rectangular numerical box, $0\le y\le L$, $0\le z\le H$ (see figure~\ref{cdefo}), on which we solve (\ref{gov-eq-y}). The true problem associated with the boundary conditions (\ref{bcs}) is on an otherwise unbounded domain $L\to\infty$, $H\to\infty$. Evidently $L$ and $H$ must be sufficiently large that any approximate boundary conditions applied on both the right-hand boundary $y=L$ and the top boundary $z=H$ do not seriously influence the nature of the solution for $x=\OR(1)$, $z=\OR(1)$.

Our choice of domain dimensions is guided by the analytic results for $z\gg 1$. For the true unbounded domain, an Ekman layer forms against the $y=0$ boundary, outside which the amplitude of the motion is rather small and tends to zero as $z\to\infty$. In reality, the decay to zero is slow in the \red{similarity sublayer} on the tangent cylinder $\CC$: $x=0$, where  $\psi=\OR(z^{-1/12})$ (see (\ref{EEL-similarity}$b$)). Recall too that the thinning Ekman boundary layer of width $\OR(z^{-1/2})$, the thickening \red{similarity sublayer} of width $\OR(z^{1/3})$ and the remaining geostrophic (or mainstream) flow domain are only distinguishable for $z\gg 1$. Accordingly, a minimal requirement is that the top right-hand corner $(L,H)$ of our $(y,z)$--rectangle should lie well outside and to the right of the \red{similarity sublayer}, i.e., $x(L,H)=L-\tfrac12 H^2 \gg \OR(H^{1/3})$, so as not to interfere with it.

We discretise the differential system (\ref{gov-eq-y}) on a regular grid by means of finite differences, which for all the $y$-derivatives utilise a symmetric, second-order scheme, while to approximate the $z$-derivatives of the stream function, $\psi$ (azimuthal velocity, $v$) we use a third-order backward (forward) scheme.

We solve the discretised system subject to no-slip, impermeable boundary conditions \red{$\psi=0$, $\partial \psi/\partial y=0$, $v=1$ on $\SC$: $y=0$ (implemented in Appendix~\ref{iteration}: (\ref{symbc}$a$)), matching with the outer flow velocity, which essentially vanishes with $\psi=0$ (equivalently $\int_0^L w\,\dR y=0$), $\partial \psi/\partial y=0$, $v=0$ on $y=L$ (implementation (\ref{symbc}$b$)) and equatorial symmetry: $\psi=0$ (implying $\partial v/\partial z=0$) on $z=0$ (implementation (\ref{symbc}$c$)).} The correct (or rather practical) choice of the remaining top boundary condition for $v$ is however more challenging, because the natural prescription of the homogeneous Dirichlet boundary condition $v=0$ on $z=H$ (correct as $H\to\infty$) needs a very large value of $H$ to give a solution for $z=\OR(1)$ independent of $H$, i.e., a domain size far too large for numerical computation. Indeed, the most obvious difficulty is manifest by the \red{similarity sublayer}. For, if the value of $v$ on the top boundary albeit small, is not the value realised by the true unbounded solution, then it will act as the source of a reflected disturbance. This spurious reflection is due to the propagative nature of the advection-like terms on the left-hand side of (\ref{gov-eq-y}$a$,$b$) along geostrophic cylinders, $x\;$constant, in both the positive and negative $z$-directions. Unfortunately, even improving the accuracy by replacing  $v(x,H)=0$ with the Stewartson similarity solution $v(x,z)=H^{-5/12}V_0(x/H^{1/3})$ (see (\ref{EEL-similarity-H}$a$)) helps little to eliminate the spurious reflection. A better option is to apply the non-local boundary condition (\ref{top-bc-0-new}) that essentially provides the needed non-reflective (i.e., ``soft") boundary condition.  However, a number of difficulties are immediately apparent, the most obvious being that (\ref{top-bc-0-new}) is a mainstream boundary condition which ignores the Ekman boundary layer on $\SC$. It also involves an integral over $-\infty<x<\infty$, whereas the numerical domain size is finite. Neither of these shortcomings lead to unresolvable difficulties and the implementation of (\ref{top-bc-0-new}) using the iterative procedure described \red{in Appendix~\ref{iteration}} appears to be surprisingly effective.

Finally we emphasise that the top boundary condition for $v_{0}(y,H)$ neither includes nor anticipates the Stewartson similarity form $v\approx z^{-5/12}V_0(\Phi)$,  $\Phi=x(y,z)/z^{1/3}$ (see (\ref{EEL-similarity}$a$) with (\ref{Ethird-sol-similarity}$c$)), valid for $z\gg 1$. However in the next \S\ref{sl-ff} we generate the series (\ref{EEL-series}$a$)) involving higher order terms each of similarity form (see (\ref{EEL-series-summary}$a$)) with an arbitrary constant. Consideration of the Ekman boundary layer in the later \S\ref{El-s} determines the constants, which are contained in each of the formulae  (\ref{Yp-Xi-Psi-1-Summary}$a$) and (\ref{Yp-Xi-Psi-2-Summary}$a$) for  $z^{-5/6}V_1(\Phi)$ and $z^{-15/12}V_2(\Phi)$ that comprise the sum (\ref{EEL-series-summary}$a$). The numerical solution for $v(y,H)$ and $\psi(y,H)$ is compared with the asymptotic expansions (\ref{EEL-series-summary}$a$,$b$) in \S\ref{Numerical-results}. 


\section{\red{The far-field similarity sublayer:} $x=\OR(z^{1/3})$, $z\gg 1$\label{sl-ff}}

In the far-field, as $z\to \infty$, the \red{similarity sublayer} $x=\OR(z^{1/3})$ and the Ekman layer $y=\OR(z^{-1/2})$ ($y-x=\tfrac12 z^2$) become distinguishable as separate regions. In the \red{similarity sublayer}, we anticipate that, on increasing $z$, the solution merges with the Stewartson solution (\ref{EEL-similarity}) driven by the Ekman jump condition at the boundary $\SC$: $v=1$ on the sphere and $v=0$ just outside. However, the realised non-zero value $v=z^{-5/12}V_0(-z^{5/3}/2)$ just outside will itself drive yet smaller meridional circulation and zonal flow contributions. 

In preparation for such an implicit expansion we utilise results of \cite{MS69}, who considered a variety of similarity solutions of the form
\bse
\label{W-Zp}
\begin{align}
\wS(x,z)\,=\,\WS(x,z;p)\,\bigl(=\,V+\iR W\bigr)\,&=\,\aS(p)\,z^{-p/3}\ZS(\Phi;p)\,,\\
\aS(p)\,&=\,-\,2^{-1/2}(p-1)2^{-(p-1)}\,,
\end{align}
\ese
dependant on the real constant $p\,(>1)$, which satisfies (\ref{velocity-new-complex}) when
\be
\label{Zp-equ}
\ZS^{\,\prime\prime\prime}\,+\,\tfrac23\iR\bigl(\Phi\ZS^{\,\prime}\,+\,p\ZS\bigr)=\,0
\ee
(here the $^\prime$ denotes the $\Phi$-derivative). The solution
\be
\label{Zp-sol}
\ZS(\Phi;p)\,=\,\dfrac{\exp\bigl(-\,\iR p\pi/2\bigr)}{\Gamma(p)}\int_0^\infty\varpi^{(p-1)}\,\exp\Bigl(\iR \varpi\Phi\,-\,\tfrac12\varpi^3\Bigr)\,\dR\varpi
\ee
with the symmetry property 
\be
\label{Zp-symm}
\ZS(-\Phi;p)\,=\,\eR^{-\iR p\pi}\,\ZS^*(\Phi;p)
\ee
(the $^*$ denotes complex conjugate) satisfies the boundary conditions (\ref{bcs-complex}$a$): $w=0$, $\partial v/\partial z=0$ on $x>0$, $z=0$ (see (\ref{VPhi}$a$,$b$) and Appendix~\ref{appA}). The $\Phi$-derivative of (\ref{Zp-sol}) shows that $\ZS(\Phi;p)$  satisfies the recurrence relation
\be
\label{Zp-rec}
\ZS^{\,\prime}(\Phi;p-1)\,=\,-\,(p-1)\,\ZS(\Phi;p)
\ee
for $p>1$, a result that is useful in the construction of 
\begin{align}
\Psi(x,z;p)=\,&-\,\int_x^\infty W(p)\,\dR x\,=\,-\,\aS(p)z^{-(p-1)/3}\int_\Phi^\infty\Im\bigl\{\ZS(p)\bigr\}\,\dR\Phi\nonumber\\
=\,&-\,\dfrac{\aS(p-1)}{2(p-2)}z^{-(p-1)/3}\Im\bigl\{\ZS(p-1)\bigr\} =\,-\,\dfrac{W(x,z,p-1)}{2(p-2)}\qquad(p>1)\quad
\label{psip}
\end{align}
$\bigl(W=\partial\Psi/\partial x\bigr)$. Then from (\ref{W-Zp}$a$,$b$), (\ref{Zp-sol}) and (\ref{psip}) we obtain
\bse
\label{VPhi}
\begin{align}
{\hskip -2mm}V(x,z;p)=\,&\dfrac{2^{-p+1/2}z^{-p/3}}{\Gamma(p-1)}\int_0^\infty\varpi^{p-1}\!\cos\bigl[\varpi\Phi-\tfrac12(p-2)\pi\bigr]\exp\bigl[-\tfrac12\varpi^3\bigr]\dR\varpi\,,\\
{\hskip -2mm}\Psi(x,z;p)=\,&-\,\dfrac{2^{-p+1/2}z^{-(p-1)/3}}{\Gamma(p-1)}\int_0^\infty\!\varpi^{p-2}\cos\bigl[\varpi\Phi-\tfrac12(p-2)\pi\bigr]\exp\bigl[-\tfrac12\varpi^3\bigr]\dR\varpi\,.
\end{align}
\ese
The asymptotic forms for $|\Phi|\gg 1$ are outlined in Appendix~\ref{appA}. There it is explained that the normalisation of $\WS$ by a real (as opposed to complex) constant $\aS(p)$ is essential to guarantee $\partial V/\partial z=\Psi=0$ when $x>0$, as required by the symmetry condition (\ref{bcs}$a$).
 
The formulae (\ref{VPhi}$a$,$b$), for the case $p=5/4$, recover Stewartson's solution (\ref{Ethird-sol-similarity}$c$,$d$):
\bme
\label{EEL-similarity_0}
\be
V(x,z;5/4)\,=\,z^{-5/12}V_0(\Phi)\,,\qquad\qquad\qquad  \Psi(x,z;5/4)\,=\,z^{-1/12}\Psi_0(\Phi)\,.
\ee
\eme
In the next section we show that matching with the Ekman layer ($z\gg 1$) on the sphere leads to the asymptotic expansions 
\bme
\label{EEL-series}
\be
v\,=\,v^{sl}\,\approx\,\overset{2}{\underset{q=0}{\textstyle\sum}}a_q V\bigl(x,z;p(q)\bigr),\qquad\qquad
\psi\,=\,\psi^{sl}\,\approx\,\overset{2}{\underset{q=0}{\textstyle\sum}}a_q \Psi\bigl(x,z;p(q)\bigr)
\ee
\be\se
\wS\,=\,\wS^{sl}\,=\,v^{sl}+\iR w^{sl}\,\approx\,\overset{2}{\underset{q=0}{\textstyle\sum}}a_q\WS\bigl(x,z;p(q)\bigr),
\ee
where
\be
p(q)=5(q+1)/4 \qquad\qquad \mbox{and}\qquad\qquad   a_0=1\,;
\ee
\eme
while the remaining real constants $a_1$ and  $a_2$ are as yet unknown

The consistency of the ansatz (\ref{EEL-series}) relies on the values of $\wS^{sl}$  and $\psi^{sl}$ on the sphere boundary, $\SC$: $z=(-2x)^{1/2}(\gg 1)$. For $\wS^{sl}$, we simply retain the two leading terms
\bse
\label{W-SC}
\be
\bigl.\wS^{sl}\bigr|_\SC\,\approx\,\WS\bigl(-\,\tfrac12 z^2,z;5/4\bigr)\,+\,a_1\WS\bigl(-\,\tfrac12 z^2,z;5/2\bigr)
\ee
($q=0$, $1$) of (\ref{EEL-series}$c$), into which we substitute only the leading term of the large $-\Phi$ expansion (\ref{VPhi-asym}$c$) for $\WS$ $\big($recall that $-x=\tfrac12 z^2\gg z^{1/3}$ on $\SC$$\big)$ giving
\be
\bigl.\wS^{sl}\bigr|_\SC\,\approx\,\dfrac{1-\iR}{4}z^{-5/2}\,+\,\dfrac{3\iR}{\sqrt{2}}a_1z^{-5}\,.
\ee
\ese
However, for $\psi^{sl}$, we require all three terms
\bse
\label{Psi-SC}
\be
\bigl.\psi^{sl}\bigr|_\SC\,\approx\,\Psi\bigl(-\,\tfrac12 z^2,z;5/4\bigr)\,+\,a_1\Psi\bigl(-\,\tfrac12 z^2,z;5/2\bigr)\,+\,a_2\Psi\bigl(-\,\tfrac12 z^2,z;15/4\bigr) 
\ee
($q=0$, $1$, $2$)  of (\ref{EEL-series}$b$), and in the case of the first $p(0)=5/4$ term we also need the second term of (\ref{VPhi-asym}$b$):
\be
\bigl.\psi^{sl}\bigr|_\SC\,\approx\,-\,\frac12 z^{-1/2}\,+\,\frac{a_1}{\sqrt2}z^{-3}
\,-\,\biggl(\dfrac{45}{32}\,+\,\dfrac{a_2}{2}\biggr)z^{-11/2}\,.
\ee
\ese

Consequent upon $\Delta p\equiv p(q+1)-p(q)=5/4$ (see (\ref{EEL-series}$d$)), the series (\ref{W-SC}$a$) and (\ref{Psi-SC}$a$) both involve the expansion parameter $z^{-2\Delta p}=z^{-5/2}$, a parameter that also emerges from the Ekman layer equation (\ref{W-hat-eq}) studied in the following \S\ref{El-s}. The ensuing blend accounts for the curious choice $\Delta p=5/4$. In this way, the results (\ref{a1-val}) and (\ref{a2-val}) for $a_1$ and $a_2$ respectively complete the far-field \red{($z\gg 1$) similarity sublayer} solution (\ref{EEL-series}):
\bse
\label{EEL-series-summary}
\begin{align}
v^{sl}\,\approx&\overset{2}{\underset{q=0}{\textstyle\sum}}z^{-p(q)/3}V_q(\Phi)=\,z^{-5/12}V_0(\Phi)\,+\,z^{-5/6}V_1(\Phi)\,+\,z^{-15/12}V_2(\Phi)\,,\\
\psi^{sl}\,\approx&\overset{2}{\underset{q=0}{\textstyle\sum}}z^{-[p(q)-1]/3}\Psi_q(\Phi)\,=\,z^{-1/12}\Psi_0(\Phi)\,+\,z^{-1/2}\Psi_1(\Phi)\,+\,z^{-11/12}\Psi_2(\Phi)\,,
\end{align}
\ese
involving the expansion parameter $z^{-\Delta p/3}=z^{-5/12}$, where $-V_0(\Phi)$ and $-\Psi_0(\Phi)$ are the Stewartson functions (\ref{EEL-similarity_0}$a$,$b$) (i.e., (\ref{Ethird-sol-similarity}$c$,$d$)) and
\bse
\label{Yp-Xi-Psi-1-Summary}
\begin{align}
V_1(\Phi)\,=&\,\,\dfrac{9}{16\sqrt{2\pi}}\,\int_0^\infty\varpi^{3/2}\sin\Bigl(\varpi \Phi\,+\,\dfrac{\pi}{4}\Bigr)\,\exp\bigl(-\tfrac12\varpi^3\bigr)\,\dR\varpi\,,\qquad\qquad\quad
\\[0.3em]
\Psi_1(\Phi)\,=&\,-\,\dfrac{9}{16\sqrt{2\pi}}\,\int_0^\infty\varpi^{1/2}\sin\Bigl(\varpi \Phi\,+\,\dfrac{\pi}{4}\Bigr)\,\exp\bigl(-\tfrac12\varpi^3\bigr)\,\dR\varpi\,,\qquad\qquad\quad
\end{align}
\ese
\bse
\label{Yp-Xi-Psi-2-Summary}
\begin{align}
V_2(\Phi)=&-\,\dfrac{143\times 2^{1/4}\,\Gamma(1/4)}{21 \times 256\,\pi}\int_0^\infty\varpi^{11/4}\cos\Bigl(\varpi \Phi\,+\,\dfrac{\pi}{8}\Bigr)\,\exp\bigl(-\tfrac12\varpi^3\bigr)\,\dR\varpi\,,
\\[0.3em]
\Psi_2(\Phi)\,=&\,\dfrac{143\times 2^{1/4}\,\Gamma(1/4)}{21 \times 256\,\pi}\int_0^\infty\varpi^{7/4}\cos\Bigl(\varpi \Phi\,+\,\dfrac{\pi}{8}\Bigr)\,\exp\bigl(-\tfrac12\varpi^3\bigr)\,\dR\varpi\,,
\end{align}
\ese
in which we have simplified using Gamma function properties \citep[see][http://dlmf.nist.gov/5.4 and /5.5]{AS10}.


\section{The Ekman layer on the sphere\label{El-s}}

The failure of $\bigl.\wS^{sl}\bigr|_\SC$ and $\bigl.\psi^{sl}\bigr|_\SC$ (see (\ref{W-SC}$b$) and (\ref{Psi-SC}$b$)) to satisfy the wall boundary conditions $\bigl.\wS\bigr|_\SC=1$, $\bigl.\psi\bigr|_\SC=0$ is the raison d'\^etre for the Ekman layer, which we consider in this section. To that end, we write
\bme
\label{sl-bl}
\be
\wS\,=\,\wS^{sl}\,+\,\wS^{bl}\,,\qquad\qquad \psi\,=\,\psi^{sl}\,+\,\psi^{bl}
\ee
\eme
where $\wS^{bl}$ and $\psi^{bl}$ are the Ekman layer contributions, which satisfy the governing equation (\ref{velocity-new-complex}), and boundary conditions
\bme
\label{sl-bl-bcs}
\begin{align}
\qquad\wS^{bl}\,&=\,1\,-\bigr.\wS^{sl}\bigr|_\SC \,,&  \psi^{bl}\,&=\,-\bigr.\psi^{sl}\bigr|_\SC &\mbox{on} \qquad \zeta=0\qquad\\
\intertext{(see (\ref{bcs-complex}$b$)) and matching conditions}
\qquad\wS^{bl}\,&\to \,0 \,,& \psi^{bl}\,&\to\, 0 &\mbox{as} \qquad \zeta\to\infty\qquad
\end{align}
\eme
\red{(see (\ref{bcs-complex}$c$,$f$))}.

We consider solutions of the form
\bme
\label{bl-form}
\be\te
\wS^{bl}\,=\,\WC(\zeta,z) \,,\qquad  \psi^{bl}\,=\,\PC(\zeta,z)\,,
\qquad  \PC(\zeta,z)\,=\,-\,z^{-1/2}\int_\zeta^\infty\Im\{\WC\}\,\dR\zeta\,,
\ee
\eme
recall that $\zeta=z^{1/2}\bigl(x+\tfrac12 z^2\bigr)$ (see (\ref{EL-coord}$a$,$b$)), where $\WC$ solves (\ref{velocity-new-complex}) when
\be
\label{W-hat-eq}
\pd{^3\WC}{\zeta^3}\,-\,2\iR\pd{\WC}{\zeta} \,=
\,\dfrac{2\iR}{z^{5/2}}\biggl(z\pd{\WC}{z}\,+\,\dfrac{\zeta}{2}\pd{\WC}{\zeta}\biggr).
\ee
Evidently a power series solution exists of the form
\bse
\begin{align}
\WC(\zeta,z)\,\approx&\,\,\,\WC_0(\zeta)\,+\,z^{-5/2}\WC_1(\zeta)\,+\,z^{-5}\WC_2(\zeta)\,,\\
\PC(\zeta,z)\,\approx&\,z^{-1/2}\PC_0(\zeta)\,+\,z^{-3}\PC_1(\zeta)\,+\,z^{-11/2}\PC_2(\zeta)\,,
\end{align}
\ese
as anticipated by the \red{similarity sublayer} expansion (\ref{EEL-series}) and the resulting boundary conditions implied by (\ref{W-SC}$b$) and (\ref{Psi-SC}$b$). From (\ref{W-hat-eq}) each $\WC_n$ satisfies
\bse
\label{Wn-hat-eq}
\be
\WC^{\,\prime\prime\prime}_n\,-\,2\iR\WC^{\,\prime}_n \,=\,\left\{\begin{array}{ll}
\,0& \qquad (n=0)\,,\\[0.4em]
\,\iR\bigl[\zeta\WC^{\,\prime}_{n-1}\,-\,5(n-1)\WC_{n-1}\bigr]& \qquad (n=1\,,2)\end{array}\right.
\ee
(here the $^\prime$ denotes the $\zeta$-derivative) and from (\ref{bl-form}$c$)
\be
\PC_n(\zeta)\,=\,-\,\int_\zeta^\infty\Im\{\WC_n\}\,\dR\zeta\qquad\qquad (n=0\,,1\,,2)\,.
\ee
\ese
In the following subsections, we outline the solutions at each order $n=0\,,1\,,2$.

\subsection{Zeroth order problem for $\WC_0$, $\PC_0$\label{El-s0}}

The $n=0$ solution of (\ref{Wn-hat-eq}$a$) subject to $\WC_0(0)=1$ is
\bme
\label{EL-comp-sol-0}
\be
\WC_0(\zeta)\,=\,E(\zeta)\,,\qquad\quad\mbox{where}\qquad E(\zeta)\,=\,\exp\bigl[-(1+\iR)\zeta\bigr]\,,
\ee
which from (\ref{Wn-hat-eq}$b$) determines
\be
\PC_0(\zeta)\,=\,\Im\Bigl\{-\,\tfrac12 (1-\iR)E(\zeta)\Bigr\}\,.
\ee
\eme
At $\zeta=0$ it gives $\PC_0(0)=1/2$ consistent with the boundary condition (\ref{sl-bl-bcs}$b$) and the leading order term in (\ref{Psi-SC}$b$). It should be emphasised that this result is not automatic but the reason why \cite{S66} chose his normalisation for his zeroth order \red{similarity sublayer} solution (\ref{EEL-similarity}$a$,$b$).\

\subsection{First order problem for $\WC_1$, $\PC_1$\label{El-s1}}

With $\WC_0(\zeta)=E(\zeta)$ the first order ($n=1$) equation (\ref{Wn-hat-eq}$a$) becomes  
\be
\label{W1-hat-eq}
\WC^{\,\prime\prime\prime}_1\,-\,2\iR\WC^{\,\prime}_1 \,=\,\iR\WC^{\,\prime}_0\,=\,(1-\iR)\zeta E(\zeta)\,.
\ee
The $\zeta=0$ boundary condition determined by the $\OR(z^{-5/2})$ term in (\ref{W-SC}$b$) and (\ref{sl-bl-bcs}$a$) is $\WC_1(0)=-\tfrac14(1-\iR)$. Subject to that and $\WC_1(\zeta)\to 0$ as $\zeta\to \infty$, the solution of (\ref{W1-hat-eq}) is
\bse
\label{EL-comp-sol-1}
\be
\WC_1(\zeta)\,=\,-\,\tfrac14\bigl[(1-\iR)\,+\,\tfrac32\zeta\,+\,\tfrac{1}{2}(1+\iR)\zeta^2\,\bigr]E(\zeta)\,.
\ee
Then the $n=1$ integral (\ref{Wn-hat-eq}$b$) determines
\be
\PC_1(\zeta)\,=\,\Im\biggl\{\dfrac{1}{16}\bigl[-9\iR\,+\,5(1-\iR)\zeta\,+\,2\zeta^2\bigr]E(\zeta)\biggr\}.
\ee
\ese
At $\zeta=0$, it gives $\PC_1(0)=-9/16$, which by (\ref{sl-bl-bcs}$b$) and the $\OR(z^{-3})$ term in (\ref{Psi-SC}$b$) yields
\be
\label{a1-val}
a_1\,=\,\dfrac{9}{8\sqrt{2}}\,.
\ee

\subsection{Second order problem for $\WC_2$, $\PC_2$\label{El-s2}}

With $\WC_1(\zeta)$ given by (\ref{EL-comp-sol-1}$a$) the second  order ($n=2$) equation (\ref{Wn-hat-eq}$a$) becomes  
\begin{align}
\WC^{\,\prime\prime\prime}_2\,-\,2\iR\WC^{\,\prime}_2 \,=&\,\iR\bigl(\zeta\WC^{\,\prime}_1\,-\,5\WC_1\bigr)\nonumber\\
=&\,\tfrac14\bigl[5(1+\iR)\,+\,8\iR\zeta\,-\,3(1-\iR)\zeta^2\,-\,\zeta^3\bigr]E(\zeta)\,.
\label{W2-hat-eq}
\end{align}
The $\zeta=0$ boundary condition determined by the $\OR(z^{-5})$ term in (\ref{W-SC}$b$) and (\ref{sl-bl-bcs}$a$) is $\WC_2(0)=-\bigl(3\iR/\sqrt2\,\bigr)a_1=-(27/16)\iR$ (see (\ref{a1-val})). The solution of (\ref{W2-hat-eq}) that decays to zero as $\zeta\to \infty$ is
\bse
\label{EL-comp-sol-2}
\be
\WC_2(\zeta)\,=\,-\,\dfrac{1}{16}\biggl[27\iR\,-\,\dfrac{145}{16}(1-\iR)\zeta\,-\,\dfrac{89}{8}\zeta^2\,-\,\dfrac{7}{4}(1+\iR)\zeta^3\,-\,\dfrac{\iR}{4}\zeta^4\biggr]E(\zeta)\,.
\ee
Then the $n=2$ integral (\ref{Wn-hat-eq}$b$) determines
\be
\PC_1(\zeta)\,=\,\Im\biggl\{\dfrac{1}{64}\biggl[\dfrac{863}{8}(1+\iR)\,+\,\dfrac{431}{4}\iR\zeta\,-\,\dfrac{143}{4}(1-\iR)\zeta^2\,-\,{9}\zeta^3\,-\,\dfrac{1+\iR}{2}\zeta^4\biggr]E(\zeta)\biggr\}.
\ee
\ese 
At $\zeta=0$ it gives $\PC_2(0)=-863/512$, which by (\ref{sl-bl-bcs}$b$) and the $\OR(z^{-11/2})$ term in (\ref{Psi-SC}$b$) yields
\be
\label{a2-val}
a_2\,=\,\dfrac{863}{256}\,-\,\dfrac{45}{16}\,=\,\dfrac{143}{256}\,.
\ee


\section{Numerical results\label{Numerical-results}}

We stress from the outset that the non-dimensionalisation (\ref{dim-vel}) of $\uv$ only retains the sign convention of the dimensional velocity $\uv^\star$, when $\epsilon<0$. We adopt that negative sign of $\epsilon$ in our discussion of the numerics with the consequence that the inner sphere is rotating faster than the surrounding fluid. Accordingly positive (negative) values for $v$ denote eastward (westward) velocities relative to the far-field rigid rotation $\Omegav+\epsilon\Delta \Omegav$ (see the remarks below (\ref{dim-vel}$a$)). In the broader context of the Stewartson's full geometry, our choice of negative $\epsilon$ implies that the inner sphere rotates faster than the outer.

As stated in the penultimate paragraph of \S\ref{num-model}, for the numerical box (see figure~\ref{cdefo}) with dimensions $[0,L]\times [0,H]$ in the $y$-$z$ plane, our choice $L=60$, $H=7$ suffices, i.e., the solutions are independent of any further increase in box size. With those values the tangent cylinder $\CC$: $y=\tfrac12 z^2$ intersects the top boundary $z=7$ at $y=24.5$ well clear of both the sphere surface $\SC$: $y=0$ and right-hand edge $y=60$ of the computational domain (see figure~\ref{cdefo}). We found clear evidence of both $v$ and $\psi$ decaying to zero long before the edge $y=60$ is reached. Near the top $z=7$, the effectiveness of the non-local boundary condition (\ref{symbc}$e$) was manifest by the absence of any reflected disturbance, an idea that we expand upon and quantify in the following subsections.

\subsection{Comparison with full spherical shell DNS\label{more-DNS}}

\begin{figure}
  \centerline{\raisebox{26mm}{\small (a)}\includegraphics[width=0.47\textwidth,clip=true,trim=15 25 0 0]{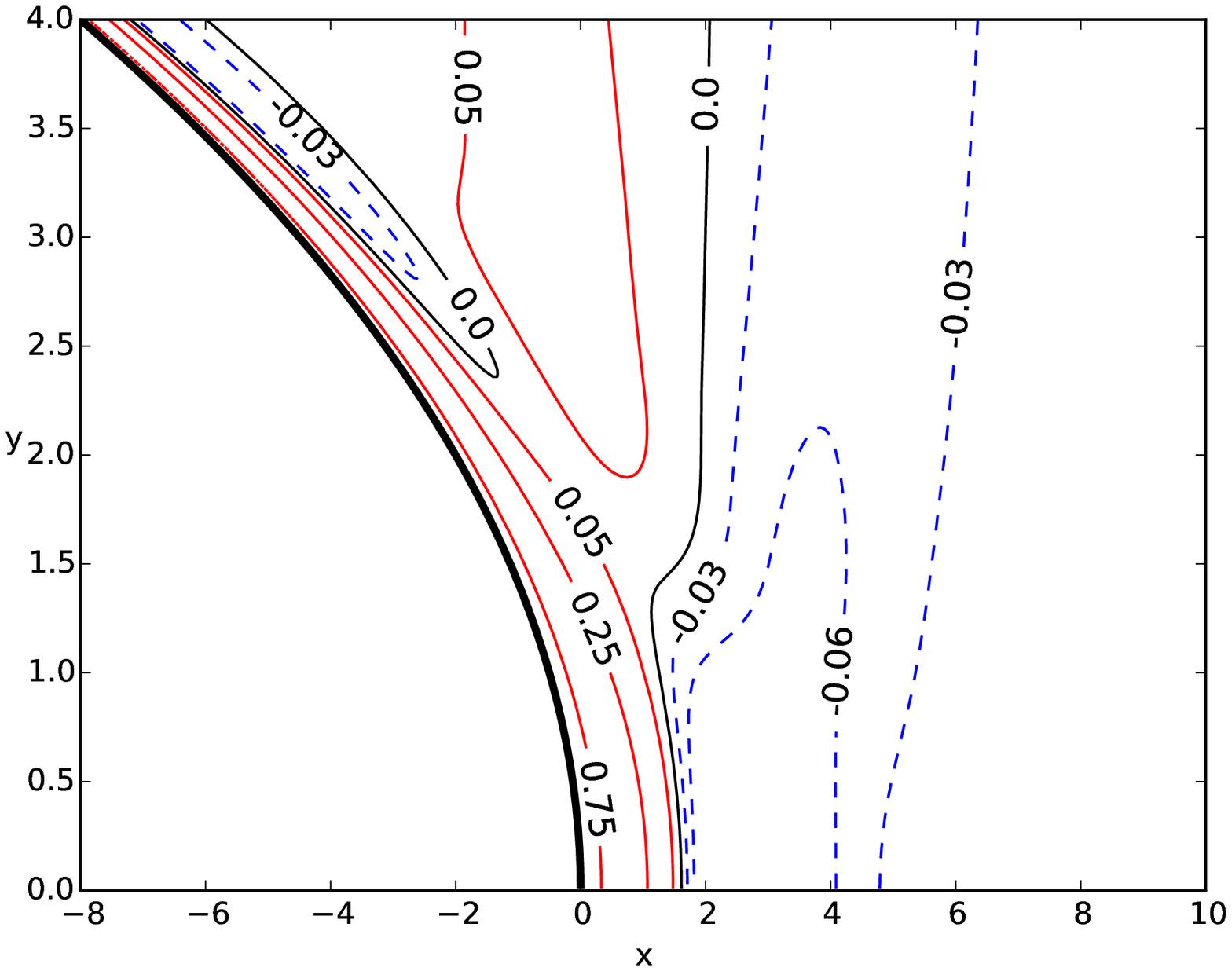}\ 
\raisebox{26mm}{\small (b)}\includegraphics[width=0.47\textwidth,clip=true,trim=15 25 0 0]{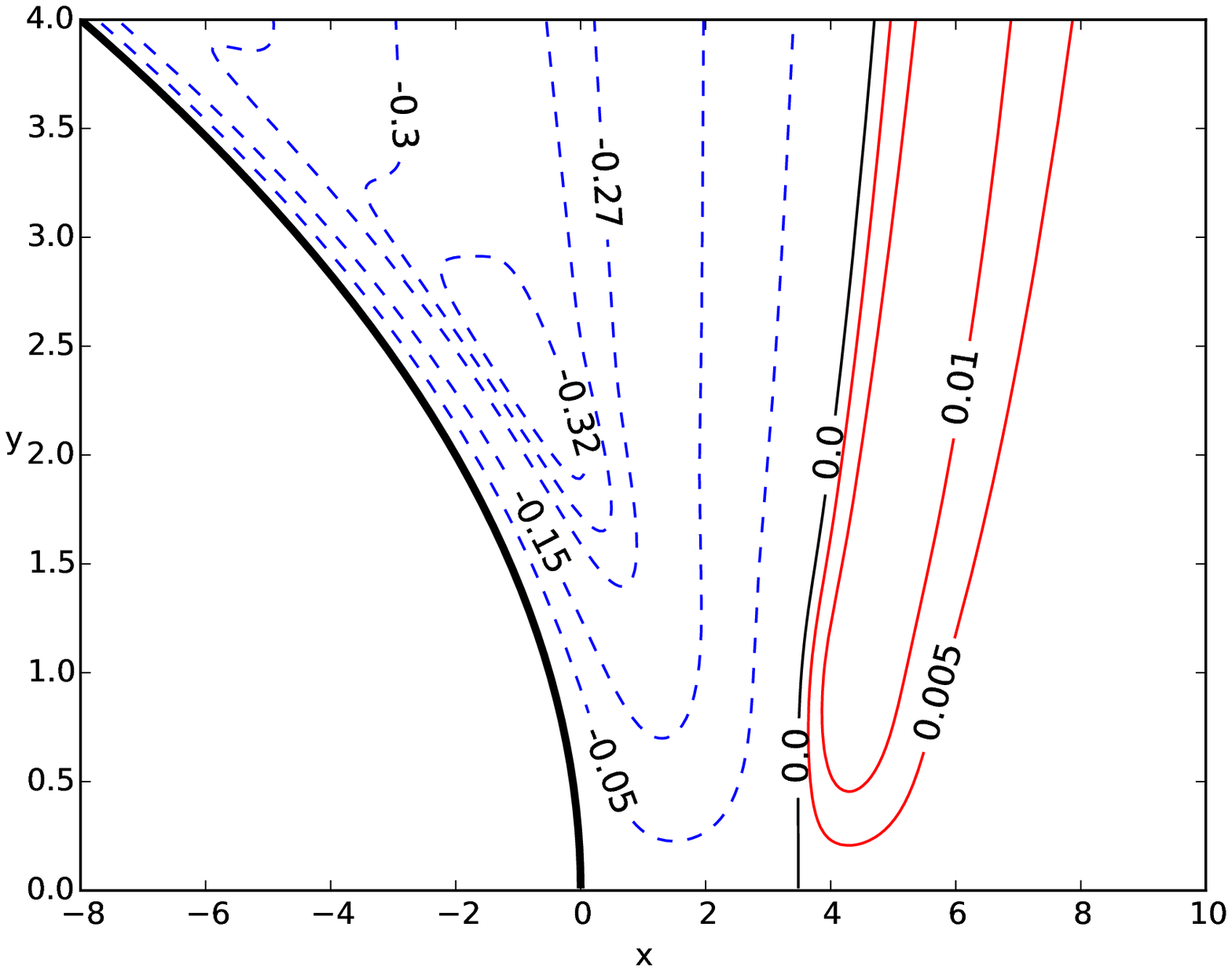}}
  \caption{\label{contours}Contours \red{(negative and non-negative values identified by broken and continuous curves respectively)} of ($a$) azimuthal velocity $v$ and ($b$) streamfunction $\psi$ in undeformed $x,\,z$ coordinates (see figure~\ref{cdefo}), in the vicinity of the equator $(x,\,z)=(0,\,0)$. The inner sphere boundary $\SC$ is approximated by the parabolic contour $x=-\tfrac12 x^2$ (thick line).}
\end{figure}

\begin{figure}
  \centerline{\raisebox{26mm}{\small (a)}\includegraphics[width=0.47\textwidth,clip=true,trim=15 25 0 0]{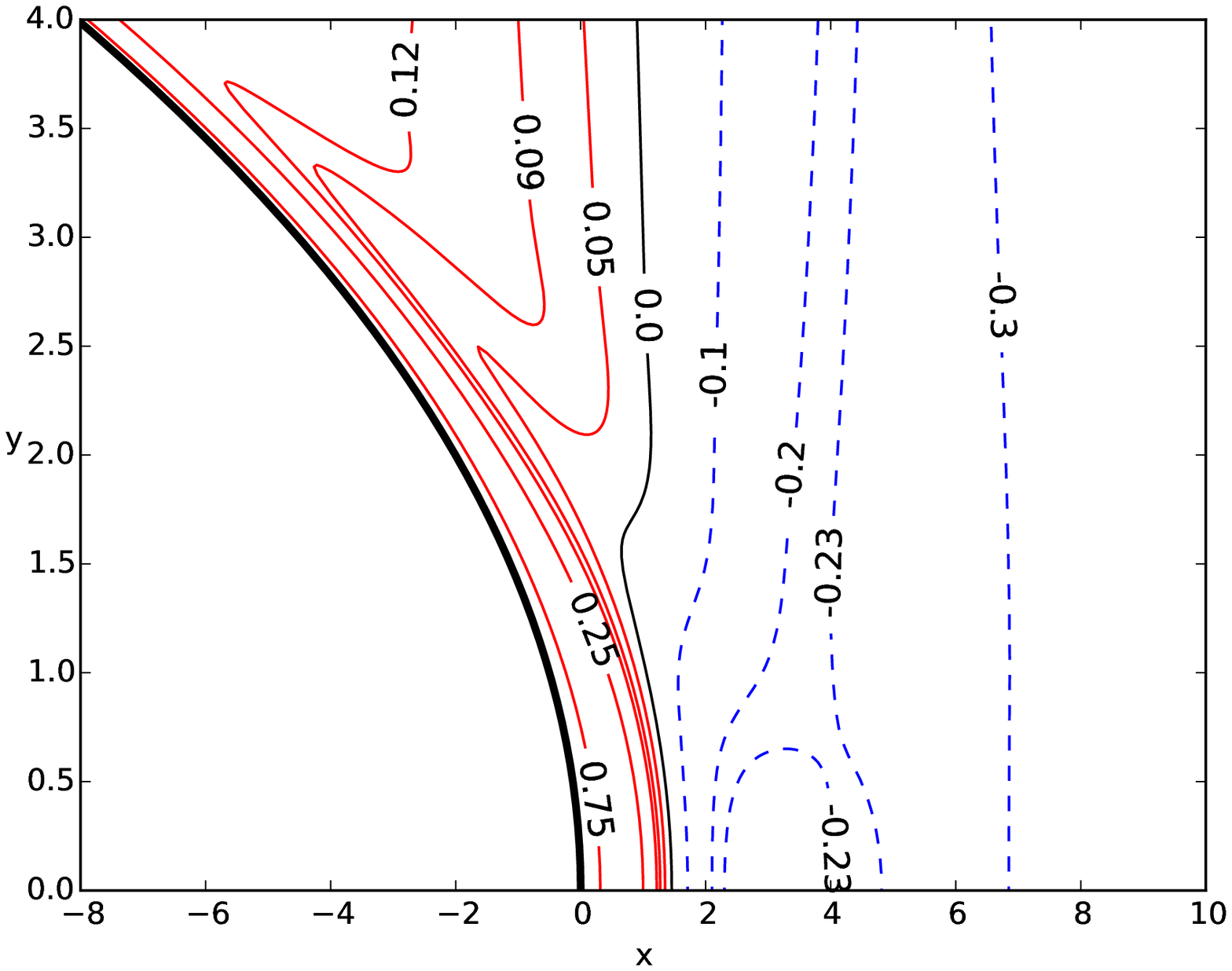}\ 
  \raisebox{26mm}{\small (b)}\includegraphics[width=0.47\textwidth,clip=true,trim=15 25 0 0]{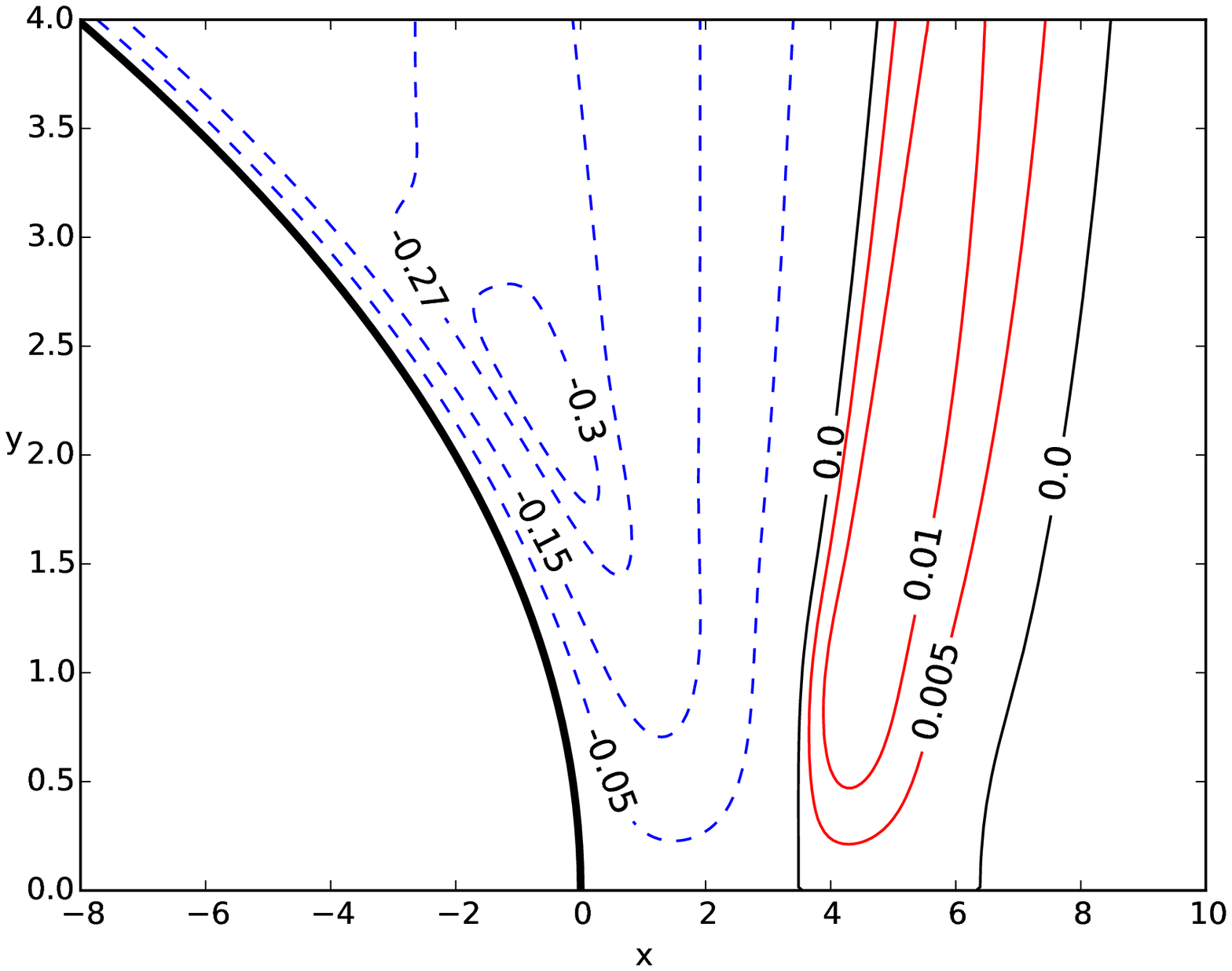}}
   \caption{\label{full_shell_num-blow_up}As in figure~\ref{contours} but for the full shell DNS in the case $\alpha^{-1}=0.35$,  $E=10^{-7}$ blown up near the equator.} 
\end{figure}

Much of the solution obtained resembles the DNS-results already portrayed on figures~\ref{full_shell_num}($a$,$b$). To emphasise the behaviour associated with the equatorial Ekman layer, we provide contour plots of our solution for the azimuthal velocity $v$ and stream-function $\psi$ on a small $[0,10]\times [0,4]$ box in figures~\ref{contours}($a$,$b$). We also show results for a DNS-solution of the full shell problem in figure~\ref{full_shell_num-blow_up}($a$,$b$) for $\alpha^{-1}=0.35$, as in figure~\ref{full_shell_num} but now for $E=10^{-7}$, with contour values expressed in the units of $v$ and $\psi$ introduced in (\ref{dim-vel}).  

In view of the extremely small powers of $E$ in our expansions  and the approximations based on them leading to our equatorial Ekman layer problem, detailed quantitative agreement between figures~\ref{contours}~and~\ref{full_shell_num-blow_up} is not to be expected. Nevertheless the $\psi$-comparison (figures~\ref{contours}($b$), \ref{full_shell_num-blow_up}($b$)) is rather good. In contrast, the $v$-comparison (figures~\ref{contours}($a$), \ref{full_shell_num-blow_up}($a$)) shows good qualitative agreement well inside the equatorial Ekman layer \red{(say, $0\le z\lessapprox 2$)}, in the sense that the closed contour on the equator is visible in both figures, but detailed comparison for large \red{$z\,(\gtrapprox 2)$} is poor. This does not mean the asymptotics is incorrect but rather that higher order terms ignored in the asymptotics are significant at the finite $E$ used to produce figure~\ref{full_shell_num-blow_up}($a$). 

\begin{figure}
  \centerline{\includegraphics[width=0.6\textwidth,clip=true]{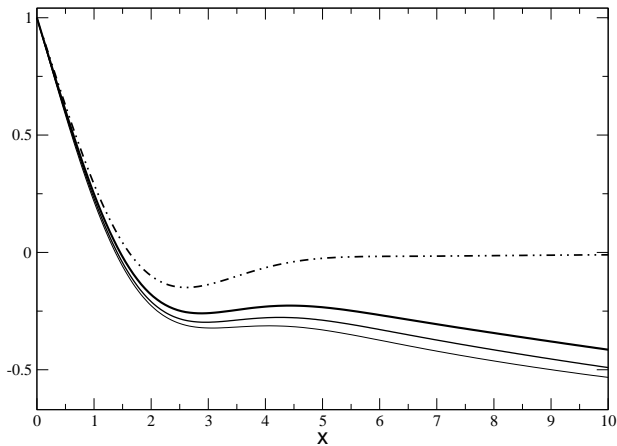}}
  \caption{\label{shear_equator} Profiles, for various values of $E$, of the azimuthal velocity $v$ versus $x$ at $z=0$. The dash-dotted line, corresponds to the equatorial Ekman layer solution portrayed in figure~\ref{contours}($a$). The DNS-results are identified by the continuous lines of increasing width: $E=10^{-5}$ (lower), $10^{-6}$ (middle), $10^{-7}$ (upper). The thickest line ($E=10^{-7}$) is determined by the equatorial cross section through figure~\ref{full_shell_num-blow_up}($a$).}
\end{figure}

The culprit is readily identified from the complete form $\omegaB=\omegaB_G+\omegaB_{\!A}$ (see (\ref{approx-third}$a$)), in which $\omegaB_G = \omegaBC\,+\,\omegaBCP\,\xB$ (see (\ref{lin-shear}$a$)). In formulating our equatorial Ekman layer problem, we have approximated $\omegaB_G$ by $\omegaBC$ and ignored the linear shear \red{$\omegaBCP\,\xB$}. This is justified by \red{(\ref{om-inf})} which provides the error estimate $\OR\bigl(E^{4/35}x\bigr)$; not particularly small at finite $E$. Indeed the ever increasing size of the linear shear for large \red{$|x|$} is overwhelming. The consequent limitation on the validity of the ensuing asymptotics in the far-field is clearly highlighted by (\ref{approx-third}$c$), which indicates that $\bigl|\omegaB_{\!A}\bigr|\ll \bigl|\omegaBCP\,\xB\bigr|$ in the bulk of  the $E^{1/3}$-layer. \red{Only within the very thin similarity sublayer region $\xB=\OR\bigl((E\zB)^{1/3}\bigr)$ for $\zB\ll O\bigl(E^{1/21}\bigr)$, whose width shrinks with decreasing $\zB$, is the ratio $\bigl|\omegaBCP\,\xB\bigr|\,\bigl/\,{|\omegaB_{\!A}|}$ small (see (\ref{linear-shear-estimate})) reaching its smallest size $\OR(E^{4/35})$ at the equatorial Ekman layer $\zB=\OR(E^{1/5})$ (see (\ref{EEL-dimensions})).} This is borne out by the DNS-results illustrated by the figure~\ref{shear_equator} plots of $v$ versus $x$ on the equatorial plane $z=0$. The curves at each value of $E$ all exhibit the linear shear $\omegaBCP\,\xB$ at large $x$. Not surprisingly, therefore, this shear significantly modifies the contours of constant-$v$ on figure~\ref{full_shell_num-blow_up}($a$) and for that matter figure~\ref{full_shell_num}($a$). 

\begin{figure}
  \centerline{\includegraphics[width=\textwidth,clip=true]{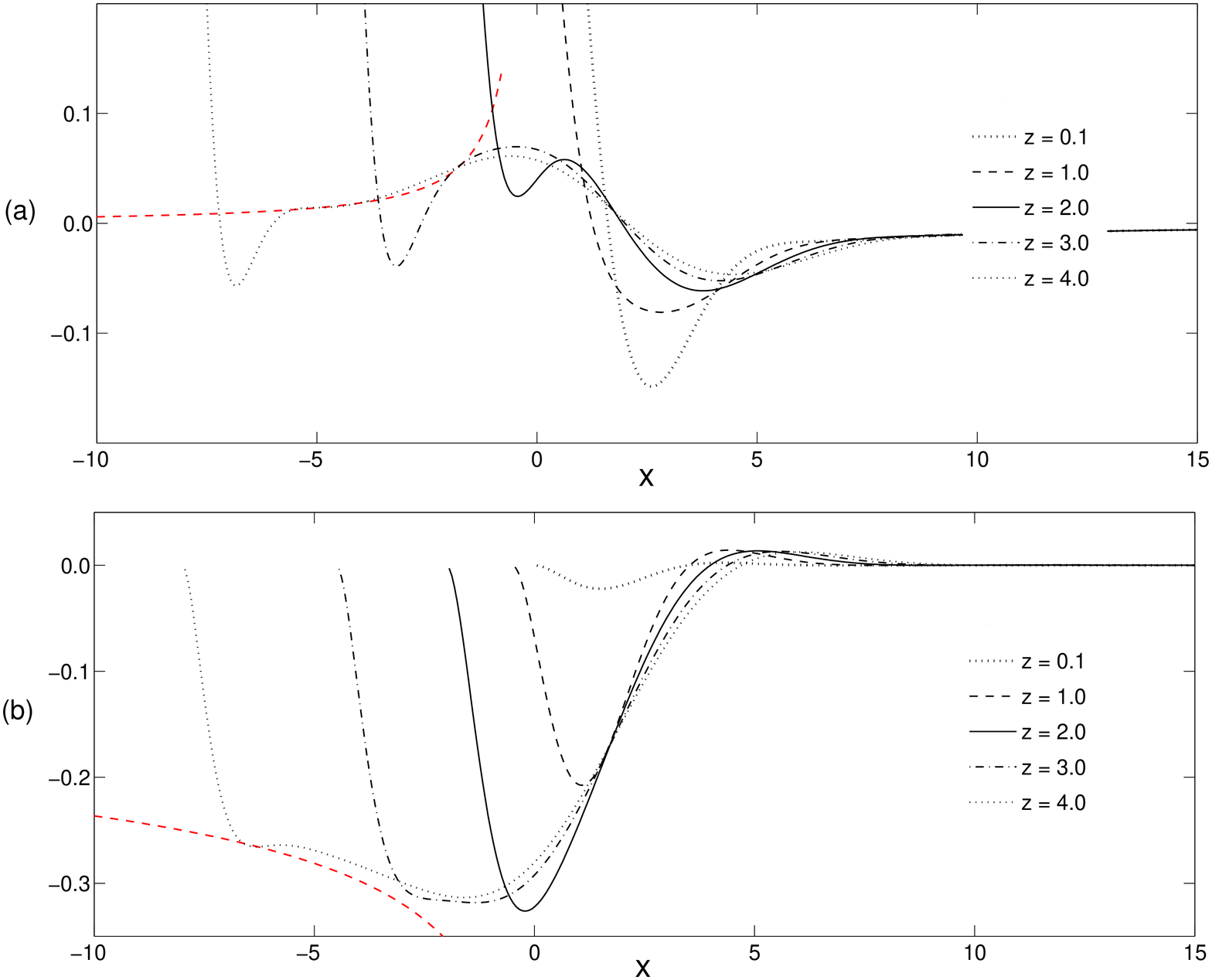}}
  \caption{\label{profiles_equator}Profiles of ($a$) azimuthal velocity $v$ and ($b$) streamfunction $\psi$ versus $x$ for various values of $z=\;$const.~identified by the inset key. Their $v^{sl}$ and $\psi^{sl}$ asymptotes (\ref{VPhi-asym-Stew}) for $-x\gg z^{1/3}$ are illustrated by the red dashed line on the left-hand sides.}
\end{figure}

\begin{figure}
\centerline{\raisebox{26mm}{\small (a)}\includegraphics[width=0.8\textwidth,clip=true,trim=105 0 0 0]{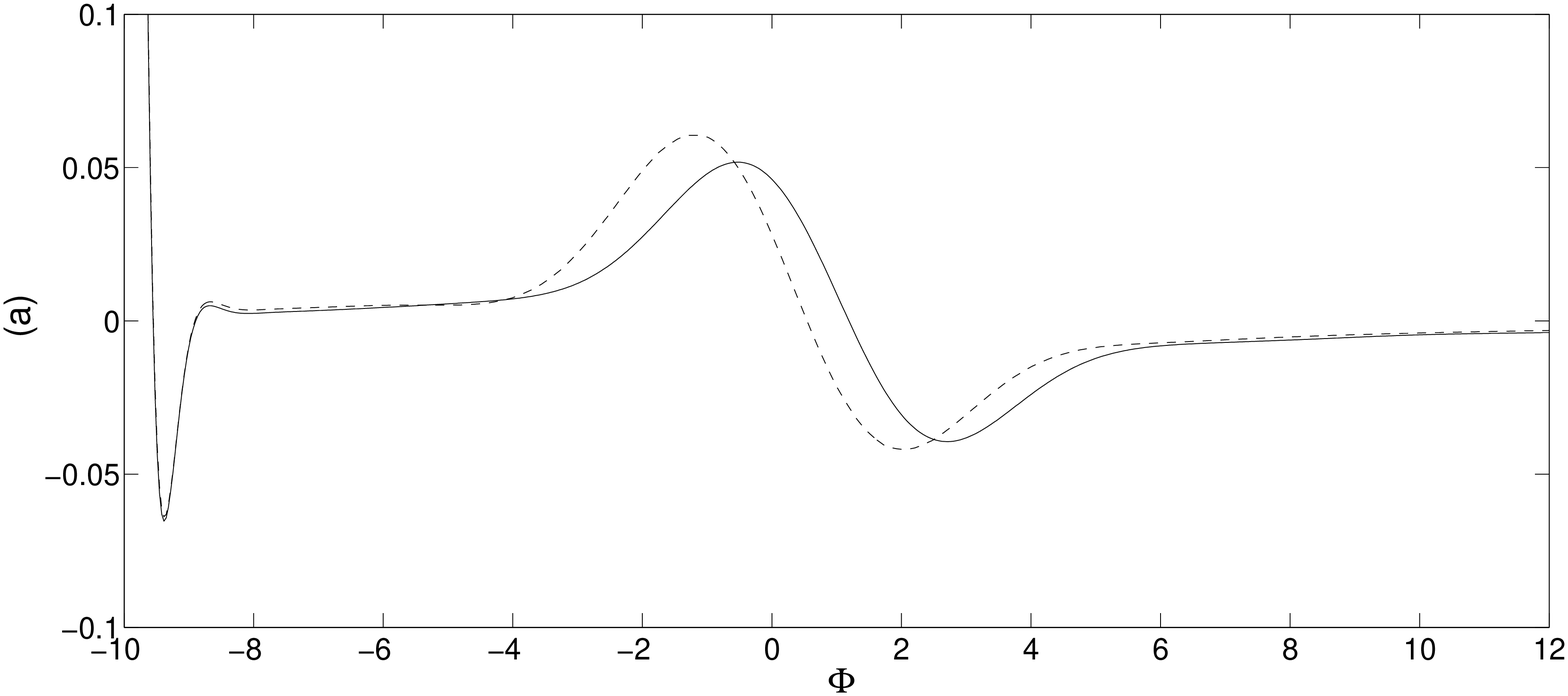}}
  \centerline{\raisebox{26mm}{\small (b)}\includegraphics[width=0.8\textwidth,clip=true,trim=105 0 0 0]{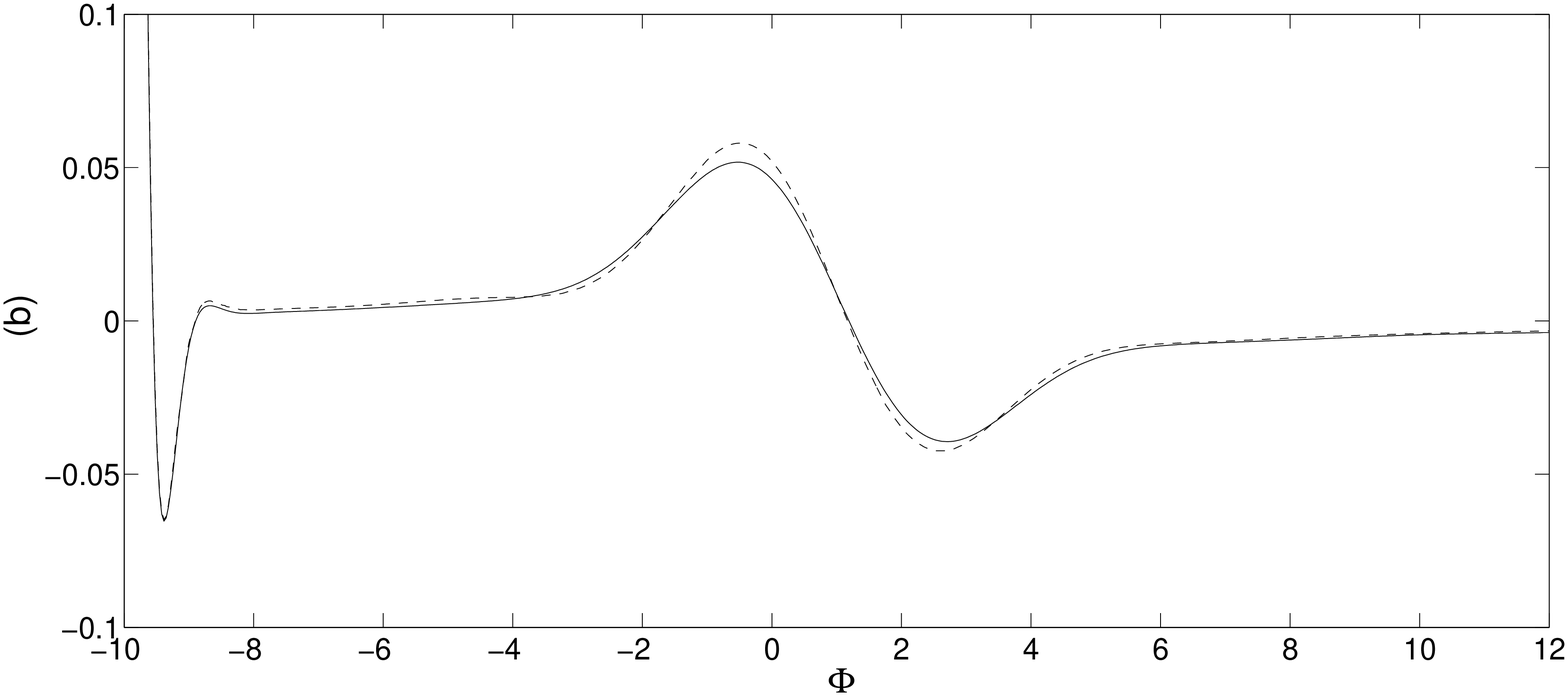}}
  \centerline{\raisebox{26mm}{\small (c)}\includegraphics[width=0.8\textwidth,clip=true,trim=105 0 0 0]{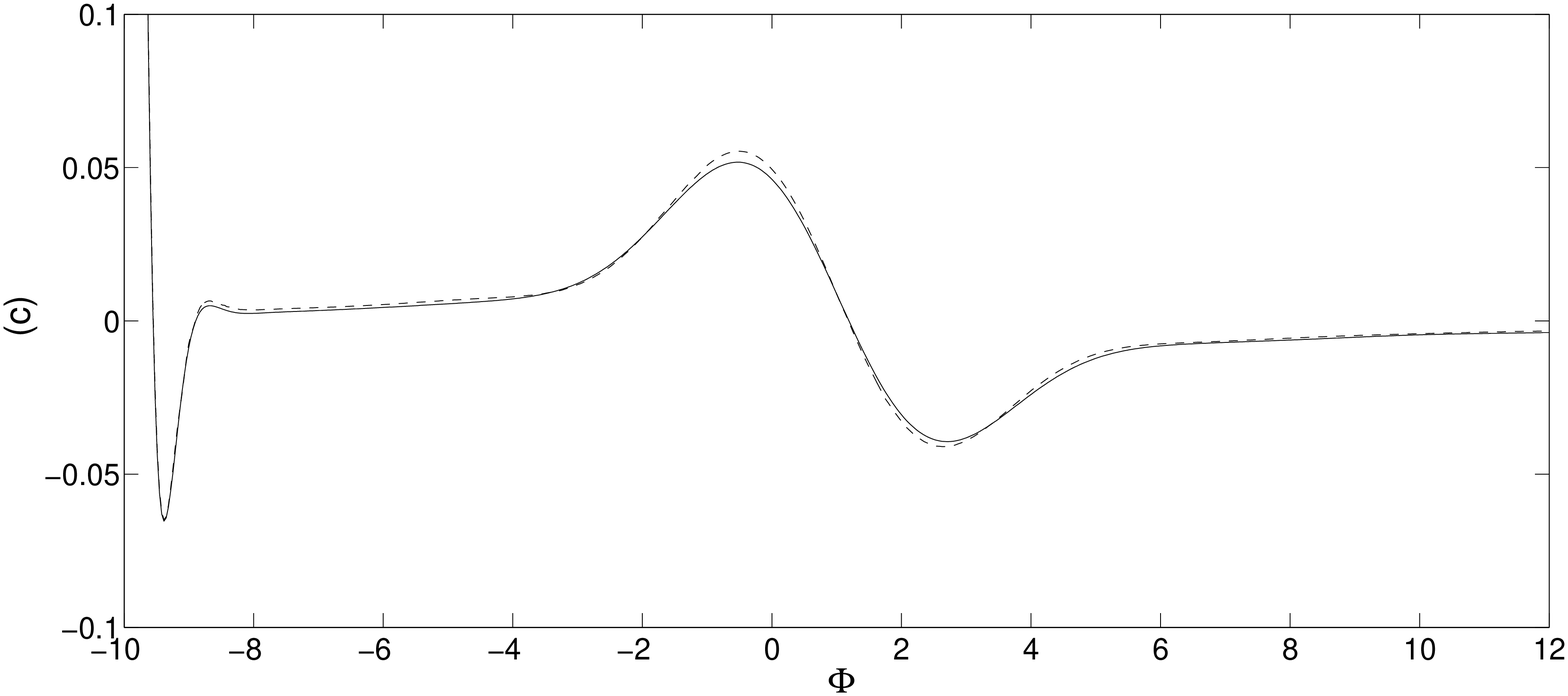}}
  \caption{\label{profile1}Profiles of azimuthal velocity $v$ at $z=6.0$ versus the self-similar variable $\Phi$. The numerical solutions for $v$ in the finite box (solid line) are compared to the asymptotic solutions $v^{sl}$ (see (\ref{EEL-series-summary}$a$)) (dashed lines): ($a$) First order, $z^{-5/12}V_0(\Phi)$; ($b$) plus the second order, $z^{-5/6}V_1(\Phi)$; ($c$) plus the third order, $z^{-15/12}V_2(\Phi)$.}
\end{figure}

\begin{figure}
 \centerline{\raisebox{26mm}{\small (a)}\includegraphics[width=0.8\textwidth,clip=true,trim=105 0 0 0]{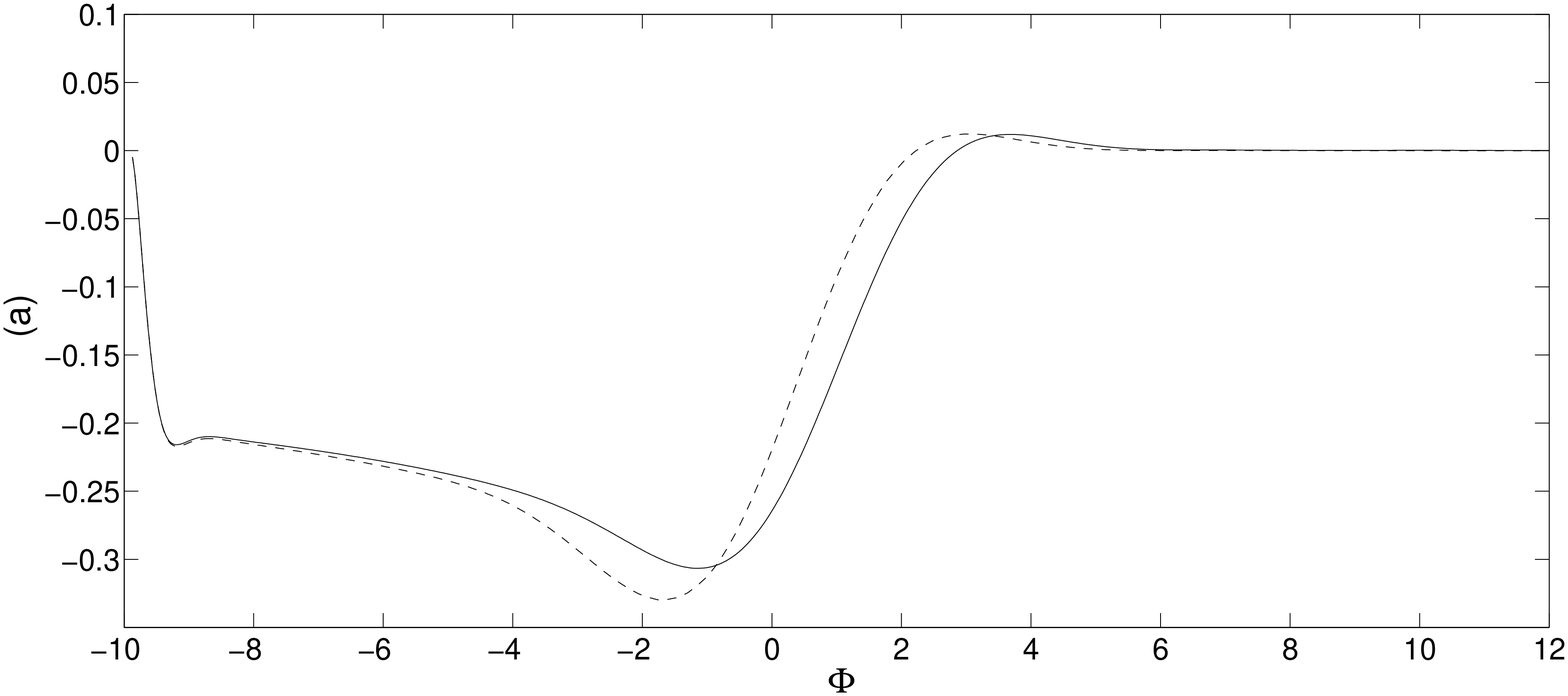}}
 \centerline{\raisebox{26mm}{\small (b)}\includegraphics[width=0.8\textwidth,clip=true,trim=105 0 0 0]{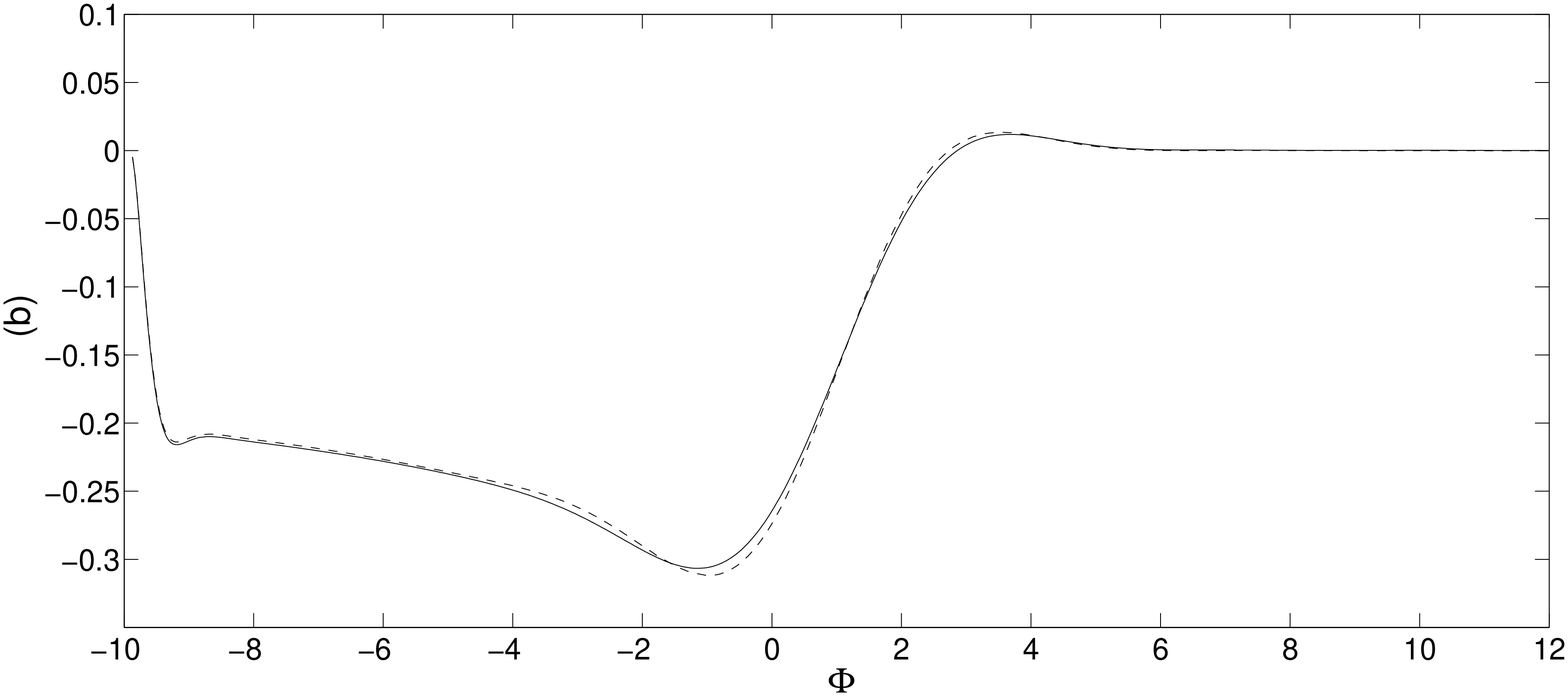}}
 \centerline{\raisebox{26mm}{\small (c)}\includegraphics[width=0.8\textwidth,clip=true,trim=105 0 0 0]{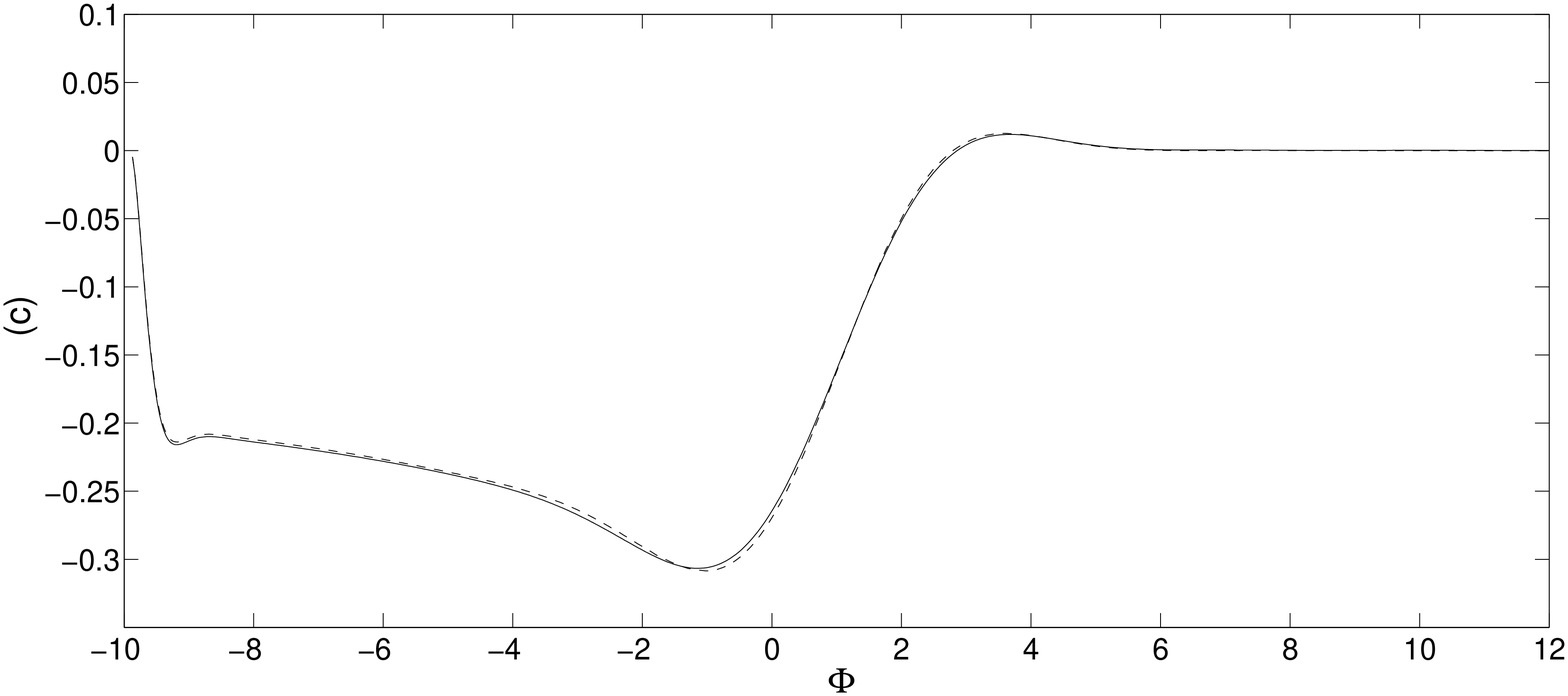}}
  \caption{\label{profile2}As in figure~\ref{profile1} but for the streamfunction $\psi$. The asymptotic solutions $\psi^{sl}$ (see (\ref{EEL-series-summary}$b$)) used are ($a$) the first order, $z^{-1/12}\Psi_0(\Phi)$; ($b$) plus the second order, $z^{-1/2}\Psi_1(\Phi)$; ($c$) plus the third order, $z^{-11/12}\Psi_2(\Phi)$.}
\end{figure}

\subsection{The far-field solution $z\gg1$ \label{far-field}}

The nature of the far-field asymptotic behaviour as both $x(y,z)=y-\tfrac12 z^2 \to  \infty$ and $z \to  \infty$ has already been discussed at length in the earlier sections. Still one feature of the leading order \cite{S66} similarity solution $v^{sl}=V(x,z;5/4)=z^{-5/12}V_0(\Phi)$, $\psi^{sl}=\Psi(x,z;5/4)=z^{-1/12}\Psi_0(\Phi)$ (see (\ref{EEL-similarity_0})) worth stressing here is that, for $|x|\gg z^{1/3}$ (equivalently $|\Phi|\gg 1$), (\ref{VPhi-asym}$a$,$b$) determines
\bme
\label{VPhi-asym-Stew}
\be
v^{sl}\,\approx\,\left\{\begin{array}{ll}
2^{-11/4}x^{-5/4}\,,\\[0.3em]
2^{-15/4}(-x)^{-5/4}\,,\qquad
\end{array}\right.
\psi^{sl}\,\approx\,\left\{\begin{array}{ll}
 0 &(x>0)\,,\\[0.3em]
-2^{-3/4}(-x)^{-1/4}&(x<0)\quad
\end{array}\right.
\ee
\eme
for the mainstream flow outside both \red{the Ekman layer and similarity sublayer}. Indeed, even by $z=4$, there is clear evidence of the tendency toward this $z$-independent QG-behaviour in the figure~\ref{profiles_equator} plots, which include the $v^{sl}$ and $\psi^{sl}$ asymptotes, lower entries of (\ref{VPhi-asym-Stew}), valid for $-x\gg z^{1/3}$. \red{A reminder of the footnote comment in \S\ref{Outline}, that the equatorial Ekman layer and its immediate surrounds lie within the $E^{1/3}$-layer, is pertinent having two implications. Firstly, the term ``mainstream'' is used here in the context of the equatorial Ekman layer (\S\ref{formulation}) rather than the Proudman  (\S\ref{Proud-prob}) problem. Secondly, it simply happens that the leading order approximation (\ref{VPhi-asym-Stew}) to the  $E^{1/3}$-layer flow, outside the similarity sublayer ($|\Phi|\gg 1$), is almost geostrophic.}

We display $v$ and $\psi$ versus the self-similar variable $\Phi=x/z^{1/3}$ at $z=6$ (beneath the top boundary $z=7$ of our numerical box but above the top $z=4$ of figures~\ref{contours}~and~\ref{full_shell_num-blow_up}) by the solid lines in figures~\ref{profile1}($a$) and \ref{profile2}($a$) respectively. They appear to feature a clearly defined Ekman layer at the sphere boundary: $\Phi=-\tfrac12 z^{5/3}$ ($\approx -9.9$ for $z=6$). For though, following our non-dimensionalisation (\ref{dim-dist})--(\ref{dim-vel}), the Ekman number is removed from our problem, it is evident from (\ref{gov-eq-y}) that, when $z\gg 1$ and $z$-derivatives are neglected, the coefficient $z$ plays the role of the inverse Ekman number $E^{-1}$. Away from the Ekman layer, motion is largely slow except in the \red{similarity sublayer} $\Phi=\OR(1)$, where according to the Stewartson similarity solution (\ref{EEL-similarity}$a$,$b$) we expect $v=\OR(z^{-5/12})$, $\psi=\OR(z^{-1/12})$,  $w=\OR(z^{-5/12})$. Indeed the comparison of the numerical solution and (\ref{EEL-similarity}$a$,$b$) in figures~\ref{profile1}($a$) and \ref{profile2}($a$), though not perfect, is encouraging.

Numerical results for the full box height $H=7$ show that the intense Ekman layer visible on figures~\ref{profile1}~and~\ref{profile2} bordering the inner sphere $\SC$  does indeed thicken $\bigl(\propto z^{-1/2}\bigr)$ with decreasing latitude. From the constant-$v$ contours at the top, $z=4$, of figure~\ref{contours}($a$) we see the normal Ekman layer feature that, on moving away from the boundary, the strong eastward azimuthal velocity associated with the moving boundary is rapidly damped and reverses sign towards the fluid interior. At about $z=3$ the Ekman layer starts to merge with the thinning \red{similarity sublayer}, width $\OR(z^{1/3})$, the small reverse (westward) flow eventually vanishing at $z=2.4$. For $z \le 1.5$ approximately, the shear and Ekman layers are certainly no longer distinguishable. Figure $\ref{contours}$($b$) displays the double-eddy structure of the equatorial meridional flow, with a strong, counterclockwise-rotating eddy located roughly inside the tangent cylinder $\CC$ and a much weaker, clockwise-rotating eddy outside; features that are already evident from the DNS-results portrayed on figures~\ref{full_shell_num}($b$)~and~\ref{full_shell_num-blow_up}($b$).

\subsection{The relevance of the Moore {\rm \&} Saffman similarity solutions\label{MS-sols}}

With respect to the comparisons of the far-field similarity and numerical solutions, two matters need to be emphasised. Firstly, these similarity solutions were not used in the specifications of the boundary condition at $z=H$. Rather the non-local Green's function integral (\ref{top-bc-0-new}) was implemented numerically by use of the form (\ref{symbc}$e$). Thus our finding that the numerical solution compares so well with the asymptotics is most reassuring. Secondly, we must be  cautious about how reliable the far-field asymptotics actually are. In deriving the asymptotic series (\ref{EEL-series-summary}$a$,$b$) we have only applied a limited set of boundary conditions. Within that class, we show in Appendix~\ref{appA} by reference to (\ref{VPhi-asym-integer}) that any similarity solution 
\be
\label{W-Zp-again}
V+\iR W\,=\,\WS(x,z;p)\,\equiv\,-\,2^{-1/2}(p-1)2^{-(p-1)}\,z^{-p/3}\ZS\bigl(x/z^{1/3};p\bigr)
\ee
(see (\ref{W-Zp}a,b)) will satisfy the lowest order homogeneous sphere boundary condition $W\mbox{ or }\Psi=0$ on $x<0$, $z\approx  0$ (or more precisely as $x/z^{1/3}\downarrow -\infty$)  whenever $p$ is an integer greater or equal to unity. These are essentially ``complementary functions'' that can be added in arbitrary proportions to the ``particular'' Stewartson solution (\ref{EEL-similarity_0}). To assess relative sizes, we note that the complementary  $\Psi$-functions are functions of $\Phi$ multiplied by $z^{-(p-1)/3}$ (see (\ref{VPhi}b)), whereas the Stewartson form $\Psi(x,z;5/4)\,=\,z^{-1/12}\Psi_0(\Phi)$ (see (\ref{EEL-similarity_0}$b$)) is proportional to $z^{-1/12}$. Of course, the application of the true sphere boundary condition on $\SC$ generates an asymptotic series similar to (\ref{EEL-series-summary}) for each integer $p$.

The important exceptional case $p=1$ requires some care as the integral on the right-hand side of (\ref{VPhi}b) is not properly defined. Nevertheless \cite{MS69} explained that it corresponds to an infinite boundary $z=0$ on which $W=0$ but with a line source of fluid at $x=0$. Such a configuration is relevant to many situations such as the split discs considered by \cite{S57}. Even so, this $p=1$ solution with $\psi=0$ both on $\SC$ and as $x\to \infty$ is excluded by our boundary conditions (\ref{bcs}$b$,$c$). Since the $p=2$ solution is generated by differentiating the $p=1$ 2D-point source solution with respect to $x$ (see (\ref{Zp-rec}) with $p=2$) the $p=2$ solution corresponds to a 2D-dipole source. Larger integer $p$ likewise produce 2D-multipoles. All these multipole solutions $p\ge 2$ must be present to a greater or lesser extent and are generated by the flow in the equatorial Ekman layer region $\bigl|(x,z)\bigr|=\OR(1)$. As that flow is unknown to us from an analytic point of view, so a priori are the magnitudes of the 2D-multipoles. Instead their magnitudes are determined, in principle, by the full numerical solution, a task that we have not attempted, because the discrepancies between the numerical and high order asymptotic solutions illustrated in figures~\ref{profile1}($b$,$c$) and \ref{profile2}($b$,$c$) appear to show apparent convergence without the addition of any multipole contribution. This finding is most intriguing and deserves further comment. The explicit powers of $z$ for $\psi$-multipoles are $-1/3$ ($p=2$), $-2/3$ ($p=3$), $-1$ ($p=4$), $\cdots\,$. Fortunately the modulus of dipole index $1/3$ for $p=2$ is larger than the corresponding $1/12$ for the Stewartson similarity solution $z^{-1/12}\Psi_0(\Phi)$, which must reassuringly dominate as $z\to \infty$. This is not the case for the higher order corrections $z^{-1/2}\Psi_1(\Phi)$ and  $z^{-11/12}\Psi_2(\Phi)$ to it with index moduli $1/2$ and $11/12$ respectively, which interlock the free mode sequence $1/3$, $2/3$, $1$. So, for sufficiently large $z$, the  $p=2$ free mode must dominate over the $\Psi_1$ correction to the Stewartson solution. We therefore conclude that, for our plots at $z=6$ in figures~\ref{profile1} and \ref{profile2}, the magnitudes of the multipoles are relatively small and on increasing $z$ their contributions must continue to decrease in magnitude relative to the dominant Stewartson solution. 

The above remarks prompt us to review the status of figures~\ref{profile1}($a$--$c$) and \ref{profile2}($a$--$c$). Evidently the lowest order asymptotic similarity profiles portrayed in ($a$)  are reliable. The marked improvement in the comparison of the asymptotics and the numerics in ($b$) is a bonus which from a formal mathematical point of view is surprising as we would reasonably expect the $p=2$ dipole contribution to influence the numerical result. Indeed the apparent continuing small improvement in ($c$) is remarkable because both the $p=2$ dipole and the $p=3$ quadrupole ought (by any reasonable expectation) to be influencing the numerical results by amounts at least comparable to the small improvement mentioned. As a further technical point, the power $z^{-4/3}$ of the next $p=5$ multipole coincides with the next order term $q=3$ in our expansion of the Stewartson solution with power $z^{-(p-1)/3}$ $\,\bigl(p=5(q+1)/4\bigr)$ (see (\ref{EEL-series}$d$) and (\ref{EEL-series-summary}$b$)). It leads to a resonance generating a similarity mode proportional to $z^{-4/3}\ln z$, which, in view of our comments and being of such high order, is largely an irrelevance and not considered here. 

\subsection{Related studies\label{rel-stud}}

All the features mentioned about figures~\ref{contours}($a$,$b$) are very similar to those also found numerically by \cite{vdV93} and illustrated in his figures~4 and 3 respectively for the simpler rotating disc problem. His geometry differs from ours through the simple expedient of replacing the sphere by a flat disc lying in the equatorial plane $z=0$ (of our sphere) with its outer edge on $x<0$ thus having its surface at $z=0$ rather than $z=-\tfrac12 x^2$. Indeed close inspection of the two sets of figures reveals that we replicate his illustrated topological features, including the noticeable kink of the $v=0$ contour, which occurs about $z=1.5$ on figure \ref{contours}($a$). This is not particularly surprising when one recalls that, in his far-field, $-x\gg 1$, the Ekman jump condition determines the mainstream boundary condition $\psi\;$const.~on the disc, in contrast to our $\psi$ proportional to $(-x)^{1/4}$. So whereas our Stewartson \red{similarity sublayer} solution in complex form is $\WS(\Phi,z;p)$ (see (\ref{W-Zp-again})) with $p=5/4$, the disc self-similar solution is the 2D-point source version $p=1$. As the integrands in the two integrals defining each $\ZS(\Phi;p)$ (see (\ref{Zp-sol})) only differ by a factor  $\varpi^{-1/4}$, we do not expect much qualitative difference in the resulting functions in the far-field. With similar far-field features and topologically similar flow structures visible for $x=\OR(1)$,  $z=\OR(1)$, we may speculate that flow topology in the small equatorial Ekman layer responds to the flow in the much larger region outside it, rather than the local geometry of the boundary.

At this point it is helpful to briefly digress to a related detached MHD shear layer in spherical Couette flow studied by \cite{DJS02}. This layer also exhibits a flow region known as the equatorial Hartmann layer, which in many respects resembles our equatorial Ekman layer. As the equations governing the equatorial Hartmann layer are by comparison simpler, \cite{DJS02} were able to find an analytic solution based on a method previously developed by \cite{R67}. Knowledge of this analytic solution enables its asymptotic form at a large distance to be recovered as a similarity solution, c.f.~our Stewartson solution $v\approx z^{-5/12}V_0(\Phi)$ and $\psi\approx z^{-1/12}\Psi_0(\Phi)$ (see (\ref{EEL-similarity}$a$,$b$)).

The common features do not end there. Inspection of figures~\ref{profile1}($a$) and \ref{profile2}($a$) reveals that the Stewartson similarity solution is offset by a small, but readily perceived, amount from the full numerical boundary layer solution. A similar shift appears in the corresponding MHD problem, where a higher order approximation of the analytic solution reveals similarity function corrections involving larger inverse powers of the distance, c.f.~our $z^{-5/6}V_1(\Phi)$ and $z^{-1/2}\Psi_1(\Phi)$ in (\ref{EEL-series-summary}$a$,$b$). Those extra terms significantly improved comparison with the numerical solution for the full MHD free shear layer. Unfortunately, as we have no analytic solution for the  equatorial Ekman layer, we have needed to resort to asymptotic solutions of the governing equations valid for $z\gg 1$. Interestingly when the $V_1$ and $\Psi_1$ corrections, just mentioned, are included, we too find considerable (indeed remarkable)  improvement in the comparison of the asymptotic solution (see figures~\ref{profile1}($b$) and \ref{profile2}($b$)) with the numerical one. \cite{DJS02} noted that the sum of the the zeroth and first order terms could be expressed to the same order of accuracy by the zeroth order solution alone provided the coordinate origin is shifted by a small amount ${\bar \delta}$ (say) from $(x,z)=(0,\,0)$ to $\bigl({\bar \delta},\,0\,\bigr)$. Here the origin shift has the classical boundary layer interpretation of a ``displacement'' thickness. However, even though our figures \ref{contours}($a$,$b$) clearly display such a \red{similarity sublayer} displacement and figures~\ref{profile1}($b$) and \ref{profile2}($b$) hint at a similar interpretation, this is not formally possible. Indeed, no origin shift ${\bar \delta}$ effected by redefining $\Phi$ as $\bigl(x-{\bar \delta}\,\bigr)\big/z^{1/3}$ can absorb the first order corrections $z^{-5/6}V_1(\Phi)$ and $z^{-1/2}\Psi_1(\Phi)$ into the zeroth order terms $z^{-5/12}V_0(\Phi)$ and $z^{-1/12}\Psi_0(\Phi)$. The reason for the difference can be traced to the far-field Ekman layer, across which the Ekman suction produces a $\psi\propto (-x)^{1/4}$ (significantly dependant of $x$), so that the flow never forgets the $x=0$ location of the origin. In  figures~\ref{profile1}($c$) and \ref{profile2}($c$) we include the second order corrections $z^{-15/12}V_2(\Phi)$ and $z^{-11/12}\Psi_2(\Phi)$ in (\ref{EEL-series-summary}$a$,$b$).


\section{Conclusions \label{Conclusions}}

By means of a combined analytic and numerical approach, we have addressed the steady, axisymmetric problem of the merging at the inner sphere equator of the Ekman layer with the \red{similarity sublayer} encompassing the tangent cylinder. Despite the fact that the analytic structure of the standard Ekman layer on spherical boundaries is well-known, its terminal shape at the equator has remained unresolved ever since Stewartson's pioneering tour de force \citep[][]{S66}, half a century ago. Although the appropriate governing equations (\ref{gov-eq}$a$,$b$) are clear, Stewartson was cautious about the far-field boundary conditions (see the sentence which includes (\ref{EEL-similarity})). Our unequivocal position is that the zeroth-order situation outside the Ekman layer is simply $v\to 0$, $\psi\to 0$, as $z\to \infty$. This is implied by our boundary condition (\ref{bcs}$d$), which we implement indirectly in our numerics using the non-local (``soft'') boundary condition (\ref{symbc}$e$).

We cannot overemphasise that, under non-dimensionalisation and scaling (\ref{dim-dist})--(\ref{dim-vel}), we have removed the Ekman number $E$ from our problem. So when we refer to Ekman layers, we mean that, at large-$z$,  boundary layers can be identified with Ekman layer structure. From a mathematical point of view, the equivalence can be traced to (\ref{W-hat-eq}), whose series solution in inverse powers of $z$, recovers Ekman layers with respect to an effective Ekman number $z^{-1}$. Likewise the similarity structure can be thought of as $E^{1/3}$-type shear layer based on the local axial length $\zB$ rather than the height $\sqrt{\alpha^2-1}$ of the inner sphere tangent cylinder.

Our use of the ``soft'' top boundary condition (\ref{symbc}$e$) in the numerical method explained in \S\ref{num-model} may appear at first sight to be an unnecessary complication. In fact, its use proved necessary in order to obtain good numerical solutions. The point is that for large $z$, our results for the numerical box with dimensions $[0,L]\times [0,H]$ ($L=60$, $H=7$) in the $y$-$z$ plane exhibit a trend towards the $z$-independency, which is beginning to become evident towards the top of figures~\ref{contours}($a$,$b$) (but see also figures~\ref{profiles_equator}($a$,$b$)). So even at moderately large $z$, where (loosely interpreted) $z^{-1}$  plays the role of an Ekman number, the flow throughout responds very sensitively to what happens at $z=H$. Put another way, unless we make the top boundary $z=H$ of the finite box invisible, our results will resemble those for a finite container rather than the unbounded region of interest to us. Be that as it may, the reader may wonder ``Why not simply increase the height $H$ of the box until the results in the vicinity of the equator reach their limiting form?'' Though theoretically sound, we found the strategy to be unrealistic from the numerical point of view, essentially because of the vast number of grid points in the $z$-direction needed to preserve numerical convergence. Even for our largest manageable values of $H$, we  could not realise the required asymptotic behaviour by that direct approach.

An analytic solution of the problem posed by (\ref{gov-eq}) and (\ref{bcs}) has proved to be elusive. \cite{D72} proposes an intriguing solution for the outer sphere equatorial Ekman layer problem by integral transform  methods but with different constant stress boundary conditions. The status of his mathematical problem and solution is unclear to us. Our attempt to use his method failed even to recover Stewartson's free shear layer. Aficionados have attempted to apply an alternative integral method developed by \cite{R67} \citep[see also][where the paper is reproduced together with a historical comment on p.~XV]{R03} in a related MHD context but without success. The analytic solution of Stewartson's problem remains an outstanding mathematical challenge. 

The best, that we have been able to do analytically, is to obtain large-$z$ ``far-field'' similarity solutions (see \S\S\ref{sl-ff},~\ref{El-s}) that extend Stewartson's lowest order \red{similarity sublayer} solution. As we have stressed in these conclusions, they are not used as part of the top boundary condition at $z=H=7$, even though by construction they necessarily satisfy (\ref{top-bc-0-new}). The convergence of the asymptotics with the numerics at $z=6$, obtained by including higher order asymptotic terms (see successively subfigures~($a$--$c$) of \red{figures~}\ref{profile1}, \ref{profile2}), is striking. Though not a closed-form analytic solution of the  problem posed by (\ref{gov-eq}) and (\ref{bcs}), the combination of our numerical and asymptotic results appear to provide a robust alternative.

Finally we recap other evidence, described in \S\ref{Numerical-results}, that our results are reasonable. Firstly, we recall that the two meridional eddy structure of figure~\ref{contours}($b$) agrees well with the full shell DNS of figure~\ref{full_shell_num-blow_up}($b$). Secondly, we note the remarkable similarity of our flow topology (even down to some detailed structure in our figures~\ref{contours}($a$,$b$)) with that observed by \cite{vdV93} in his numerical investigation of the corresponding isolated disc \citep[building on the split disc problem of][]{S57}. In both cases, the structure of the layer, that occurs when the Ekman layer and the $E^{1/3}$ \red{similarity sublayer} of \cite{S66} merge, appears to be passive in the sense that the local flow responds to far-field conditions rather than influencing them. 

\section*{Acknowledgements}

F.M.~and E.D.~have been partially funded by the ANR project Dyficolti ANR-13-BS01-0003-01. F.M.~acknowledges a PhD mobility grant from Institut de Physique du Globe de Paris. A.M.S.~visited ENS, Paris (19--25 October 2014), while F.M.~and E.D.~visited the School of Mathematics \& Statistics, Newcastle University (respectively, 7--25 September 2015 and 25--30 November 2015); the authors wish to thank their respective host institutions for their hospitality and support.


\appendix

\section{The $E^{2/7}$-solution}\label{Bessel}

\subsection{First order problem: $\GC_0(\xs)$\label{GC0}}

The solution of the lowest order (\ref{sl-G-eq}) problem, $\GC_0^{\,\prime\prime}-(-\xs)^{-1/4}\GC_0\bigl(=-f(\xs)\bigr)=-1$ with $\GC_0(0)=0$ and $\GC_0\approx (-\xs)^{1/4}$ as $\xs \to-\,\infty$, is obtained by the method of variation of parameters:
\be
\label{G0x}
\GC_0(\xs)\,=\,\JS[\,\HC;\xs\,]\,\equiv\,\KS[\,\HC;\xs\,]\,+\,\LS[\,\HC;\xs\,]\,,\qquad\qquad \HC(\xs)=1\,,
\ee
where the functional $\JS[f;\xs\,]$ is defined by
\bse
\label{sl-ints}
\begin{align}
\JS[f;\xs\,]\,=\,&\,({8}/{7})^{2/7}(-\xs)^{1/2}\biggl[\KR_{4/7}(\sigma)\!\int_0^\sigma\!{\sigma'}^{5/7}\IR_{4/7}(\sigma')\,f(-\xs')\,\dR\sigma'\biggr.\nonumber\\
&\qquad\qquad\qquad\biggl.\,\,+\,\IR_{4/7}(\sigma)\!\int_\sigma^\infty\!{\sigma'}^{5/7}\KR_{4/7}(\sigma')\,f(-\xs')\,\dR\sigma'\biggr]
\intertext{and the partition  $\JS[f;\xs\,]\equiv\KS[f;\xs\,]+\LS[f;\xs\,]$ is realised by}
\KS[f;\xs\,]\,=\,&\,({8}/{7})^{2/7}(-\xs)^{1/2}\biggl[\int_0^\infty\!{\sigma'}^{5/7}\KR_{4/7}(\sigma')\,f(-\xs')\,\dR\sigma'\biggr]\IR_{4/7}(\sigma)\,,\\
\LS[f;\xs\,]\,=\,&\,\tfrac12 ({8}/{7})^{2/7}\Gamma(4/7)\Gamma(3/7)(-\xs)^{1/2}
\biggl[\IR_{-4/7}(\sigma)\!\int_0^\sigma\!{\sigma'}^{5/7}\IR_{4/7}(\sigma')\,f(-\xs')\,\dR\sigma'\biggr.\nonumber\\
&\qquad\qquad\qquad\qquad\qquad\qquad\quad\biggl.\biggl.\,\,-\,\IR_{4/7}(\sigma)\!\int_0^\sigma\!{\sigma'}^{5/7}\IR_{-4/7}(\sigma')\,f(-\xs')\,\dR\sigma'\biggr].
\end{align}
\ese

It is easy to verify that $\GC_0(0)=\JS[\,\HC;0\,]=0$ on use of 
\bme
\label{I-nr-0}
\be
\IR_n(\sigma)\,\approx\,\dfrac{\sigma^{n}}{2^{n}\Gamma(n+1)}\,,\qquad\quad
\KR_n(\sigma)\,\approx\,2^{n-1}\Gamma(n)\,\sigma^{-n} 
 \qquad\quad\mbox{as}\quad \sigma\downarrow 0
\ee
\eme
\citep[see][http:/$\!$/dlmf.nist.gov/10.25.E2 and /10.31.E1]{AS10} holding for positive non-integer $n$, while
\be
\label{sl-int-K}
\int_0^\infty\!{\sigma}^{5/7}\KR_{4/7}(\sigma)\,\dR\sigma\,=\,2^{-2/7}\Gamma(8/7)\Gamma(4/7)
\ee
\citep[see][http:/$\!$/dlmf.nist.gov/10.43.E19]{AS10} determines
\be
\label{sl-K-1}
\KS[\,\HC;\xs\,]\,=\,({4}/{7})^{2/7}\Gamma(8/7)\Gamma(4/7)\,(-\xs)^{1/2}\IR_{4/7}(\sigma)\,.
\ee
For small $-\xs\,(\ge 0)$ it is readily shown that $\LS[\,\HC;\xs\,]\approx\tfrac12 \xs^2$ so that $\JS[\,\HC;\xs\,]$ is dominated by the $\KS[\,\HC;\xs\,]$ contribution:
\be
\label{sl-G0x}
\GC_0(\xs)\,\approx\,\KS[\,\HC;\xs\,]\,\approx\,-\,(7/4)^{1/7}\Gamma(8/7)\,\xs\qquad\quad\mbox{as}\qquad \sigma\downarrow 0\quad\bigl(\xs \uparrow 0\bigr),
\ee
from which the value (\ref{sl-deriv0}$a$) for $\GC^{\,\prime}_0(0)$ is readily derived.

\subsection{Second order problem: $\FC_1(\xs)$\label{FC1}}

The solution of the first order (\ref{sl-F-eq}) problem, $\FC_1^{\,\prime\prime}-(-\xs)^{-1/4}\FC_1\bigl(=-f(\xs)\bigr)=\FC_0$ with $\FC_1(0)=0$ and $\FC_1\to 0$ as $\xs\to-\infty$, is 
\be
\label{F1x}
\FC_1(\xs)\,=\,-\,\JS[\,\FC_0;\xs\,]\,\equiv\,-\,\KS[\,\FC_0;\xs\,]\,-\,\LS[\,\FC_0;\xs\,]\,,
\ee
as in (\ref{G0x}) and (\ref{sl-ints}$a$-$c$) above. Noting that $\FC_0(\xs)=\bigl[2^{3/7}\big/\Gamma(4/7)\bigr]\sigma^{4/7}\KR_{4/7}(\sigma)$ (see (\ref{sl-sim-sol}$a$,$d$)) and use of the integral 
\be
\label{sl-int-K-new}
\int_0^\infty\!{\sigma}^{9/7}\,\bigl[\KR_{4/7}(\sigma)\bigr]^2\,\dR\sigma\,=\,2^{-5/7}\,\,\dfrac{\Gamma(12/7)\,\bigl[\Gamma(8/7)\bigr]^2\,\Gamma(4/7)}{\Gamma(16/7)}
\ee
\citep[see][\S6.576, eq.~(4), noting $F(12/7,\,8/7;\,16/7;\,0)=1$]{GR07} determines
\be
\label{sl-int-more}
\KS[\,\FC_0;\xs\,]\,=\,(4/7)^{2/7}\,\dfrac{\Gamma(12/7)\,\bigl[\Gamma(8/7)\bigr]^2}{\Gamma(16/7)}\,(-\xs)^{1/2}\IR_{4/7}(\sigma)\,.
\ee
This gives the dominant behaviour of $\JS[\,\FC_0;\xs\,]$ for small $-\xs$:
\be
\label{sl-F1x}
\FC_1(\xs)\,\approx\,-\,\KS[\,\FC_0;\xs\,]\,\approx\,(7/4)^{1/7}\,\dfrac{\Gamma(12/7)\,\bigl[\Gamma(8/7)\bigr]^2}{\Gamma(16/7)\Gamma(4/7)}\,\xs \qquad\mbox{as}\quad \sigma\downarrow 0\quad\bigl(\xs \uparrow 0\bigr),
\ee
from which the value (\ref{sl-deriv0}$b$) for $\FC_1^{\,\prime}(0)$ is readily derived.

\section{\red{The iteration used to solve the \S\ref{num-model} numerical model}\label{iteration}}

\red{We overcome the non-local nature of the boundary condition (\ref{top-bc-0-new}) on $z=H$ by solving a sequence ($n=0,\,1\,2\,\cdots$) of problems with solutions $\psi_{n},v_{n}$. For the $n=0$ problem, we take the homogeneous Dirichlet boundary condition $v_{0}(y,H)=0$ (exact for $H\to\infty$). For $n\ge 1$, the known result for $\psi_{n-1}(y,H)$ is inserted into the convolution integral on the right-hand side of (\ref{top-bc-0-new}) in order to define the new value $v_{n}(y,H)$ on the left (see (\ref{symbc}$e$)). The subsequent iteration, hopefully, reduces the reflection and converges onto the non-reflecting solution. However, the finite extent of the computational domain together with the failure of (\ref{top-bc-0-new}) to take account of Ekman layers requires careful treatment of the convolution integral. To begin, we divide the integral up into various intervals. Then, we avoid the Ekman layers and integrate over the range $a\le y\le b$ with constants $a$, $b$ chosen such that $a\gg H^{-1/2}$,  $L-b\gg H^{-1/2}$. Unfortunately, the value of $v_{n}(y,H)$, so obtained, diverges weakly like $\ln(y-a)$ and $\ln(b-y)$ towards the respective end-points $y=a$ and $b$ of the integration range $a\le y\le b$. This in itself is not serious, as $v_{n}(y,H)$ is very small in the vicinity of $y=a$ and $b$. However, any discontinuity in the value $v_{n}(y,H)$ will trigger disturbances on the geostrophic cylinders $\CC(y,H)$ (see (\ref{geo-cyl})) through it and so must be avoided. Accordingly we employ the integral result on the narrower range $y_-\le y\le y_+$ inside $a\le y\le b$ ($a<y_-$, $\,\,\,y_+<b$) and simply linearly interpolate $v_{n}(y,H)$ on the remaining intervals $a\le y\le y_-$, $\,\,\,y_+\le y\le b$ to avoid any discontinuity. We choose the values of $y_\pm$ so as to contain most of the mainstream and certainly the similarity sublayer: $\tfrac12 H^2-y_-\ll H^{1/3}$, $\,\,y_+-\tfrac12 H^2 \ll H^{1/3}$ (see figure~\ref{cdefo}).}

\red{In summary, we solve (\ref{gov-eq-y}) iteratively for each successive solution $\psi_{n}(y,z),v_{n}(y,z)$ ($n\ge 0$) subject to the boundary conditions
\bse
\label{symbc}
\begin{align}
\psi_n=\,&0\,,\qquad \partial \psi_n/\partial y\,=\,0\,,\qquad v_n=1 &\qquad\mbox{on}&& y=\,&0\,,\qquad\\[0em]
\psi_n=\,&0\,,\qquad\partial \psi_n/\partial y\,=\,0\,,\qquad v_n=0 &\qquad\mbox{on}&& y=\,&L\,,\qquad\\[0em]
\psi_n=\,&0\,,&\mbox{on}&& z=\,&0\,.\qquad
\end{align}
In addition, on the top $z=H$ we apply 
\begin{align}
v_0&=\,0\\
\intertext{for $n=0$, and}
v_n&=\left\{\begin{array}{lrrl}
 0\, , &\qquad\mbox{on}&\quad &y \leq a\,,\\[0.2em]
v_n(y_-,H)\dfrac{y-a}{y_- - a}\, ,   & \qquad\mbox{on}&\quad a<\!\!\!\!&y<y_{-}\,,\\[1.0em]
-\,\dfrac{1}{\pi} {\ds\dashint}_{\!\!\!a}^{\,b}{\dfrac{1}{y-y'}\pd{\psi_{n-1}}{y}(y',H)\,\dR y'}\, ,& \qquad\mbox{on} &\quad y_{-} \!\leq\!\!\!\!& y \leq y_{+}\,,\\[1.0em]
v_n(y_+,H)\dfrac{b-y}{b- y_+}\, , & \qquad\mbox{on}  &\quad y_+<\!\!\!\!&y<b\,,\\[0.3em]
 0\, , &\qquad\mbox{on}&\quad b\leq \!\!\!\!&y 
\end{array}\right.
\end{align}
\ese
for $n\ge 1$. We remark that the top boundary condition $v_{0}(y,H)=0$ for ($y \leq a$) (\ref{symbc}$e$)$_1$ ignores the fact that $v_{0}(y,H)$ jumps to unity at $y=0$ across a thin Ekman layer. Elsewhere such a discontinuity at $y={\bar y}\,(>0)$, $z=H$ would trigger a disturbance on the geostrophic cylinder $\CC\bigl({\bar y}, H\bigr)$. However, for ${\bar y}=0$, the cylinder lies entirely outside ($z>H$) our numerical box. So any corner effects are localised in the vicinity of $(y,z)=(0, H)$ and have no effect on our numerical solution elsewhere.}

\red{The robustness of the procedure's convergence was verified by varying the choice of the initial condition, the values of $a$, $y_{-}$, $y_{+}$, $b$ (while keeping these reasonably far from the similarity sublayer) and the nature of the interpolation over $a \le y \le y_{-}$ and $y_{+} \le y \le b$ (linear, cubic or other power law). We have also checked that the domain size and grid resolution used ($H=7$, $L=60$, with 700 $\times$ 950 grid points) are sufficiently large to attain accurate numerical solutions.}

\section{The asymptotic form of $\ZS(\Phi;p)$ ($p>1$) for $|\Phi|\gg 1$}\label{appA}

For $|\Phi|\gg 1$, the integral on the right-hand side of (\ref{Zp-sol}) for $p>1$ may be evaluated asymptotically by the method of steepest descent to obtain
\be
\label{Wp-asym}
\ZS(\Phi;p)\,\sim\,
\left\{\begin{array}{ll}
\dfrac{1}{\Phi^p}\,\overset{\infty}{\underset{k=0}{\ds\sum}}\biggl(\dfrac{\iR}{2}\biggr)^{\!k}\dfrac{\bigl(\,p\,\bigr)_{\!3k}}{k!}\,\dfrac{1}{\Phi^{3k}}\qquad\qquad& (\Phi\gg 1)\,,\\[1.5em]
\dfrac{\eR^{-\iR p\pi}}{(-\Phi)^p}\,\overset{\infty}{\underset{k=0}{\ds\sum}}\biggl(\dfrac{-\iR}{2}\biggr)^{\!k}\dfrac{\bigl(\,p\,\bigr)_{\!3k}}{k!}\,\dfrac{1}{(-\Phi)^{3k}}\qquad\qquad& (-\Phi\gg 1)\,,
\end{array}\right.
\ee
compatible with (\ref{Zp-symm}), where
\be
\bigl(\,p\,\bigr)_{\!\ell}\,\equiv\,\bigl.{\Gamma(p+\ell\,)}\bigr/{\Gamma(p)}\,=\,p(p+1)(p+2)\cdots(p+\ell-1)
\ee
 is Pochhammer's Symbol 
\citep[see][http:/$\!$/dlmf.nist.gov/5.2.E5]{AS10}. We substitute (\ref{Wp-asym}) into (\ref{W-Zp-again}) to obtain
\bse
\label{VPhi-asym}
\begin{align}
V(x,z;p)\sim&
\left\{\begin{array}{ll}
\!\!-\,\dfrac{\sqrt2(p-1)}{(2x)^p}\,\overset{\infty}{\underset{k=0}{\ds\sum}}\dfrac{(-1)^k\bigl(\,p\,\bigr)_{\!6k}}{2^{2k}(2k)!}\dfrac{z^{2k}}{x^{6k}}\!\!\quad &\bigl(x\gg z^{1/3}\bigr)\,,\\[1.5em]
\!\!-\,\dfrac{\sqrt2(p-1)}{(-2x)^p}\,\overset{\infty}{\underset{k=0}{\ds\sum}}\dfrac{\cos\bigl[\bigl(p+\tfrac12 k\bigr)\pi\bigr]\bigl(\,p\,\bigr)_{\!3k}}{2^{k}(k)!}\dfrac{z^{k}}{(-x)^{3k}}\!\!\quad& \bigl(-x\gg z^{1/3}\bigr),\!\!\!\!
\end{array}\right.\\[1.0em]
\Psi(x,z;p)\sim&
\left\{\begin{array}{ll}
\!\!\dfrac{1}{\sqrt2(2x)^{p-1}}\,\overset{\infty}{\underset{k=0}{\ds\sum}}\dfrac{(-1)^k\bigl(\,p-1\,\bigr)_{\!6k+3}}{2^{2k+1}(2k+1)!}\dfrac{z^{2k+1}}{x^{6k+1}}& \bigl(x\gg z^{1/3}\bigr)\,,\\[1.5em]
\!\!\dfrac{1}{\sqrt2(-2x)^{p-1}}\,\overset{\infty}{\underset{k=0}{\ds\sum}}\dfrac{\sin\bigl[\bigl(p+\tfrac12 k\bigr)\pi\bigr]\bigl(p-1\bigr)_{\!3k}}{2^{k}k!}\dfrac{z^{k}}{(-x)^{3k}}& \bigl(-x\gg z^{1/3}\bigr), \!\!\!\!
\end{array}\right.\\[-0.5em]
\intertext{while also helpful is the form}
\WS(x,z;p)\sim&-\,\dfrac{\sqrt2(p-1)}{(-2x)^p}\,\overset{\infty}{\underset{k=0}{\ds\sum}}\dfrac{\exp\bigl[-\iR\bigl(p+\tfrac12 k\bigr)\pi\bigr]\bigl(\,p\,\bigr)_{\!3k}}{2^{k}(k)!}\dfrac{z^{k}}{(-x)^{3k}}\quad \bigl(-x\gg z^{1/3}\bigr).\!\!\!\!
\end{align}
\ese

We note that for $x\gg z^{1/3}$, the series for $\partial V/\partial z$ and $\Psi$ determined by (\ref{VPhi-asym}$a$,$b$) respectively involve odd powers of $z$ (i.e.~$z$, $z^3$, $z^5$, $\cdots$) so that at $z=0$ we have $(\partial V/\partial z)(x,z;p)=0$ and $\Psi(x,z;p)=0$ on $x>0$. As the series expansions for $-x\gg z^{1/3}$ show, this vanishing is due the absence of even powers of $z$ (i.e.~$z^0\,(=1)$, $z^2$, $z^2$, $\cdots$). Specifically the $z^0$ terms lead to non-zero values. The absence of the even powers was guaranteed by normalising $\WS$ (see (\ref{W-Zp-again})) with a constant $\aS(p)$ (see (\ref{W-Zp}$b$)) which is real.

For integer $p (\ge 1)$, separate expansions for negative $x$ are no longer needed and we may simply express  (\ref{VPhi-asym}$c$) for both signs in the form
\be
\label{VPhi-asym-integer}
\dfrac{\WS(x,z;p)}{p-1}\sim-\,\dfrac{\sqrt2}{(2x)^p}\,\overset{\infty}{\underset{k=0}{\ds\sum}}\,\dfrac{\iR^k\bigl(\,p\,\bigr)_{\!3k}}{2^{k}(k)!}\dfrac{z^{k}}{x^{3k}}\qquad\qquad \bigl(|x|\gg z^{1/3}\bigr).
\ee



\begin{thebibliography}{} 
  \bibitem[Abramowitz \& Stegun (2010)]{AS10}
    {\sc Abramowitz, M. \& Stegun, I. A.} 2010
    {\em NIST Handbook of Mathematical Functions}. (ed.\ F.W.J.~Olver, D.W.~Lozier, R.F.~Boisvert and C.W.~Clark), CUP, NY (Available online http:/$\!$/dlmf.nist.gov/)
  \bibitem[Aurnou {\itshape et al.} (2003)]{Aetal03}
   {\sc Aurnou,~J.,~Andreadis,~S.,~Zhu,~L.~\& Olson,~P.} 2003
   Experiments on convection in Earth’s core tangent cylinder.
   {\em Earth Planet.~Sci.~Lett.} {\bf 212}, 119--134.
   \bibitem[Dormy \& Soward (2007)]{DS07}  
    {\sc Dormy,~E.~\& Soward,~A.M.} 2007
    {\em Mathematical aspects of natural dynamos}
    in {\em The fluid mechanics of astrophysics and geophysics} (series ed.~A.~M.~Soward \& M.~Ghil), 
    Vol.~13, pp.~120--136. Chapman \& Hall.
  \bibitem[Dormy {\itshape et al.} (1998)]{DCJ98}
    {\sc Dormy,~E.,~Cardin,~ P.~\& Jault,~D.} 1998
   MHD flow in a slightly differentially rotating spherical shell, with conducting inner core, in a dipolar magnetic field. {\em Earth Planet.~Sci.~Lett.} {\bf 160}, 15--39. 
  \bibitem[Dormy {\itshape et al.}(2002)]{DJS02}
    {\sc Dormy,~E., Jault,~D.~\& Soward,~A.M.} 2002
    A super-rotating shear layer in magnetohydrodynamic spherical Couette flow.  {\em J. Fluid Mech.} {\bf 452}, 263--291.
   \bibitem[Dormy {\itshape et al.} (2004)]{Detal04}
    {\sc Dormy,~E., Soward,~A.M.,~Jones,~C.A.,~Jault,~D.~\& Cardin,~ P.} 2004  
    The onset of thermal convection in rotating spherical shells. {\em J. Fluid Mech.} {\bf 501}, 43--70.
   \bibitem[Dowden (1972)]{D72}
    {\sc Dowden,~J.M.} 1972
    An equatorial boundary layer. {\em J. Fluid Mech.} {\bf 56}, 193--200.
  \bibitem[Gill (1971)]{G71}
    {\sc Gill,~A.E.} 1971
   The equatorial current in a homogeneous ocean. {\em Deep Sea Res.} {\bf 18}, 421--431.
  \bibitem[Glatzmaier (2014)]{G14}
    {\sc Glatzmaier,~G.A.} 2014
    {\em Introduction to Modeling Convection in Planets and Stars}. Princeton University Press, Princeton and Oxford.
  \bibitem[Gradshteyn and Ryzhik (2007)]{GR07}
    {\sc Gradshteyn,~I.S.~\& Ryzhik,~ I.M.}
    {\em Table of Integrals, Series, and Products} (ed.~ A.~Jeffrey \& D.~Zwillinger). Elsevier
  \bibitem[Greenspan (1968)]{G68}
    {\sc Greenspan,~H.P.} 1968
    {\em The Theory of Rotating Fluids}. Cambridge University Press, U.K.
  \bibitem[Hide \& Titman (1967)]{HT67}
    {\sc Hide,~R.~\& Titman,~C.W.} 1967
    Detached shear layers in a rotating fluid.
    {\em J.~Fluid Mech.} {\bf 29}, 39--60. 
  \bibitem[Hollerbach (2003)]{H03}
    {\sc Hollerbach,~R.} 2003
     Instabilities of the Stewartson layer. Part 1. The dependence on the sign of $Ro$. 
    {\em J.~Fluid Mech.} {\bf 492}, 289--302.
  \bibitem[Hollerbach \& Proctor (1993)]{HP93}
   {\sc Hollerbach,~R.~\& Proctor,~M.R.E.} 1993 
    Non-axisymmetric shear layers in a rotating spherical shell. In {\em Solar and Planetary Dynamos} Eds. M.R.E.~Proctor \& A.D.~Gilbert, pp.~145--152. Cambridge University Press.
  \bibitem[Hollerbach {\itshape et al.} (2004)]{Hetal04}
   {\sc Hollerbach,~R., Futterer,~B., More,~T. \& Egbers,~C.} 2004
    Instabilities of the Stewartson layer. Part 2. Supercritical mode transitions.
    {\em Theor.~Comp.~Fluid Dyn.} {\bf 18}, 197--204.
  \bibitem[Kerswell (1995)]{K95}
    {\sc Kerswell,~R.R.} 1995
    On the internal shear layers spawned by the critical regions in oscillatory Ekman boundary layers.
    {\em J.~Fluid Mech.} {\bf 298}, 311--325.
  \bibitem[Koch  {\itshape et al.} (2013)]{Ketal13}
    {\sc Koch,~S.,~Egbers,~C.~\& Hollerbach,~R.} 2013
    Inertial waves in a spherical shell induced by librations of the inner sphere: experimental and numerical results.
    {\em Fluid Dyn.~Res.} {\bf 45}, 035504 (19pp).
  \bibitem[Le~Bars {\itshape et al.} (2015)]{LBetal15}
    {\sc Le Bars,~M., C\'ebron, D.~\& Le Gal,~P.} 2015
    Flows driven by libration, precession, and tides.
    {\em Annu.~Rev.~Fluid Mech.} {\bf 47} 163--193.
  \bibitem[Livermore \& Hollerbach (2012)]{LH12}
    {\sc Livermore,~P.W.~\& Hollerbach,~R.} 2012
    Successive elimination of shear layers by a hierarchy of constraints in inviscid spherical-shell flows.
    {\em J.~Math.~Phys.} {\bf 53}, 073104.
   \bibitem[Moore \& Saffman (1969)]{MS69}
    {\sc Moore,~D.W.~\& Saffman,~P.G.} 1969
    The structure of free vertical shear layers in a rotating fluid and the motion produced by a slowly rising body.
    {\em Phil.~Trans.~R.~Soc.~Lond.}~A.{\bf 264} {\rm{(1156)}} 597--634
  \bibitem[Pedlosky (1979)]{P79}
    {\sc Pedlosky, J.} 1979
    {\em Geophysical Fluid Dynamics}. Springer-Verlag New York, USA
  \bibitem[Philander (1971)]{P71}
    {\sc Philander, S.G.H.} 1971
    On the flow properties of a fluid between concentric spheres
    form. {\em J. Fluid Mech.} {\bf 47}, 799--809.
   \bibitem[Proudman (1956)]{P56}
    {\sc Proudman, I.} 1956
    The almost-rigid rotation of viscous fluid between concentric spheres. {\em J. Fluid Mech.} {\bf 1}, 505--516.
  \bibitem[Roberts (1967)]{R67}
    {\sc  Roberts,~P.H.} 1967
     Singularities of Hartmann layers. {\em Proc.~R.~Soc.~Lond.}~A{\bf 300}, 94--107.
  \bibitem[Roberts (2003)]{R03}  
    {\sc Roberts,~P.H.} 2003
    {\em Magnetohydrodynamics and the Earth's Core: Selected works of Paul Roberts}
    in {\em The fluid mechanics of astrophysics and geophysics} (series ed.~ A.~M.~Soward \& M.~Ghil), 
    Vol.~10, Taylor \& Francis, London, New York. 
  \bibitem[Roberts \& King (2013)]{RK13}
    {\sc  Roberts,~P.H.~\& King,~E.M.} 2013
    On the genesis of the Earth's magnetism.
    {\em Rep.~Prog.~Phys.} {\bf 76}, 096801 (55pp.).
  \bibitem[Roberts \& Stewartson (1963)]{RS63}
    {\sc  Roberts,~P.H.~\& Stewartson,~K.} 1963
    On the stability of a Maclaurin spheroid with small viscocity.
    {\em Astrophys.~J.} {\bf 137}, 777--790.
  \bibitem[Rousset (2007)]{R07}
    {\sc Rousset,~R.} {2007}
    Asymptotic behavior of geophysical fluids in highly rotating balls.
    {\em Z.~angew.~Math.~Phys.} {\bf 58}, 53--67.
  \bibitem[Sakuraba \& Roberts (2009)]{SR09}
    {\sc  Sakuraba, A.~\& Roberts,~P.H.} 2009
    Generation of a strong magnetic field using uniform heat flux at the surface of the core.
    {\em Nature Geoscience} {\bf 2}, 802--805.
   \bibitem[Stewartson (1957)]{S57}
    {\sc Stewartson,~K.} 1957
    On almost rigid rotations. {\em J.~Fluid Mech.} {\bf 3}, 17--26.
   \bibitem[Stewartson (1966)]{S66}
    {\sc Stewartson,~K.} 1966
    On almost rigid rotations. Part 2 {\em J.~Fluid Mech.} {\bf 26}, 131--144.
   \bibitem[Stewartson~\& Rickard (1970)]{SR70}
    {\sc Stewartson,~K.~\& Rickard,~J.R.A.} 1970
    Pathological oscillations of a rotating fluid. {\em J.~Fluid Mech.} {\bf 35}, 759--73.
   \bibitem[Taylor (1963)]{T63}
    {\sc Taylor,~J.B.} 1963 
    The magnetohydrodynamics of a rotating fluid and the Earth's dynamo problem.
 {\em Proc.~R.~Soc.~Lond.} {\bf A274}, 27--283.
   \bibitem[van~de~Vooren (1993)]{vdV93}
    {\sc van~de~Vooren,~A.I.} 1993
    The connection between Ekman and Stewartson layers for a rotating disk. 
    {\em J.~Eng.~Math.} {\bf 27}, 189-207.
   \bibitem[Vo {\itshape et al.} (2015)]{Vetal15}
    {\sc Vo,~T.,~Montabone,~L.,~Read,~P.L.~\& Sheard,~G.J.} 2015
    Non-axisymmetric flows in a differential-disk rotating system.
    {\em J.~Fluid Mech.} {\bf 775}, 349--386.
   \bibitem[Wei \& Hollerbach (2008)]{WH08}
    {\sc Wei,~X.~\& Hollerbach, R.} 2008
    Instabilities of Shercliffe and Stewartson layers in spherical Couette flow.
    {\em Phy.~Rev.~E} {\bf 78}, 026309, pp.~1--5.

\end{thebibliography}

\end{document}